\documentclass[seceq]{ptptex}
\usepackage{latexsym}
\usepackage{mathptmx}
\usepackage{amssymb}
\usepackage{bm}
\usepackage{amsmath}
\usepackage{graphicx}
\usepackage{dcolumn}
\usepackage{array}
\usepackage{bm}

\title{The auxiliary field method in quantum mechanics}

\author{Bernard \textsc{Silvestre-Brac}$^{1,}$\footnote{E-mail: silvestre@lpsc.in2p3.fr}, 
Claude \textsc{Semay}$^{2,}$\footnote{F.R.S.-FNRS Senior Research Associate. E-mail: claude.semay@umons.ac.be},  
Fabien \textsc{Buisseret}$^{2,}$\footnote{F.R.S.-FNRS Postdoctoral Researcher. E-mail: fabien.buisseret@umons.ac.be}
}

\inst{$^{1}$LPSC Universit\'{e} Joseph Fourier, Grenoble 1,
CNRS/IN2P3, Institut Polytechnique de Grenoble, 53 Avenue des 
Martyrs, 38026 Grenoble-Cedex, France \\
$^{2}$Service de Physique Nucl\'{e}aire et Subnucl\'{e}aire,
Universit\'{e} de Mons - UMONS, 
Acad\'{e}mie universitaire Wallonie-Bruxelles, 
Place du Parc 20, 7000 Mons, Belgium 
}

\abst{
The auxiliary field method is a new technique to obtain closed formulae for the solutions of eigenequations in quantum mechanics. The idea is to replace a Hamiltonian $H$ for which analytical solutions are not known by another one $\tilde H$, including one or more auxiliary fields, for which they are known. For instance, a potential $V(r)$ not solvable is replaced by another one $P(r)$ more familiar, or a semirelativistic kinetic part is replaced by an equivalent nonrelativistic one. If the auxiliary fields are eliminated by an extremization procedure, the Hamiltonian $\tilde H$ reduces to Hamiltonian $H$. The approximation comes from the replacement of these fields by pure real constants. The approximant solutions for $H$, eigenvalues and eigenfunctions, are then obtained by the solutions of $\tilde H$ in which the auxiliary parameters are eliminated by an extremization procedure for the eigenenergies, which takes the form of a transcendental equation to solve. If $H=T(\bm p)+V(r)$ and if $P(r)$ is a power law, the approximate eigenvalues can be written $T(p_0)+V(r_0)$, where the mean impulsion $p_0$ is a function of the mean distance $r_0$ and where $r_0$ is determined by an equation which is linked to the generalized virial theorem. The general properties of the method are studied and the connections with the envelope theory presented. Its mean field and (anti)variational characters are also discussed. This method is first applied to nonrelativistic and semirelativistic two-body systems, with a great variety of potentials (sum of power laws, logarithm, exponential, square root). Closed formulae are produced for energies, eigenstates, various observables and critical constants (when it is relevant), with sometimes a very good accuracy. The method is then used to solve nonrelativistic and semirelativistic many-body systems with one-body and two-body interactions. For such cases, analytical solutions can only be obtained for systems of identical particles, but several systems of interest for atomic and hadronic physics are studied. General results concerning the many-body critical constants are presented, as well as duality relations existing between approximate and exact eigenvalues.
}
\notypesetlogo
\begin{document}
\maketitle

\tableofcontents

\section{Introduction: Analytical methods in quantum mechanics}
\label{sec:intro}

The aim of this work is to present the auxiliary field method (AFM), which is a new and remarkably simple method to find analytical approximate solutions of eigenequations corresponding to Hamiltonians admitting bound states. The quest for exactly solvable problems in quantum physics is actually as old as quantum mechanics itself, and it is worth making general comments about this topic before focusing on the auxiliary field method. 

The most famous systems for which the Schr\"odinger equation admits analytical solutions are probably the harmonic oscillator and the hydrogen atom, \textit{i.e.} Hamiltonians of the form ${\bm{p}^2}+r^2$ and ${\bm{p}^2}-1/r$ respectively, where $\bm p$ and $\bm r$ are conjugate variables. The spectrum of the one-dimensional harmonic oscillator is actually known since 1925 thanks to Heisenberg's pioneering work \cite{histo2}, while the spectrum of the nonrelativistic hydrogen atom was found in Schr\"odinger's paper \cite{histo1} as a first and crucial test of his celebrated equation. Notice that the same problem was solved the same year by Pauli. \cite{histo3}

Since those early results, there has been a considerable amount of works devoted to the computation of analytical solutions of the Schr\"odinger equation, especially in bound state problems. The Schr\"odinger equation has been found to be exactly solvable in many one-dimensional cases: Dirac comb, exponential and linear potentials, \ldots\ However, only a few analytical three-dimensional solutions are known for any value of the angular momentum (the $S$-wave channel is very similar to a simpler one-dimensional equation). Besides the aforementioned harmonic oscillator and Coulomb cases, one can mention the Kratzer potential, the particle in a box and the symmetric top, which are of quite limited use in modeling realistic quantum systems. An exhaustive discussion of quantum problems admitting analytical solutions would be out of the scope of the present introduction; therefore we refer the interested reader to the textbooks \cite{QM0,QM1,NB1,QM2,QES,flug}, in which a wide range of problems is covered.   

When exact analytical solutions cannot be found, physicists can either resort to numerical computations or to methods leading to approximate analytical solutions. Even if the computational power that we have at our disposal nowadays allows to solve accurately nearly every eigenequation for few body systems, finding approximate analytical formulae is always useful in physics, not only because of the intrinsic mathematical interest of such a task, but also in view of obtaining informations about the dependence of the observables on the parameters of a model and on the quantum numbers of the states. Calculating an analytical expression being always much less time consuming than solving numerically the corresponding eigenequation, the use of closed formulae can be a great advantage when one tries to fit the parameters of a given model to some experimental data, or even to interpret new data. We find worth mentioning here three approximation schemes that are all ``textbook material" and able to lead to analytic results: the WKB or semiclassical approximation, the variational method, and the perturbation theory. 

The principle of the WKB approximation has been found by Wentzel, Kramers and Brillouin around 1926~\cite{WKB}, but it is also known that Jeffreys brought important contributions to the field in 1924. \cite{J}\ It basically consists in a Taylor expansion of the Hamiltonian and the wavefunction in powers of $\hbar$. At dominant order for a one-dimensional problem, such an expansion leads for example to the Bohr-Sommerfeld quantization rule $\int^{a(E)}_{b(E)} p(E,x)\, dx=\pi\hbar(n+1/2)$, where $p$ is the classical momentum expressed as a function of the energy and of the position $x$, and where $a(E)$ and $b(E)$ are the turning points of the classical trajectory. The energy spectrum can then be extracted from this integral and is generally accurate, but the explicit expression for $E(n)$ is not always possible to find. One has instead often to numerically compute the energy from the Bohr-Sommerfeld quantization rule. Further informations about the semiclassical approximation can be found in \textit{e.g.} Refs.~\citen{QM0,QM2}.

While the WKB method is particularly efficient for highly excited states, where $\hbar$ clearly becomes a small parameter, the variational method due to Ritz \cite{varia} allows to gain relevant informations about the ground state (for a modern presentation
see Ref.~\citen{reed78}). Indeed, let us consider a Hamiltonian $H$ and a normalized arbitrary function $\psi$ belonging to $L^2$ space. Then, the Ritz theorem states that $E_0\leq \langle\psi|H|\psi\rangle$, where $E_0$ is the exact ground-state energy. Provided that $\psi$ is chosen such that $\langle\psi|H|\psi\rangle$ is analytical, one can then obtain an analytical upper bound on the ground-state energy. Typically, one chooses a gaussian- or exponential-type wavefunction; many examples of analytical upper bounds can be found in Ref.~\citen{flug}. Notice also the MacDonald's theorem stating that, given a set of $n$ orthonormalized functions $\psi_i$ belonging to $L^2$ space, the eigenvalues of the matrix $\langle\psi_i|H|\psi_j\rangle$ provide upper bounds on the $n$ first eigenenergies of $H$. \cite{macd33}\ However, analytical results can hardly be obtained from such a generalization due to the complexity of the underlying calculations.

A last technique to be mentioned is the perturbation theory. \cite{pertu}\ Let us consider that the Schr\"odinger equation with a potential $V$ is exactly solvable. Then the Schr\"odinger equation with potential $V+\xi\, w$ can be solved thanks to a Taylor expansion in $\xi$ provided that this parameter is small. At first order in particular, the computation of $\xi \langle w\rangle $ with the lowest-order states provides the dominant correction to the whole energy spectrum. Again, a wide range of quantum problems can be studied by using perturbation theory, and we refer to \textit{e.g.} \cite{QM1,flug} for explicit examples in which $\langle w\rangle $ is analytical.

Those three methods being presented, it appears that a major challenge remains: To find approximate analytical solutions of eigenequations for the \textit{complete} spectrum (not only the ground state as with the variational method) \textit{without} using any Taylor expansion as done with the WKB method or the perturbation theory. The AFM, that will be presented in detail in the rest of this work, is an attempt to address that issue.    

\section{Generalities on the auxiliary field method}
\label{sec:genAFM}

\subsection{Historical aspects}
\label{sec:hist}

Auxiliary fields, also known as einbein fields, are used in various domains of
theoretical physics. Historically, they have been introduced in order to get
rid of the square roots typically appearing in relativistic
Lagrangians. \cite{dir66,brink77}\ The most obvious example is the free
relativistic particle, described by the Lagrangian ${\cal L}=-m\, 
\sqrt{\dot{\rm x}^2}$, where $\dot{\rm x}$ is the world-velocity. An auxiliary
field $\mu$ can be introduced in this expression so that one is led to the more
convenient form ${\cal L}(\mu)= \dot{\rm x}^2/(2\mu)+m^2\mu/2$. This last
expression is formally simpler than Lagrangian ${\cal L}$, but it is equivalent
if the equations of motion of the auxiliary field are considered. The auxiliary
field is indeed given on shell by $\mu_0=-\sqrt{\dot{\rm x}^2}/m$, and it is
readily checked that ${\cal L}(\mu_0)={\cal L}$. That feature can be seen as a
definition of an auxiliary field: It is a field whose equation of motion is not
dynamical but leads to algebraic relations that allow to express it in terms of
the other degrees of freedom of the problem. 

Besides pointlike particles, auxiliary fields have soon become a cornerstone
of string theory because of the systematic replacement of the Nambu-Goto
Lagrangian ${\cal L}_{NG}=-a\int d\theta\sqrt{(\dot{\rm{w}}\rm{w}')^{2}-
\dot{\rm{w}}^{2}{\rm{w}'}^{2}}$, in which an unconvenient square root appears,
by the Polyakov Lagrangian ${\cal L}_P=-(a/2)\int d\theta\ \sqrt{-\det \; \rm{h}}\  \, 
h^{bc}\, \partial_{b}w^{\mu}\, \partial_{c}w_{\mu}$, where the induced metric
$h^{bc}$ is made of auxiliary fields. \cite{deser76,howe77,polya81,str}\  In
connection with string theory, supersymmetric field theories, linking bosons
and fermions through supersymmetric transformations, also generally demand the
introduction of extra fields -- auxiliary fields -- in order to close the
supersymmetric algebra. More information concerning the building of
\textit{e.g.} consistent supergravity theories can be found for example 
in Refs.~\citen{ferrara78,vannie81,berg82,susy}.  

More generally, auxiliary fields have become widely used in field theory.
Indeed, they allow to replace kinetic terms of the form $\sqrt{\dot{\rm x}^2}$
by a formally nonrelativistic term such as $\dot{\rm x}^2/(2\mu)$. Computations
in the path integral formulation of field theories are then simplified because
one is then formally led to Gaussian path integrals, about which analytical
results are known. \cite{kash97,schmidt93}\ In quantum chromodynamics (QCD) 
for example, the introduction of auxiliary fields has
allowed to gain relevant informations about confinement in hadrons within the 
framework of potential models \cite{simo89a,simo89b} 
and in particular to give support to the phenomenological QCD
string model. \cite{morgunov99,kalash00,sema04}\ It is worth mentioning that
renormalization problems \cite{jako08} as well as numerical computations in
$N-$body systems \cite{honma95,nakada97} can also be addressed using auxiliary
fields as a computational tool.

A particular interest for our purpose is to come back to effective models of QCD.
It is known that, in the strong coupling limit, the Wilson loop formulation of QCD
supports a linear confinement between a heavy quark and antiquark. \cite{Wils}\ That
result can be generalized to light quarks also; it then appears that the effective
confining interaction in mesonic systems can be modeled by a Nambu-Goto string
linking the quark with the antiquark \cite{QCDstring2}, whose dominant contribution
is indeed a linear potential of type $a \, r$. An idea that is already present in
Refs.~\citen{simo89a,simo89b} is to introduce an additional auxiliary field, $\nu$,
in a way similar to what is done for relativistic kinetic terms: $a\, r$ can be
replaced by $a^2r^2/(2\nu)+\nu/2$, formally reducing the linear potential to a
harmonic oscillator, for which analytical solutions are known. Of course the
remaining auxiliary field has eventually to be eliminated, but analytical approximate mass
formulae for mesons and baryons can be found following this philosophy. The
question underlying the present report is nothing else but an attempt to
generalize such a procedure, namely: Could any arbitrary potential appearing
in an eigenequation be approximated by an expression involving
an auxiliary field and another potential for which an analytical solution is
known? As we shall see in the following, an affirmative answer can be given
to that question, leading to a method allowing to compute approximate
analytical energy formulae of various eigenequations. Moreover the method can
be generalized to treat problems involving more than 1 or 2 particles.

\subsection{The auxiliary field method}
\label{sec:AFM}

The most general form that we consider for the eigenvalue equation of a one- or two-body problem in
first quantized quantum mechanics is 
\begin{equation}
\label{eq:formpartH}
H|\psi\rangle = \left[T(\bm{p}) + V(r)\right]|\psi\rangle = E |\psi\rangle.
\end{equation}
The potential $V$ in the Hamiltonian $H$ depends on the variable $r=|\bm{r}|$ which is the
distance from  the center of forces in a one-body problem and the relative
distance between the two particles in a two-body problem. The kinetic part $T$ depends on
the momentum operator $\bm{p}$ which is the conjugate variable of $\bm r$. 
Practically, we will consider a nonrelativistic form $\bm{p}^2/(2m)$
($m$ is the particle mass for a one-body problem or the reduced mass of a two-body system) 
as in the Schr\"odinger equation
and a semirelativistic one $\sigma \sqrt{\bm{p}^2+m^2}$ 
($\sigma=1$ for one-body problems and $\sigma=2$ for two identical particles)
as in a spinless Salpeter equation. 

An exact analytical expression for \textbf{all the eigenvalues} is known
explicitly only for very specific potentials $P(r)$ with a nonrelativistic
kinetic parts: the
quadratic interaction $P(r)=r^2$ (harmonic oscillator) and the Coulomb potential
$P(r)=-1/r$ (hydrogen-like system) are the most familiar for a practical use. An exact
expression is known for a number of other potentials, but only for $S$-waves (see
Ref.~\citen{flug} for a more detailed discussion) or for one-dimensional problems.

Our goal is the search for approximate analytical solutions for the Hamiltonian 
of type (\ref{eq:formpartH}), relying on Hamiltonians for which solutions are well-known. In
other words, we assume that we are able to obtain an analytical solution
for the equation
\begin{equation}
\label{eq:schrodinit}
h(a) | a \rangle = \left[ T(\bm{p})+ a P(r) \right]
\left| a \right\rangle = e(a) \left| a \right\rangle,
\end{equation}
in which, at this stage, $a$ is a real parameter.
This method was formulated and applied for the first time in Ref.~\citen{bsb08a}.

We summarize here the principle of the method introducing an auxiliary field, which is
\emph{a priori} an operator. 
It consists in four steps:
\begin{enumerate}
\item We calculate the function $K(r)$ (the prime denotes the derivative
with respect to the argument of the function) 
\begin{equation}
\label{eq:funcK}
K(r) = \frac{V'(r)}{P'(r)}.
\end{equation}
\item We denote by $J=K^{-1}$ the inverse function of $K$. Thus, one has
\begin{equation}
\label{eq:definvK}
K(r) = z \quad \textrm{and} \quad J(z) = r.
\end{equation}
Since both $V(r)$ and $P(r)$ do exhibit an analytical form, the same property
holds for $K(r)$. But it is by no means sure that $J(z)$ can be expressed
analytically. We will see in the following that an explicit analytical
expression for $J$ is not necessary to write down the basic equations.
\item We introduce the ``bridge function" $B$ defined by
\begin{equation}
\label{eq:bridg}
B(y)=V(J(y)) - y P(J(y)).
\end{equation}
The name of this function comes from the fact that it makes a bridge between the
potential $P(r)$ for which an analytical expression for the energies is known
and the potential $V(r)$ for which the corresponding analytical expression
is \emph{a priori} not known.
Lastly, it is interesting to define the AFM potential $\tilde{V}$ through
\begin{equation}
\label{eq:AFMV}
\tilde{V}(r,y)=yP(r) + B(y).
\end{equation}
It depends not only on the position operator as the original potential $V(r)$,
but also on the $y$ variable, which is undetermined at this stage. The AFM
Hamiltonian is now defined as
\begin{equation}
\label{eq:AFMH}
\tilde{H}(y) = T(\bm{p})+\tilde{V}(r,y) = h(y)+B(y).
\end{equation}
The construction of $\tilde{H}$ seems rather artificial and one can wonder
what could be the justification of such a procedure. In fact, there is no
mystery.

Let us suppose that the $y$ variable is an operator denoted $\hat{\nu}$. It
plays the role of an auxiliary field, not present in the original Hamiltonian
$H$, but which is an essential ingredient of the AFM Hamiltonian $\tilde{H}$.
Among the infinite number of possibilities for the auxiliary field $\hat{\nu}$,
let us choose a very specific one $\hat{\nu}_0$ defined by
\begin{equation}
\label{eq:defchaux}
\hat{\nu}_0=K(r).
\end{equation}
A more explicit notation would be $\hat{\nu}_0(r)$ indicating clearly that it is
an operator depending on the position only.
The very important property is that $\hat{\nu}_0$, coming from (\ref{eq:defchaux}),
is an extremum the AFM Hamiltonian, \emph{i.e}.
\begin{equation}
\label{eq:defchaux2}
\left. \frac{\delta \tilde{H}(\hat{\nu})}{\delta \hat{\nu}} \right |_{\hat{\nu} =
\hat{\nu}_0} = 0. 
\end{equation}

Moreover, one has the additional property that the value of the AFM Hamiltonian
taken for this operator coincides with the original Hamiltonian $H$
\begin{equation}
\label{eq:valHnu0}
\tilde{H}(\hat{\nu}_0)= H. 
\end{equation}
These properties are easy to show and follow from the definitions (\ref{eq:funcK}),
(\ref{eq:definvK}) and (\ref{eq:defchaux}). Thus, considering the auxiliary field
as an operator and affecting it the value given by (\ref{eq:defchaux}) is just an
alternative method to solve the original problem.

\item The philosophy of the AFM is to consider the $y$ variable, no longer
as an operator $\hat{\nu}$, but as a pure real number $\nu$. In this case
the AFM potential $\tilde{V}(r,\nu)$ is of the form $\nu P(r)+B(\nu)$ where
$\nu$ and $B(\nu)$ are no longer operators but must be considered as arbitrary
constants. Taking into account (\ref{eq:schrodinit}),
the eigenvalues of $\tilde{H}(\nu)$ are
\begin{equation}
\label{eq:enerprop}
E(\nu)=e(\nu)+ B(\nu),
\end{equation}
where $e(\nu)$ are the eigenvalues of $h(\nu)$ which are supposed to be
known whatever the radial and orbital quantum numbers $n,l$. Provided that
the $J$ function is calculable, the same property is true for $E(\nu)$.
Then, we determine the value $\nu_0$ that extremizes $E(\nu)$:
\begin{equation}
\label{eq:detnu0}
\left. \frac{\partial E(\nu)}{\partial \nu} \right|_{\nu=\nu_0}=0.
\end{equation}
We propose to consider $E(\nu_0)$ as the approximate form of the
exact energy $E$ of the Hamiltonian $H$
\begin{equation}
\label{eq:energAFM}
E \approx E_{\textrm{AFM}} = E(\nu_0).
\end{equation}
In order to obtain an analytical expression for the eigenvalues
$E_{\textrm{AFM}}$, we must then fulfill a second necessary condition: to be able
to determine $\nu_0$ and, then, $E(\nu_0)$ in an analytical way. Denoting $|\nu\rangle$
the eigenstate of $\tilde H(\nu)$ corresponding to the eigenvalue $E(\nu)$, we consider
$|\nu_0\rangle$ as an approximation of the corresponding genuine eigenstate of $H$.

\end{enumerate}

Presented as such, this procedure appears to be an empirical recipe. 
Arguments given in the third item show that this prescription makes sense
and that the AFM must be considered in essence as a mean field approximation.
This point will be developed more deeply below.

Another interesting property concerning the AFM potential~(\ref{eq:AFMV}) is
the following. Let us denote by $r_0$ the value of the radius defined by
\begin{equation}
\label{eq:defr0}
\nu_0 = K(r_0) \quad \textrm{and} \quad r_0 = J(\nu_0).
\end{equation}
It is just a matter of simple calculation to verify that the AFM potential
taken for the special value $y=\nu_0$ can be recast as
\begin{equation}
\label{eq:AFMpotnu0}
\tilde{V}(r,\nu_0) = \nu_0 [P(r) - P(r_0)] + V(r_0).
\end{equation}
Owing to the definition of the $K$ function~(\ref{eq:funcK}), one can easily
show that the AFM potential coincides both to the exact potential and its
derivative at the point $r=r_0$, namely
\begin{equation}
\label{eq:pottang}
\tilde{V}(r_0,\nu_0) = V(r_0) \quad \textrm{and} \quad \tilde{V}'(r_0,\nu_0) = V'(r_0).
\end{equation}
In other words, at this particular point $r_0$, the AFM potential and the exact
potential are tangent curves. This property will be exploited in Sect.~\ref{sec:ulb} in connection
with the envelope theory (see Appendix~\ref{sec:et}). 

Let us note by $\left|\nu\right\rangle$ an eigenstate of $\tilde H(\nu)$.  
The Hellmann-Feynman theorem \cite{feyn} states that 
\begin{equation}
\frac{\partial E(\nu)}{\partial \nu}=
\left\langle \nu\left| \frac{\partial \tilde H(\nu)}{\partial \nu} \right|\nu\right\rangle.
\end{equation}
Using this relation, one can show that \cite{bsb08a}
\begin{equation}
\label{pr0}
\left\langle \nu_0\right| P(r) \left|\nu_0\right\rangle = P(r_0).
\end{equation}
This means that $r_0$ is a kind of ``average point" for the potential $P(r)$. That is why it
will be often called ``the mean radius" in the following.
Using this last relation with the definitions above, we get
\begin{equation}\label{Zrho}
\langle \nu_0 | Z(\hat \nu_0) | \nu_0 \rangle = Z(\nu_0)
\quad \textrm{with} \quad Z(x)=P(J(x)).
\end{equation}
So, our method can actually be considered as a ``mean field approximation" with respect to a
particular auxiliary field which is introduced to simplify the calculations: $\nu_0$ is the
mean value of the operator $\hat \nu_0=K(r)$ through a function $Z$ which can be quite simple.
For example, $Z(x)=x$ if $V(x)=P(x)^2/2+V_0$ where $V_0$ is a constant. 

It is in the step of passing from an operator to a constant interpretation for
the auxiliary field that lies the approximation. It is clear
that the quality of the approximate results strongly depends on the choice of
the function $P(r)$. The cleverness of the physicist relies in his ability to
guess a form for $P(r)$ as close as possible of $V(r)$ while leading to a
manageable form of the corresponding eigenvalues $e(\nu)$. A big bulk
of this report is devoted to the discussion of the quality of this
approximation.

This method is completely general and \emph{a priori} valid for any potential
$V(r)$. Of course, all the four steps mentioned above can be done numerically so
that $E_{\textrm{AFM}}$ can be computed numerically. But it seems that
the corresponding numerical treatment is as heavy as solving directly the
eigenvalue equation to get lastly a poorer result. This is true if we
start with an arbitrary function $P(r)$. However we have the complete freedom
for this choice.

Let us assume that, for a kinetic part $T$, we do start with a 
potential $P$ allowing to obtain analytical
results for its eigenvalues $e(\nu)$ ($e(\nu)$ is a shorthand notation for the
most correct expression $e(\nu;n,l)$); the same is true of course for its
derivative with respect to $\nu$ : $e'(\nu)$. Using the definition of the $K$
and $J$ functions defined previously, it is easy to show that
\begin{equation}
\label{eq:derivE}
E'(\nu) = e'(\nu) - P(J(\nu)).
\end{equation}
The determination of $\nu_0$ results from the condition $E'(\nu_0) = 0$,
that is to say
\begin{equation}
\label{eq:trans1}
e'(\nu_0) = P(J(\nu_0)).
\end{equation}
It appears that obtaining the $\nu_0$ value requires only to solve a
transcendental equation. Once this value is obtained, the AFM energy is very
easy to calculate
\begin{equation}
\label{eq:Egen2c}
E_{\textrm{AFM}} =E(\nu_0)= e(\nu_0) + B(\nu_0).
\end{equation}
Thus, provided that we have an analytical expression for $e(\nu)$, the AFM
results only need solving a transcendental equation, a much easier procedure
than solving exactly the genuine eigenvalue equation.

The practical usefulness of the AFM method is to get a final solution which
is completely analytical. In order to do that, we must fulfill,
as we saw, three conditions:
\begin{enumerate}
\item to choose a $P$ function leading to analytical expression for $e(\nu)$; 
\item to be able to invert relation (\ref{eq:defchaux}) in order to have
access to the function $J$ defined by (\ref{eq:definvK})
(it appears into the $B$ function);
\item to be able to determine $\nu_0$ (through (\ref{eq:trans1})) and
to calculate the corresponding value $E(\nu_0)$ in an analytical way.
\end{enumerate}
In fact, the second condition is formally not necessary. Indeed, let us
introduce the mean radius $r_0 = J(\nu_0)$. Then $\nu_0 = K(r_0)$ so that
the bridge function is expressed as
\begin{equation}
\label{eq:bridgec}
B(\nu_0) = V(r_0) - K(r_0) P(r_0) = B(r_0),
\end{equation}
where, for simplicity, we still maintain the notation $B$: $B = V - KP$.
The transcendental equation (\ref{eq:trans1}) is transformed into
\begin{equation}
\label{eq:trans2}
e'(K(r_0)) = P(r_0),
\end{equation}
while the energy is given by
\begin{equation}
\label{eq:Egen3c}
E_{\textrm{AFM}} =E(\nu_0)= e(K(r_0)) + B(r_0).
\end{equation}
Using the expression (\ref{eq:bridgec}) for $B(r_0)$ and the definition of $e(\nu)$, 
this last formula can be
recast under the form
\begin{equation}
\label{eq:Egen4c}
E_{\textrm{AFM}} = T_0 + V(r_0),
\end{equation}
where, according to (\ref{pr0}),
\begin{equation}
\label{eq:cinr0}
T_0 = e(K(r_0)) - K(r_0) P(r_0)
= \left\langle \nu_0\right| T(\bm p) \left|\nu_0\right\rangle
\end{equation}
is an average kinetic energy.

In this new formulation, any reference to the $J$ function has disappeared
so that it is not necessary to try to get it. However, solving analytically
the transcendental equation (\ref{eq:trans2}) is more or less of the same
difficulty than getting an analytical expression for the $J$ function.
Anyhow, this new formulation leads very often to simpler practical
calculations and is preferred most of the time. 

It is important to stress the following point: if $P(r)$ is chosen to be
$V(r)$, it is trivial to check that $K(r) = 1$, $B(r)=0$, $r_0$ is
meaningless since obviously $P(r)$ and $V(r)$ are tangent curves everywhere,
and $E_{\textrm{AFM}}$ = $e(1)$ which is precisely the exact value
$E$. In this particular case, AFM recovers the exact result.
This property is sometimes used to check special results.  

\subsection{Scaling laws}
\label{sec:scalinglaws}

Scaling laws represent an important property for quantum mechanical systems.
They allow to give the expression for the eigenenergies (and wavefunctions) 
of the most general equation in terms of the corresponding eigenenergies
(and wavefunctions) of a reduced equation with less parameters.
In fact, the scaling laws are nothing else than a direct consequence of dimensional
analysis applied to the various dimensioned parameters entering the problem. 
Starting from a general Hamiltonian $H(\alpha_1, \ldots, \alpha_n)$ depending on $n$ 
dimensioned parameters $\alpha_i$, it is generally possible to write
\begin{equation}\label{scalHh}
H(\alpha_1, \ldots, \alpha_n) = \gamma \, h(\beta_1, \ldots, \beta_m),
\end{equation}
where $\gamma$ has the dimension of an energy and where $h$ is a Hamiltonian 
expressed in terms of dimensionless conjugate variables and
depending on $m < n$  dimensionless parameters $\beta_i$. 
The scaling properties of the nonrelativistic Schr\"{o}dinger equation have been
studied and used for a long time, but spinless Salpeter equations also benefit of scaling
properties. They will be explicitly described below.
In the following, in order to lighten the notations, the AFM will be generally applied
to dimensionless  Hamiltonians as in Refs.~\citen{bsb08a,bsb08b,bsb09c}. So, the exact and the
AFM solutions automatically share the same scaling properties. 

\subsection{Eigenstates and upper/lower bounds}
\label{sec:ulb}

If $E(\nu_0)$, which is an eigenvalue of $\tilde H(\nu_0)$, is an approximation of the exact energy $E$, 
the corresponding eigenstate $\left|\nu_0\right\rangle$ is an approximation of a genuine eigenstate
of $H$. The shape of the corresponding wavefunction $\left\langle \bm r | \nu_0\right\rangle$ depends
on the quantum numbers via the parameter $\nu_0$ or the mean radius $r_0$. Practically, only
nonrelativistic harmonic oscillator or nonrelativistic Coulomb wavefunctions can be used. Explicit
examples will be presented below. 

Since the potential $\tilde V(r,\nu_0)$ is tangent to the potential $V(r)$ at $r=r_0$, and since $H$
and $\tilde H(\nu_0)$ have the same kinetic part, the comparison theorem \cite{hall1992,hall2010,sema11}
(for both nonrelativistic and relativistic equations) implies that the approximation $E(\nu_0)$ is an
upper (lower) bound on the exact energy if $\tilde V(r,\nu_0) \ge V(r)$ ($\tilde V(r,\nu_0) \le V(r)$)
for all values of $r$. Equivalently, a function $g(x)$ can be defined by 
\begin{equation}\label{defg}
V(x)=g(P(x)). 
\end{equation}
It can then be shown that, if $g(x)$ is a concave (convex) function, that is if $g''(x)\le 0\ (g''(x)\ge 0)\ \forall x$, the approximation $E(\nu_0)$ is
an upper (lower) bound on the exact energy. This property has been demonstrated in the framework of
the envelope theory \cite{env0}, but can be applied as well to the AFM. \cite{afmenv}\ Several examples
will be presented below. 

The knowledge of lower and upper bounds on an eigenstate is a first technique to estimate the
accuracy of the AFM. It has also been shown that \cite{bsb08a}
\begin{equation}\label{eevv}
E(\nu_0) - \langle \nu_0 | H | \nu_0 \rangle = V(r_0) - \langle \nu_0 | V(r) | \nu_0 \rangle.
\end{equation}
The right-hand side of this equation is the difference between the value of potential $V$ computed
at the average point $r_0$ and the average of this potential for the AFM state $| \nu_0 \rangle$
considered here as trial state. In some favorable cases (the trial state is a ground state for
instance), $E \le \langle \nu_0 | H | \nu_0 \rangle$
and a bound on the error can be computed by 
\begin{equation}\label{relerr}
E(\nu_0) - E \ge V(r_0) - \langle \nu_0 | V(r) | \nu_0 \rangle.
\end{equation}
If the mean value of $V$ can be computed analytically, this constitutes a second procedure to estimate
the accuracy of the AFM. Several examples of this calculation are presented in Ref.~\citen{bsb08a}. At last,
the eigenstates of a Hamiltonian of type~(\ref{eq:formpartH}) can be solved numerically with an arbitrary
precision. So, as a third possibility, a direct comparison with the AFM results is always possible. So,
one can wonder why to use the AFM? Let us recall that the interest of this method is mainly to obtain
analytical information about the whole spectra (dependence of eigenenergies on the parameters of the
Hamiltonian and on the quantum numbers), without necessarily searching
a very high accuracy. Moreover, the AFM approximation can be extended to $N$-body
problems for which exact eigenenergies are not easily reachable, even numerically.

In some cases, it is possible to compute analytically the mean value
$E^*(\nu_0)=\langle \nu_0 | H | \nu_0 \rangle$, considering the AFM solution
$| \nu_0 \rangle$ as a trial state (see Sect.~\ref{sec:wfobserv} for two examples).
We have then
\begin{equation}\label{ulb1}
E^*(\nu_0) - E(\nu_0) = \langle \nu_0 | H - \tilde H(\nu_0)| \nu_0 \rangle =
\langle \nu_0 | V(r) - \tilde V(r,\nu_0)| \nu_0 \rangle = \Delta V.
\end{equation}

If $g(x)$ is concave, $E(\nu_0) \ge E$ and $\tilde V(r,\nu_0) \ge V(r)$. In this case,
$\Delta V \le 0$ and $E^*(\nu_0) \le E(\nu_0)$. By the Ritz theorem, we know that
$E^*(\nu_0) \ge E$ for the ground state. For this state, it is interesting to compute
the mean value $E^*(\nu^*)=\langle \nu^* | H | \nu^* \rangle$, where $| \nu^* \rangle$
is an eigenstate of $\tilde H(\nu^*)$ with $\nu^*$ determined in order to minimize
the energy $E^*(\nu^*)$. We have then $E^*(\nu^*) \le E^*(\nu_0)$ since the parameter
$\nu_0$ is fixed by the AFM computation. 

If $g(x)$ is convex, $E(\nu_0) \le E$ and $\tilde V(r,\nu_0) \le V(r)$. Thus,
$\Delta V \ge 0$ and $E^*(\nu_0) \ge E(\nu_0)$. In this case, we have also
$E^*(\nu_0) \ge E$ and $E^*(\nu^*) \le E^*(\nu_0)$ for the ground state. 

We can gather these results to obtain:
\begin{itemize}
  \item If $g(x)$ is concave or equivalently $\tilde V(r,\nu_0) \ge V(r)$:
	\begin{itemize}
	  \item For the ground state, $E(\nu_0) \ge E^*(\nu_0) \ge E^*(\nu^*) \ge E$;
	  \item For the other states, $E(\nu_0) \ge E$ and $E(\nu_0) \ge E^*(\nu_0)$.
	\end{itemize}
  \item If $g(x)$ is convex or equivalently $\tilde V(r,\nu_0) \le V(r)$:
	\begin{itemize}
	  \item For the ground state, $E^*(\nu_0) \ge E^*(\nu^*) \ge E \ge E(\nu_0)$;
	  \item For the other states, $E \ge E(\nu_0)$ and $E^*(\nu_0) \ge E(\nu_0)$.
	\end{itemize}
\end{itemize}

The results mentioned above are directly applicable if $H$ and $\tilde H(\nu_0)$ have the same kinetic part. This is always the case for nonrelativistic Hamiltonians but not for the spinless Salpeter Hamiltonians, whose treatment requires generally the replacement
of the square root operator by a nonrelativistic operator. The bounds on these kinds
of Hamiltonians are specifically studied in Sect.~\ref{sec:tbsalp}. 

\subsubsection{Extension to the form $a P(r) + V(r)$}
\label{sec:ext}

If the AFM is applicable
for some potentials $V_1(r)$ and $V_2(r)$ independently, there is no
certainty that it still applies for a potential which is their sum:
$V(r)=V_1(r)+V_2(r)$. However, the method can
be used in the case where one of the potentials $V_i$
can be identified with the basic $P$ potential. Thus, in this section we
consider a Hamiltonian of type 
\begin{equation}
\label{eq:anu0}
H_a=T(\bm{p})+a P(r)+ V(r),
\end{equation}
whose exact eigenvalues are denoted $E_a$.
The extension of AFM to potentials of this type was first presented in
Ref.~\citen{bsb08b}. One introduces an auxiliary field $\nu$ as before,
forgetting about the $a P(r)$ contribution. The first 3 steps of the
algorithm remain unchanged. Thus the $\hat{\nu}_0$ field is the same, as is
the same the function $B(\nu)$. The only difference arises in the expression
(\ref{eq:AFMH}) of $\tilde{H}$ and $h$ where $\nu P(r)$ has to be
replaced by $(a+\nu) P(r)$. As a consequence, the corresponding energy
(\ref{eq:enerprop}) has to be replaced by
\begin{equation}
\label{eq:newenerg}
E_a(\nu) = e(a+\nu) + B(\nu).
\end{equation}
$E_a(\nu)$ is an eigenvalue of Hamiltonian
\begin{equation}
\label{eq:anu2}
\tilde H_a(\nu) = h(a+\nu)+B(\nu),
\end{equation}
where $h$ is defined by (\ref{eq:schrodinit})
and where $e(a+\nu)$ is an eigenvalue of Hamiltonian $h(a+\nu)$.
An eigenstate of Hamiltonians $\tilde H_a$ and $h(a+\nu)$ is denoted 
$|a+\nu\rangle$, and we have $e(a+\nu) = \langle a+\nu|h(a+\nu) |a+\nu\rangle$.
If $\nu_0$ is the value of $\nu$ which extremizes (\ref{eq:newenerg}), 
then we could expect that 
\begin{equation}
\label{eq:Eanu0}
E_a(\nu_0) = e(a+\nu_0) + B(\nu_0) 
\end{equation}
is a good approximation of $E_a$, an exact eigenvalue of Hamiltonian~(\ref{eq:anu0}). 
It seems that (\ref{eq:newenerg}) is very similar to (\ref{eq:enerprop}), the
only difference being the replacement of $\nu$ by $a+\nu$ in the argument of
the function $e(\nu)$. Nevertheless, this small difference is important,
because, even if the determination of $\nu_0$ from (\ref{eq:enerprop}) is
technically easy and analytical, it may happen that its determination from
(\ref{eq:newenerg}) could be much more involved and very often not analytical.
The alternative formulation in terms of $r_0$ is also slightly modified in this
case. The transcendental equation now writes
\begin{equation}
\label{eq:minr0mod}
e'(a+K(r_0))=P(r_0)
\end{equation}
and the AFM energy is given by
\begin{equation}
\label{eq:energr0mod}
E_{\textrm{AFM}} = e(a+K(r_0))+B(r_0).
\end{equation}

Using again the Hellmann-Feynman theorem \cite{feyn}, it can be shown that
\begin{equation}
\label{eq:anu3}
\langle a+\nu_0|P(r)|a+\nu_0\rangle = P\left( J(\nu_0) \right)=P(r_0). 
\end{equation}
So, $J(\nu_0)= r_0$ is also in this case a kind of ``average point" for the potential $P(r)$. 
Recalling that $Z(x)=P(J(x))$, we have 
\begin{equation}
\label{eq:anu4b}
\langle a+\nu_0|Z(\hat \nu)|a+\nu_0\rangle = Z(\nu_0).
\end{equation}
We confirm also in this case the ``mean field approximation" with respect to a
particular auxiliary field which is introduced to simplify the calculations.

\subsection{Recursion procedure}
\label{sec:recproc}

An idea for obtaining analytical expressions of the eigenenergies for an arbitrary
potential is the following (see also Ref.~\citen{bsb08b}).
We start with a potential $P(r)=P^{[0]}(r)$ for which the energies of the
corresponding Hamiltonian $H^{[0]}$ are exactly known. We then proceed as above
to find approximate solutions for the eigenenergies of a Hamiltonian $H^{[1]}$
in which the potential is at present $V(r)=P^{[1]}(r)$. In general, a large
class of potentials can be treated in that way. Moreover, as we will see below,
by comparison with  accurate numerical results, we can even refine the expressions
so that analytical forms for the energies are very close to the exact solutions. 

Considering now these approximate expressions as the exact ones, we apply once more
the AFM with $P(r)=P^{[1]}(r)$ to obtain approximate solutions for the eigenenergies
of a Hamiltonian $H^{[2]}$ in which the potential is at present $V(r)=
P^{[2]}(r)$. Even if analytical solutions for Hamiltonian $H^{[2]}$ were not
attainable directly with $P(r)=P^{[0]}(r)$, it may occur that they indeed are
with $P(r)=P^{[1]}(r)$. Pursuing recursively such a procedure, one can
imagine to get analytical solutions for increasingly complicated potentials.
Presumably, the quality of the analytical expressions deteriorates with the
order of the recursion. 

\section{Schr\"odinger equation with a power-law potential}
\label{sec:PowerLawpot}

In this section, we discuss in detail the case of the Schr\"odinger equation with 
a power-law potential for two reasons. First, it is a simple case to illustrate 
the AFM. Second, as we will see below, this kind of potentials is in the thick of our method.
The Hamiltonian can be written
\begin{equation}
\label{eq:fgplw}
H = \frac{\bm p^2}{2 m}+\textrm{sgn}(\lambda)\, a\, r^\lambda,
\end{equation}
where $a$ is a strictly positive constant and where the sign function 
is defined by $\textrm{sgn}(\lambda)=\lambda/|\lambda|$ with $\lambda\ne 0$.
The physical values of $\lambda$ must be such that $\lambda \geq -2$, otherwise
the wave equation leads to a collapse. The two unavoidable starting potentials
$P(r)=r^2$ and $P(r)=-1/r$ are indeed two particular cases of power-law
potentials. In fact, the only interesting values studied in this paper are those
comprised between these extreme values: $-1 \leq \lambda \leq 2$. They represent
most of the physical applications. For example, the linear potential $\lambda=1$
is the traditional form of the confining potential in hadronic physics
 \cite{sema04,lucha,morg99}, the
value $\lambda=2/3$ was shown to give the good slope for Regge
trajectories in nonrelativistic treatment \cite{fab88} 
and $\lambda=0.1$ was considered by
Martin. \cite{mart80}

The energy $E$ of (\ref{eq:fgplw}) depends on the physical parameters 
$m$, $a$, $\lambda$ but also on the
quantum numbers $n$ (radial) and $l$ (orbital). It is easy to show that 
the scaling laws allow to write
\begin{equation}
\label{eq:scallpw}
E(m,a,\lambda;n,l)= 2^\frac{\lambda}{\lambda+2} 
a^{\frac{2}{\lambda+2}} m^{-\frac{\lambda}{\lambda+2}}\, \epsilon(\lambda;n,l),
\end{equation}
where $\epsilon(\lambda;n,l)$ is the eigenvalue of equation
\begin{equation}
\label{eq:Hpowred2}
H = \frac{\bm{p}^2}{4}+\textrm{sgn}(\lambda) r^\lambda.	
\end{equation}
This form is chosen in order to match the conventions of Ref.~\citen{bsb08a}.
The scaling laws are unable to say anything about the dependence of $\epsilon$
on the power $\lambda$ and on the quantum numbers $(n,l)$. It is the virtue of the
AFM to shed some light on these very important questions.

An another interesting potential is the logarithmic
potential, discussed for instance in Ref.~\citen{QR}
\begin{equation}
\label{eq:Vlog1}
V(r) = a\, \ln(b\,r).	
\end{equation}
It is in strong connection with the power-law interaction since it can be
rewritten into a similar form 
\begin{equation}
\label{lnxasxpow}
\ln\, x=\lim_{\lambda\rightarrow0}\frac{1}{\lambda}(x^\lambda-1).
\end{equation}
Using the the scaling law, we can write   
\begin{equation}
\label{eq:Elog1}
E(m,a,b) = \frac{2 b^2}{m} \epsilon(\beta)	
\quad \textrm{with} \quad \beta=\frac{ma}{2 b^2},
\end{equation}
where $\epsilon(\beta)$ is the eigenenergy of the reduced Schr\"{o}dinger
Hamiltonian
\begin{equation}
\label{eq:redHlog}
h=\frac{\bm{p}^2}{4}+\beta \ln r.
\end{equation}

We will apply the AFM for these two interactions with the two starting
potentials $P(r)=r^2$ and $P(r) = -1/r$. Let us note that a number of
results presented in the following sections have been already obtained in the framework
of the envelope theory. \cite{env0}\ Nevertheless we remind them in order to keep some
consistency in our presentation.

\subsection{Energies with $P(r)=r^2$}
\label{sec:Epw}

The harmonic oscillator (HO) Hamiltonian $h(\nu)=\bm{p}^2/4 + \nu r^2$ has the following
eigenvalues
\begin{equation}
\label{eq:vpOH}
e(\nu)=\sqrt{\nu} Q_{HO} \quad \textrm{with} \quad Q_{HO}=2n+l+3/2.
\end{equation}
In the following, a combination of quantum numbers as in $Q_{HO}$ ($Q_{HO}$ is a simplified notation for $Q_{HO}(n,l)$) will be denoted ``principal quantum number".

Let us apply the AFM for the power-law potential 
$V(r)=\textrm{sgn}(\lambda) r^\lambda$.
We have then $K(r)=|\lambda|r^{\lambda-2}/2$, $e(K(r)) =
Q_{HO}\sqrt{|\lambda|/2} r^{(\lambda-2)/2}$, $e'(K(r)) = Q_{HO}
r^{(2-\lambda)/2}/\sqrt{2|\lambda|}$ and $B(r) = ((2-\lambda)/2)V(r)$.
The transcendental equation (\ref{eq:trans2}) for the extremization
of the energy leads to the value of the mean radius
\begin{equation}
\label{eq:minr0pw}
r_0=\left[\frac{Q_{HO}^2}{2 |\lambda|}\right]^{1/(\lambda+2)}.
\end{equation}
Inserting this value in $E(r_0)$ (\ref{eq:Egen3c}) gives the AFM approximation
of the eigenenergy
\begin{equation}
\label{eq:epsAFMr2}
\epsilon_{\textrm{AFM}}(\lambda;n,l)= \frac{2+\lambda}{2 \lambda}
 |\lambda|^{\frac{2}{2+\lambda}}2^{-\frac{\lambda}{2+\lambda}}
Q_{HO}^{\frac{2 \lambda}{2+\lambda}}.
\end{equation}
It is important to emphasize that this value is an upper bound on the exact
energy for the domain of $\lambda$ we are interested in. Moreover,
putting $\lambda=2$ in the previous expression, one recovers the
exact result $\epsilon_{\textrm{AFM}}(2;n,l) = \epsilon(2;n,l) =
Q_{HO}$, as expected.

Let us consider now the logarithmic potential $V(r)=\beta \ln r$. We have then
$K(r)=\beta/(2r^2)$, $e(K(r)) = Q_{HO}\sqrt{\beta/2}/r$,
$e'(K(r))$ = $r Q_{HO}/\sqrt{2\beta}$ and $B(r) = V(r)-\beta/2$.
The transcendental equation (\ref{eq:trans2}) for the extremization
of the energy leads to the value of the mean radius
\begin{equation}
\label{eq:minr0log}
r_0=\frac{Q_{HO}}{\sqrt{2 \beta}}.
\end{equation}
Inserting this value in $E(r_0)$ (\ref{eq:Egen3c}) gives the AFM approximation
of the eigenenergy
\begin{equation}
\label{eq:elogred}
\epsilon_{\textrm{AFM}}(\beta;n,l)=\beta \, \ln \left[\sqrt{\frac{\textrm{e}}
{2 \,\beta}} Q_{HO}\right].
\end{equation}
It is well known that the eigenenergies $E(m,a,b;n,l)$ resulting from a
Schr\"{o}dinger equation with the potential~(\ref{eq:Vlog1}) satisfy the
property \cite{QR}
\begin{equation}
\label{proplogpot}
E(\alpha m,a,b;n,l)=E(m,a,b;n,l) - \frac{a}{2} \ln \alpha.
\end{equation}
The immediate consequence is that the corresponding spectrum is independent
of the mass of the particle. It is remarkable that the basic
property~(\ref{proplogpot}) still holds for our approximate
expression~(\ref{eq:elogred}).

It is worth mentioning that the energy formula~(\ref{eq:elogred}) can be understood
as a particular limit case of formula~(\ref{eq:scallpw}), as it was suggested
in Ref.~\citen{lucha}. Guided by (\ref{lnxasxpow}),
let us consider the following Hamiltonian
\begin{equation}
\label{hlog2}
H(\lambda)=\frac{\bm p^{\, 2}}{4}+\frac{\beta}{\lambda}\left[ x^{\lambda}-1\right].
\end{equation}
It reduces to the Hamiltonian~(\ref{eq:redHlog}) in the limit
$\lambda\rightarrow 0$. A simple rewriting of formula~(\ref{eq:scallpw}) for
$a \rightarrow \beta/|\lambda|$ and $m=2$ gives the eigenenergies of 
Hamiltonian~(\ref{hlog2}). We have thus
\begin{equation}
\label{eq:elogred1}
E(\lambda)=\frac{2+\lambda}{2\lambda} \frac{\beta^{\frac{2}{\lambda+2}}}
{2^{\frac{\lambda}{\lambda+2}}} Q_{HO}^{\frac{2\lambda}{\lambda+2}}-\frac{\beta}
{\lambda},
\end{equation}
and as expected
\begin{equation}
\label{liml0}
\lim_{\lambda\rightarrow0}E(\lambda)=\beta\, \ln\left[\sqrt{\frac{\textrm{e}}{2\,\beta}}
\, Q_{HO} \right],
\end{equation}
that is precisely formula~(\ref{eq:elogred}), directly obtained from the logarithmic
potential. This confirms the idea that the logarithmic potential can be seen as
the limit of a power-law potential $r^\lambda$ when $\lambda$ goes to zero. The same
conclusion was obtained with the envelope theory (see Ref.~\citen{hall2003}).  

\subsection{Energies with $P(r)=-1/r$}
\label{sec:Epw1or}

Now we consider the Coulomb (C) Hamiltonian $h(\nu)=\bm{p}^2/4 - \nu /r$
whose eigenvalues are given by ($Q_{C}$ is a simplified notation for $Q_{C}(n,l)$)
\begin{equation}
\label{eq:vpC}
e(\nu)=- \frac{\nu^2}{Q_{C}^2} \quad \textrm{with} \quad Q_{C}=n+l+1.
\end{equation}

Let us apply the AFM for the power-law potential 
$V(r)=\textrm{sgn}(\lambda) r^\lambda$.
We have then $K(r)=|\lambda|r^{\lambda+1}$, $e(K(r)) = 
-r^{2\lambda+2}(|\lambda|^2/Q_{C}^2)$, $e'(K(r)) = -r^{\lambda+1}
(2|\lambda|/Q_{C}^2)$ and $B(r)=(1+\lambda)V(r)$.
The transcendental equation (\ref{eq:trans2}) for the extremization
of the energy leads to the value of the mean radius
\begin{equation}
\label{eq:minr0pw2}
r_0=\left[\frac{Q_{C}^2}{2 |\lambda|}\right]^{1/(\lambda+2)}.
\end{equation}
Inserting this value in $E(r_0)$ (\ref{eq:Egen3c}) gives the AFM approximation
of the eigenenergy
\begin{equation}
\label{eq:epsAFMrm1}
\epsilon_{\textrm{AFM}}(\lambda;n,l)= \frac{2+\lambda}{2 \lambda}
 |\lambda|^{\frac{2}{2+\lambda}}2^{-\frac{\lambda}{2+\lambda}}
Q_{C}^{\frac{2 \lambda}{2+\lambda}}.
\end{equation}
It is important to emphasize that this value is a lower bound on the exact
energy. Putting $\lambda=-1$ in the previous expression, one recovers the
exact result $\epsilon_{\textrm{AFM}}(-1;n,l) = \epsilon(-1;n,l) =
-1/Q_{C}^2$, as expected. It is worth noting that formulae~(\ref{eq:epsAFMr2}) and (\ref{eq:epsAFMrm1}) exhibit the same formal
expression with just the exchange $Q_{C} \leftrightarrow Q_{HO}$. This important point will be discussed in Sect.~\ref{sec:invformenerg}.

The resolution of the Schr\"{o}dinger equation with the logarithmic
potential in~(\ref{eq:redHlog}) is even simpler with the present choice of
$P(r)$. Indeed, one finds in this case $K(r)=\beta\, r$, $e(K(r)) =
-(r \beta/Q_{C})^2$, $e'(K(r)) = -2 \beta r^2/Q_{C}^2$ and $B(r) = V(r)
+ \beta$.
The transcendental equation (\ref{eq:trans2}) leads  
to the value of the mean radius
\begin{equation}
\label{eq:minr0log2}
r_0=\frac{Q_{C}}{\sqrt{2 \beta}}.
\end{equation}
Inserting this value in $E(r_0)$ (\ref{eq:Egen3c}) gives the AFM approximation
of the eigenenergy
\begin{equation}
\label{eq:elogred2}
\epsilon_{\textrm{AFM}}(\beta;n,l)=\beta\, \left[\sqrt{\frac{\textrm{e}}{2 \,\beta}}
Q_{C}\right].
\end{equation}
Again, formulae~(\ref{eq:elogred}) and (\ref{eq:elogred2}) exhibit the same formal
expression with just the exchange $Q_{C} \leftrightarrow Q_{HO}$. Consequently,
the properties mentioned above concerning the spectrum and the limit of a
power-law potential when $\lambda \to 0$ still hold in this case.

\subsection{Wavefunctions and observables}
\label{sec:wfobserv}

In this section, the quality of the wavefunction $\left\langle \bm r | \nu_0\right\rangle$
given by the AFM is tested for the Schr\"odinger Hamiltonian with the linear potential
(results for logarithmic and exponential potentials are presented in Ref.~\citen{AFMeigen}).
This Hamiltonian is chosen because it can be solved analytically for a vanishing angular
momentum $l$ (see Appendix~\ref{sec:obs_Ai}). When $l\ne 0$, numerical results have been
obtained with two different methods. \cite{lag,luch99}\ In order to simplify the notation,
we will denote $| n, l \rangle$ an eigenstate of $H$ and $| n \rangle=| n, 0 \rangle$,
with $\langle \bm r | n \rangle$ given by~(\ref{psiAi}). The corresponding energies will
be denoted $E_{n,l}=\langle n,l |H|n,l\rangle$ and $E_n=E_{n,0}=\langle n |H|n\rangle$.
As we need a Hamiltonian $\tilde H$ with a central potential which is completely solvable
to apply the AFM, we can only use in practice a hydrogen-like system ($P(r)=-1/r$) or a
harmonic oscillator ($P(r)=r^2$). As a linear potential seems closer to $r^2$ than $-1/r$,
we can expect that the use of a harmonic oscillator to start the AFM will give better
results. Using the scaling properties, we can consider the following simple Hamiltonian 
\begin{equation}\label{Hlin}
H=\bm p^2 + r,
\end{equation}
in order to match the conventions of Ref.~\citen{AFMeigen}.

As it is shown in the previous section, the approximate AFM energies, denoted here by
$\epsilon_{n,l}$, are given by 
\begin{equation}\label{epsN}
\epsilon_{n,l}=\frac{3}{2^{2/3}}Q^{2/3},
\end{equation} 
with $Q=2 n+l+3/2$ for $P(r)=r^2$ and $Q=n+l+1$ for $P(r)=-1/r$. Exact energies, given
by (\ref{EAi}), reduces to $E_n= -\alpha_n$, where $\alpha_n$ is the $(n+1)^{\textrm{th}}$
zero of the Airy functions $\textrm{Ai}$. A simple approximation of $E_n$ can be obtained
using the expansion (\ref{betan}) at the first order
\begin{equation}\label{Enex}
E_n= -\alpha_n \approx \left( \frac{3 \pi}{2}\right)^{2/3} \left( n + \frac{3}{4}\right)^{2/3}
\approx 2.811 \left( n + \frac{3}{4}\right)^{2/3}.
\end{equation} 
It is worth noting that the linear potential is not only a toy model to test the AFM method.
Effective theories of QCD have proved that it is a good interaction to
take into account the confinement of quarks or gluons in potential models of hadronic physics.
\cite{sema04,lucha,morg99}

\subsubsection{AFM with $P(r)=-1/r$}\label{analytHy}

One can ask whether it is possible to obtain good approximations for the solutions of a
Schr\"odinger equation with a linear potential by means of hydrogen-like eigenfunctions. This
will be examined in this section. Using the previous results for $P(r)=-1/r$, we find
$\nu_0=r_0^2=2^{2/3}(n+l+1)^{4/3}$. Exact eigenstates are then approximated by AFM eigenstates
which are, in this case, hydrogen-like states (\ref{psiHy}) with 
\begin{equation}\label{etaHy}
\eta=\frac{\nu_0}{2}=\frac{(n+l+1)^{4/3}}{2^{1/3}}.
\end{equation} 
Such states are denoted $|\textrm{Hy};n,l\rangle$ and $|\textrm{Hy};n\rangle=|\textrm{Hy};n,0\rangle$.
Using~(\ref{rkHy}) and results above, it can be shown that (\ref{pr0}) is satisfied, with
$P(r_0)=-1/r_0=-2^{-1/3}(n+l+1)^{-2/3}$. It is also worth noting that (\ref{Zrho}) gives
$\left\langle 1/\sqrt{\hat \nu} \right\rangle = 1/\sqrt{\nu_0}$. We denote $\epsilon_{n,l}^{\textrm{Hy}}$
and $\epsilon_n^{\textrm{Hy}}=\epsilon_{n,0}^{\textrm{Hy}}$ the approximated energies which are given
by (\ref{epsN}) with $Q=n+l+1$. Since $\tilde V(r,\nu_0)-V(r)=-(r-r_0)^2/r \le 0$, $\epsilon_{n,l}^{\textrm{Hy}}$
are lower bounds on the exact energies. This can also be determined with the function $g$ defined by (\ref{defg}).
In this case, $g(y)=-1/y$ with $y<0$. The function $g''(y)=-2/y^3$ being positive, $g$ is convex as expected. 

The quantum number dependence of the scaling parameter $\eta$ corrects partly the difference between
the shapes of $\langle \bm r|n,l\rangle$ and $\langle \bm r|\textrm{Hy};n,l\rangle$. Consequently,
$\langle\textrm{Hy};n,l |\textrm{Hy};n,l'\rangle=\delta_{ll'}$ because of the orthogonality of the
spherical harmonics, but $\langle\textrm{Hy};n,l |\textrm{Hy};n',l\rangle\ne \delta_{nn'}$. Using the
definition~(\ref{ovgen}), we find 
\begin{equation}\label{ovHy}
\langle\textrm{Hy};n,l |\textrm{Hy};n',l\rangle = F^{Hy}_{n,n',l}\left(\left(\frac{n'+l+1}{n+l+1}\right)^{4/3}\right),
\end{equation} 
with $F^{Hy}$ given by (\ref{FHy}). Table~\ref{tab:recHy} gives some values of $|\langle\textrm{Hy};n,l |
\textrm{Hy};n',l\rangle|^2$. We can see that the overlap is not negligible for $n$ close to $n'$, but it
decreases rapidly with $|n-n'|$. The situation improves when $l$ increases: $|\langle\textrm{Hy};0,l |
\textrm{Hy};1,l\rangle|^2=0.43$, 0.29, 0.22, 0.18, 0.15, 0.13 for $l=0\to 5$.

\begin{table}[htb]
\caption{Results for $P_{n,n',l}=|\langle\textrm{Hy};n,l |\textrm{Hy};n',l\rangle|^2$. Values for $l=0$ ($l=1$)
are given in the lower-left (upper-right) triangle of the Table. $P_{n,n',l}=P_{n',n,l}$ and $P_{n,n,l}=1$.}
\label{tab:recHy}
\begin{center}
\begin{tabular}{rllll}
\hline\hline\noalign{\smallskip}
& $n=0$ & 1 & 2 & 3 \\ [3pt]
\hline
$n'=0$ & 1 & 0.29 & 0.028 & 0.0039 \\
1 & 0.43 & 1 & 0.36 & 0.036 \\
2 & 0.055 & 0.43 & 1 &  0.39 \\
3 & 0.0097 & 0.049 & 0.43 & 1 \\
\noalign{\smallskip}\hline
\end{tabular}
\end{center}
\end{table}

\begin{figure}
\centerline{
\includegraphics*[width=\halftext]{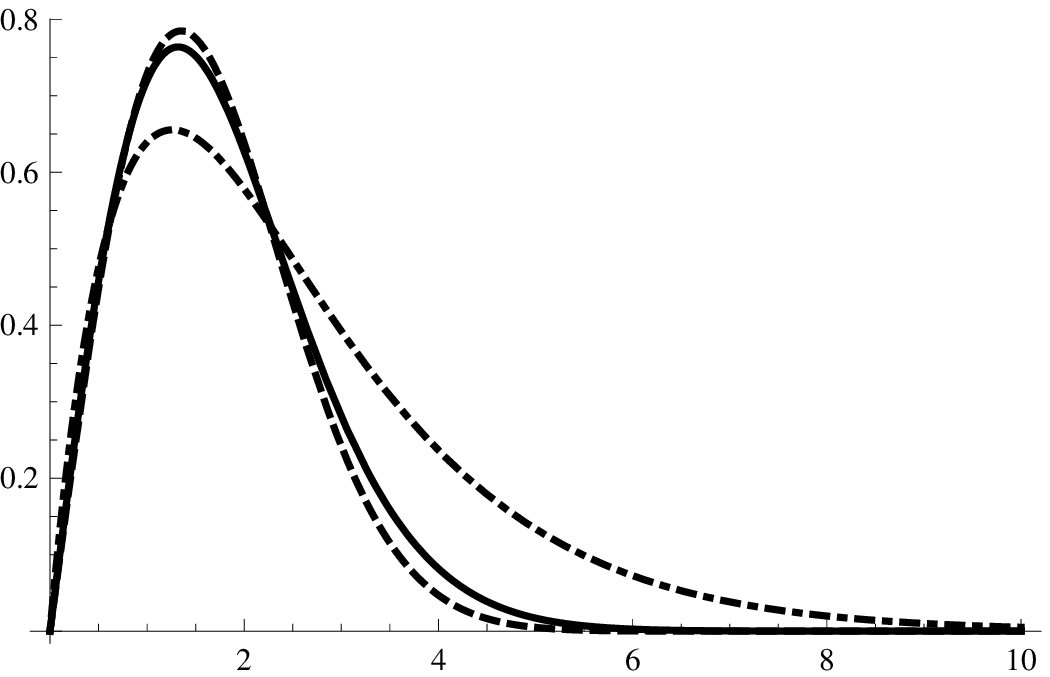}\quad
\includegraphics*[width=\halftext]{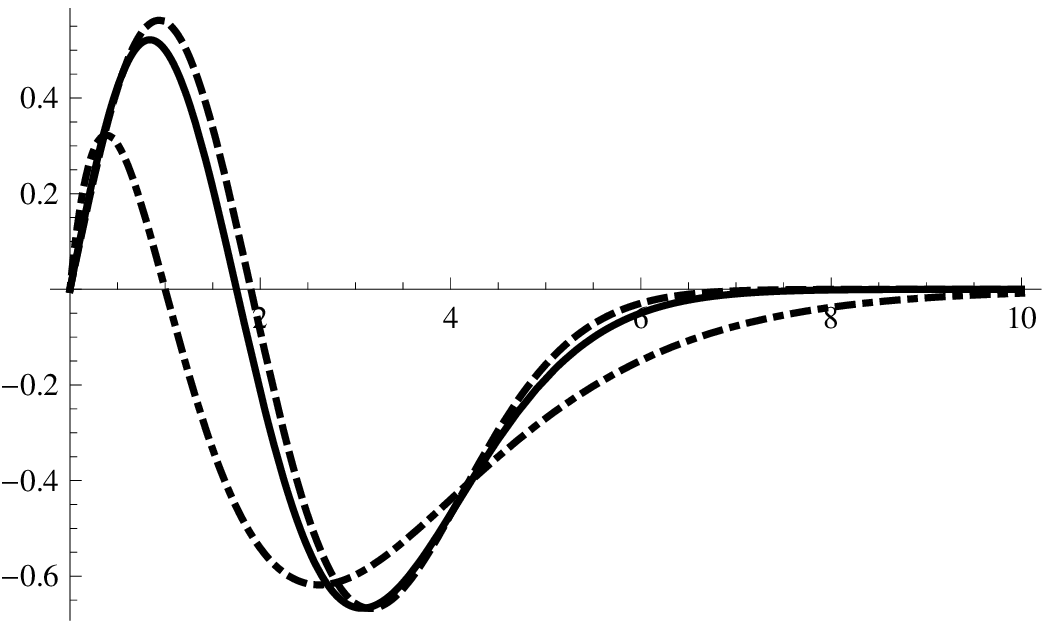}
}
\caption{Normalized wavefunctions, as a function of the variable $r$ in (\ref{Hlin}), for $n=0$
(left) and 1 (right): $\langle \bm r|n\rangle$ (solid line), $\langle \bm r|\textrm{Hy};n\rangle$
(dotted-dashed line) and $\langle \bm r|\textrm{HO};n\rangle$ (dashed line).}
\label{fig:HyHO}
\end{figure}

It is interesting to compare $\epsilon_n^{\textrm{Hy}}$ with (\ref{Enex})
\begin{equation}\label{EepsHy}
\epsilon_n^{\textrm{Hy}}= \frac{3}{2^{2/3}}(n+1)^{2/3}\approx 1.890\, (n+1)^{2/3}. 
\end{equation}
The ratio $\epsilon_n^{\textrm{Hy}}/E_n$ is respectively equal to 0.808, 0.734, 0.712 for $n=0,1,2$ and
tends rapidly toward the asymptotic value $3^{1/3}/\pi^{2/3}\approx 0.672$. As expected, these ratios
are smaller than 1 since $\epsilon_n^{\textrm{Hy}}$ are lower bounds. Two wavefunctions are given
in Fig.~\ref{fig:HyHO}. We can see that the differences between exact $\langle \bm r|n\rangle$ and
AFM $\langle \bm r|\textrm{Hy};n\rangle$ wavefunctions can be large. The overlap $|\langle n|
\textrm{Hy};n\rangle|^2$ between these wavefunctions can be computed numerically with a high accuracy.
We find respectively the values 0.934, 0.664, 0.298 for $n=0,1,2$, showing a rapid decrease of the
overlap. It is worth noting that, asymptotically, $\langle \bm r|n\rangle \propto \exp\left( -\frac{2}{3}
r^{3/2} \right)$ while $\langle \bm r|\textrm{Hy};n\rangle$ is characterized by an
exponential tail. Nevertheless, if an observable is not too sensitive to the large $r$ behavior, this
discrepancy will not spoil its mean value.

The various observables $|\psi_{n,0}(0)|^2$, $\langle r^k \rangle$ and $\langle p^k \rangle$ computed
with the AFM states can be obtained using formulae~(\ref{psi0Hy}), (\ref{rkHy}) and (\ref{pkHy}) for
$l=0$ with the parameter $\eta$ given by (\ref{etaHy}). Results are summed up in Table~\ref{tab:obsHy}.
A direct comparison between the structure of exact and AFM observables can be obtained if we remind
that the exact ones depend on $|\alpha_n|$ which can be well approximated by $\beta_n$
(see~(\ref{betan})). Let us look in detail only at the mean value $\langle r \rangle$. For the exact
and AFM solutions, we have respectively
\begin{eqnarray}\label{rHy}
\label{rHy1} 
\langle n| r |n\rangle &=& \frac{2 |\alpha_n|}{3} \approx \left( \frac{2 \pi^2}{3}\right)^{1/3}
\left( n + \frac{3}{4}\right)^{2/3} \approx 1.874 \left( n + \frac{3}{4}\right)^{2/3},\\
\label{rHy2} 
\langle \textrm{Hy};n| r |\textrm{Hy};n\rangle &=& \frac{3}{2^{2/3}}(n+1)^{2/3}
\approx 1.890\, (n+1)^{2/3} . 
\end{eqnarray}
Some observables like $|\psi_{n,0}(0)|^2$ and $\langle p^4 \rangle$ are very badly reproduced. Others
can be obtained with a quite reasonable accuracy. Despite the fact that exact and AFM wavefunctions
differ strongly when $n$ increases, their sizes stay similar. 

\begin{table}[hbt]
\protect\caption{Various observables $A$ computed with the AFM and $P(r)=-1/r$. 
}
\label{tab:obsHy}
\begin{center}
\begin{tabular}{cccccc}
\hline\hline\noalign{\smallskip}
 & $\langle\textrm{Hy};n|A|\textrm{Hy};n\rangle$ & \multicolumn{4}{c}{$\langle\textrm{Hy};n|A|\textrm{Hy};n
\rangle/\langle n|A|n\rangle$} \\
$\langle A \rangle$ & ($2 m = a = 1$) & $n=0$ & $n=1$ & $n=2$ & $n\gg 1$ \\ [3pt]
\hline
$|\psi_{n,0}(0)|^2$ & $\frac{n+1}{2\pi}$ & 2 & 4 & 6 & $2(n+1)$ \\
$\langle r \rangle$ & $\frac{3(n+1)^{2/3}}{2^{2/3}}$ & 1.212 & 1.101 & 1.068 & $1.009 + \frac{0.168}{n}+
O\left(\frac{1}{n^2}\right)$ \\
$\langle r^2 \rangle$ & $\frac{5 n^2+10 n+6}{2^{1/3}(n+1)^{2/3}}$ & 1.633 & 1.178 & 1.080 & $0.942 +
\frac{0.314}{n}+ O\left(\frac{1}{n^2}\right)$  \\
$\langle p^2 \rangle$ & $\frac{(n+1)^{2/3}}{2^{2/3}}$ & 0.808 & 0.734 & 0.712 & $0.672 + \frac{0.112}{n}+
O\left(\frac{1}{n^2}\right)$  \\
$\langle p^4 \rangle$ & $\frac{(8 n+5)(n+1)^{4/3}}{2^{4/3}}$ & 1.815 & 3.890 & 5.916 & $2.009\, n + 1.926 +
O\left(\frac{1}{n}\right)$  \\
\noalign{\smallskip}\hline
\end{tabular}
\end{center}
\end{table}

Since it is possible to compute analytically the mean value $\langle\textrm{Hy};n| \bm p^2 + r |
\textrm{Hy};n\rangle$ for an arbitrary value of the scale factor $\eta$, the relations between the bounds
given in Sect.~\ref{sec:ulb} can be checked. We can verify that 
$E=-\alpha_n \ge E(\nu_0)=\frac{3}{2^{2/3}}(n+1)^{2/3}$ and $E^*(\nu_0)=\frac{4}{2^{2/3}}(n+1)^{2/3} \ge
E(\nu_0)=\frac{3}{2^{2/3}}(n+1)^{2/3}$. Moreover, for the ground state, we have $E^*(\nu_0)=\frac{4}{2^{2/3}}
\ge E^*(\nu^*)=\frac{3^{5/3}}{4^{2/3}} \ge E=-\alpha_0 \ge E(\nu_0)=\frac{3}{2^{2/3}}$.

The behavior of observables computed with the AFM is similar for values of $l=0,1,2$. We do not expect strong
deviations for larger values of $l$. This is illustrated with some typical results gathered in 
Table~\ref{tab:lne0Hy}. Observables are generally not very well reproduced, but this is expected since eigenstates
for a linear potential are very different from eigenstates for a Coulomb potential. Actually, the agreement is not
catastrophic, except for $|\psi_{n,0}(0)|^2$ and $\langle p^4 \rangle$ as mentioned above. It is even surprising
that the AFM with $P(r)=-1/r$ could give energies and some observables for a linear potential with a quite
reasonable accuracy. Moreover, lower bounds on the energies are obtained. 

\begin{table}[htb]
\caption{Ratios between the AFM results (energies and $\langle r \rangle$) with $P(r)=-1/r$ and the
exact results, for several quantum number sets $(n,l)$.}
\label{tab:lne0Hy}
\begin{center}
\begin{tabular}{lllllll}
\hline\hline\noalign{\smallskip}
$l$ & $n=0$ &$n=1$ &$n=2$ &$n=3$ &$n=4$ &$n=5$ \\ [3pt]
\hline
\multicolumn{7}{l}{$\epsilon_{n,l}^{\textrm{Hy}}/E_{n,l}$ } \\
0 & 0.808 & 0.734 & 0.712 & 0.702 & 0.696 & 0.692 \\
1 & 0.893 & 0.805 & 0.767 & 0.746 & 0.733 & 0.724 \\
2 & 0.925 & 0.846 & 0.805 & 0.779 & 0.762 & 0.749 \\
\noalign{\smallskip}
\multicolumn{7}{l}{$\langle\textrm{Hy};n|r|\textrm{Hy};n\rangle/\langle n|r|n\rangle$} \\
0 & 1.212 & 1.101 & 1.068 & 1.053 & 1.043 & 1.037 \\
1 & 1.116 & 1.118 & 1.103 & 1.089 & 1.079 & 1.071 \\
2 & 1.080 & 1.110 & 1.110 & 1.104 & 1.096 & 1.089 \\
\noalign{\smallskip}\hline
\end{tabular}
\end{center}
\end{table}

\subsubsection{AFM with $P(r)=r^2$}\label{analytOH}

Since the quadratic potential is closer to the linear potential than a Coulomb one, one can expect
better results with harmonic oscillator wavefunctions. This will be examined in this section.
Using the previous results for $P(r)=r^2$, we find 
$\nu_0=1/(2 r_0)=2^{-4/3}(2 n+l+3/2)^{-2/3}$. Exact eigenstates are then approximated by AFM
eigenstates which are, in this case, harmonic oscillator states (\ref{psiOH}) with 
\begin{equation}\label{lambHO}
\lambda=\nu_0^{1/4}=2^{-1/3}(2 n+l+3/2)^{-1/6}.
\end{equation} 
Such states are denoted $|\textrm{HO};n,l\rangle$ and $|\textrm{HO};n\rangle=|\textrm{HO};n,0
\rangle$. Using~(\ref{rkOH}) and results above, it can be shown that (\ref{pr0}) is satisfied,
with $P(r_0)=r_0^2=2^{2/3}(2 n+l+3/2)^{4/3}$. It is also worth noting that (\ref{Zrho}) gives
$\left\langle 1/\hat \nu^2 \right\rangle = 1/\nu_0^2$. This is in agreement with (20) in Ref.~\citen{bsb09c}
(an auxiliary field $\phi=1/\nu$ is used in this last reference). We denote $\epsilon_{n,l}^{\textrm{HO}}$
and $\epsilon_n^{\textrm{HO}}=\epsilon_{n,0}^{\textrm{HO}}$ the approximated energies which are given by
(\ref{epsN}) with $Q=2 n+l+3/2$. Since $\tilde V(r,\nu_0)-V(r)=(r-r_0)^2/(2 r_0) \ge 0$,
$\epsilon_{n,l}^{\textrm{HO}}$ are upper bounds on the exact energies. In this case, $g(y)=\sqrt{y}$ with
$y>0$. The function $g''(y)=-1/(4 y^{3/2})$ being negative, $g$ is concave as expected.

Since the scaling parameter $\lambda$ depends on the quantum numbers, $\langle\textrm{HO};n,l |
\textrm{HO};n,l'\rangle=\delta_{ll'}$ because of the orthogonality of the spherical harmonics, but
$\langle\textrm{HO};n,l |\textrm{HO};n',l\rangle\ne \delta_{nn'}$. Using the definition~(\ref{ovgen}),
we find 
\begin{equation}\label{ovHO}
\langle\textrm{HO};n,l |\textrm{HO};n',l\rangle = F^{HO}_{n,n',l}\left(\left(\frac{4 n+2 l+3}
{4 n'+2 l+3}\right)^{1/6}\right),
\end{equation} 
with $F^{HO}$ given by (\ref{FOH}). Table~\ref{tab:recOH} gives some values of $|\langle\textrm{HO};n,l |
\textrm{HO};n',l\rangle|^2$. We can see that the overlap is always small and decreases rapidly with
$|n-n'|$. The situation is even better when $l$ increases:  for $l=0\to 5$, $|\langle\textrm{HO};0,l |\textrm{HO};1,l
\rangle|^2=0.029$, 0.023, 0.019, 0.017, 0.014, 0.013.

\begin{table}[htb]
\caption{Results for $P_{n,n',l}=|\langle\textrm{HO};n,l |\textrm{HO};n',l\rangle|^2$. Values for $l=0$ ($l=1$)
are given in the lower-left (upper-right) triangle of the Table. $P_{n,n',l}=P_{n',n,l}$ and $P_{n,n,l}=1$.}
\label{tab:recOH}
\begin{center}
\begin{tabular}{rllll}
\hline\hline\noalign{\smallskip}
& $n=0$ & 1 & 2 & 3 \\ [3pt]
\hline
$n'=0$ & 1 & 0.023 & 0.0026 & 0.00039 \\
1 & 0.029 & 1 & 0.026 & 0.0027 \\
2 & 0.0036 & 0.027 & 1 &  0.026 \\
3 & 0.00064 & 0.0031 & 0.027 & 1 \\
\noalign{\smallskip}\hline
\end{tabular}
\end{center}
\end{table}

Let us look at the energies $\epsilon_n^{\textrm{HO}}$ 
\begin{equation}\label{EespHO}
\epsilon_n^{\textrm{HO}}= 3 \left( n + \frac{3}{4}\right)^{2/3}. 
\end{equation}
By comparing with (\ref{Enex}), we can see immediately that the situation is more favorable than in
the previous case. The ratio $\epsilon_n^{\textrm{HO}}/E_n$ is respectively equal to 1.059, 1.066,
1.067, for $n=0,1,2$. The asymptotic value $3^{1/3}2^{2/3}/\pi^{2/3}\approx 1.067$ is rapidly approached.
As expected, these ratios are greater than 1 since $\epsilon_n^{\textrm{HO}}$ are upper bounds. Two
wavefunctions are given in Fig.~\ref{fig:HyHO}. We can see that the differences between exact $\langle
\bm r|n\rangle$ and AFM $\langle \bm r|\textrm{HO};n\rangle$ wavefunctions are quite small. The overlap
$|\langle n|\textrm{HO};n\rangle|^2$ between these wavefunctions can be computed numerically with a high
accuracy. We find respectively the values 0.997, 0.979, 0.951 for $n=0,1,2$. A value below 0.75 is reached
for $n=6$ and below 0.25 for $n=14$. A wavefunction $\langle \bm r|\textrm{HO};n\rangle$ is characterized
by an Gaussian tail, while we have asymptotically $\langle \bm r|n\rangle \propto \exp\left( -\frac{2}{3}
r^{3/2} \right)$. Again, if an observable is not too sensitive to the large $r$ behavior,
this discrepancy will not spoil its mean value.

The various observables $|\psi_{n,0}(0)|^2$, $\langle r^k \rangle$ and $\langle p^k \rangle$ computed
with the AFM states can be obtained using formulae~(\ref{psi0OH}), (\ref{rkOH}) and (\ref{pkOH}) for $l=0$
with the parameter $\lambda$ given by (\ref{lambHO}). Results are summed up in Table~\ref{tab:obsOH}. Again,
a direct comparison between the structure of exact and AFM observables can be obtained by using $\beta_n$
(see~(\ref{betan})) instead of $|\alpha_n|$. Let us look in detail only at the mean value $\langle r^2
\rangle$. For the exact and AFM solutions, we have respectively
\begin{eqnarray}\label{rHO}
\label{rHO1} 
\langle n| r^2 |n\rangle &=& \frac{8|\alpha_n|^2}{15} \approx \frac{2^{5/3} 3^{1/3} \pi^{4/3}}{5}
\left( n + \frac{3}{4}\right)^{4/3}\approx 4.214 \left( n + \frac{3}{4}\right)^{4/3},\\
\label{rHO2} 
\langle \textrm{HO};n| r^2 |\textrm{HO};n\rangle &=& 4 \left( n + \frac{3}{4}\right)^{4/3} . 
\end{eqnarray}
In contrast with the previous case, all observables are very well reproduced. This is also the case for
$n\gg 1$, while the overlap $|\langle n|\textrm{HO};n\rangle|^2$ tends towards very small values in this
limit. This is due to the fact that some observables are not very sensitive to the details of the
wavefunctions and that the sizes of exact and AFM states stay similar.

\begin{table}[htb]
\caption{Various observables $A$ computed with the AFM and $P(r)=r^2$. 
}
\label{tab:obsOH}
\begin{center}
\begin{tabular}{cccccc}
\hline\hline\noalign{\smallskip}
 & $\langle\textrm{HO};n|A|\textrm{HO};n\rangle$ & \multicolumn{4}{c}{$\langle
\textrm{HO};n|A|\textrm{HO};n\rangle/\langle n|A|n\rangle$} \\
$A$ & ($2 m = a = 1$) & $n=0$ & $n=1$ & $n=2$ & $n\gg 1$ \\ [3pt]
\hline
$|\psi_{n,0}(0)|^2$ & $\frac{2 \Gamma(n+3/2)}{\pi^2 n! \sqrt{8 n+6}}$ & 0.921 & 0.905 & 0.902 & $0.900 + \frac{0.014}{n^2}+ O\left(\frac{1}{n^3}\right)$ \\
$\langle r \rangle$ & $\frac{4 (8 n+6)^{1/6} \Gamma(n+3/2)}{\pi n!}$ & 0.976 & 0.964 & 0.962 & $0.961 + \frac{0.011}{n^2}+ O\left(\frac{1}{n^{8/3}}\right)$ \\
$\langle r^2 \rangle$ & $4(n+3/4)^{4/3}$ & 0.935 & 0.946 & 0.948 & $0.949 - \frac{0.009}{n^2}+ O\left(\frac{1}{n^{7/3}}\right)$  \\
$\langle p^2 \rangle$ & $(n+3/4)^{2/3}$ & 1.059 & 1.066 & 1.067 & $1.067 - \frac{0.005}{n^2}+ O\left(\frac{1}{n^{7/3}}\right)$  \\
$\langle p^4 \rangle$ & $\frac{3(8 n^2+12 n+5)}{4(8 n+6)^{2/3}}$ & 1.039 & 0.966 & 0.956 &  $0.949 + \frac{0.050}{n^2}+ O\left(\frac{1}{n^{7/3}}\right)$ \\
\noalign{\smallskip}\hline
\end{tabular}
\end{center}
\end{table}

The mean value $\langle\textrm{HO};n| \bm p^2 + r |\textrm{HO};n\rangle$ being analytically computable
for an arbitrary value of the scale factor $\lambda$, the relations between the bounds can also be
checked. We can verify that $E(\nu_0)=\frac{3}{2^{2/3}}(2 n+3/2)^{2/3} \ge E=-\alpha_n$ and
$E(\nu_0)=\frac{3}{2^{2/3}}(2 n+3/2)^{2/3} \ge E^*(\nu_0)=(n+3/4)^{2/3}+\frac{4}{\pi n!}(8 n+6)^{1/6}
\Gamma(n+3/2)$. Moreover, for the ground state, we have $E(\nu_0)=\frac{3^{5/3}}{2^{4/3}} \ge E^*(\nu_0)
=\frac{3^{2/3}}{2^{4/3}}+\frac{2\times 6^{1/6}}{\sqrt{\pi}} \ge E^*(\nu^*)=\frac{3^{4/3}}{(2\pi)^{1/3}}
\ge E=-\alpha_0$.

As in the previous case, the behavior of observables computed with the AFM is similar for values of $l=0,1,2$.
We do not expect strong deviations for larger values of $l$. Some typical results are presented in
Table~\ref{tab:lne0HO}. Agreement between AFM and exact results are very good, much better than for the
previous case. This is expected, since eigenstates for a quadratic potential are closer to eigenstates for
a linear potential than eigenstates for a Coulomb potential. The quantum number dependence of the scaling
parameter $\lambda$ corrects, much better than $\eta$, the difference between the shapes of AFM and exact
eigenstates. 

\begin{table}[htb]
\protect\caption{Ratios between the AFM results (energies and $\langle r \rangle$) with $P(r)=r^2$ and the
exact results, for several quantum number sets $(n,l)$.}
\label{tab:lne0HO}
\begin{center}
\begin{tabular}{lllllll}
\hline\hline\noalign{\smallskip}
$l$ & $n=0$ &$n=1$ &$n=2$ &$n=3$ &$n=4$ &$n=5$ \\ [3pt]
\hline
\multicolumn{7}{l}{$\epsilon_{n,l}^{\textrm{HO}}/E_{n,l}$ } \\
0 & 1.059 & 1.066 & 1.067 & 1.067 & 1.067 & 1.067 \\
1 & 1.036 & 1.055 & 1.060 & 1.063 & 1.064 & 1.065 \\
2 & 1.026 & 1.046 & 1.054 & 1.058 & 1.061 & 1.062 \\
\noalign{\smallskip}
\multicolumn{7}{l}{$\langle\textrm{HO};n|r|\textrm{HO};n\rangle/\langle n|r|n\rangle$} \\
0 & 0.976 & 0.964 & 0.962 & 0.962 & 0.961 & 0.961 \\
1 & 0.985 & 0.972 & 0.968 & 0.965 & 0.964 & 0.963 \\
2 & 0.990 & 0.978 & 0.972 & 0.969 & 0.967 & 0.966 \\
\noalign{\smallskip}\hline
\end{tabular}
\end{center}
\end{table}

\subsubsection{General considerations}
\label{sec:genconsfo}

Within the auxiliary field method, if the problem studied is analytically manageable, an eigenstate can then
be determined with the same computational effort for any set of quantum numbers ($n,l$). So this method is
very useful if one is interested in obtaining analytical information about the whole spectra, wavefunctions,
and observables of a Hamiltonian without necessarily searching a very high accuracy. The selection of the
potential $P(r)$ seems crucial to obtain good results for the eigenstates. For a linear potential, $P(r)=r^2$
is  clearly the best choice. In Ref.~\citen{AFMeigen}, it is shown that $P(r)=-1/r$ provides much better results
for an exponential potential, while the choice of $P(r)$ is not so crucial for the logarithmic interaction. 

An eigenvalue equation can also be solved within the variational method by expanding trial states in terms
of special basis states. The correct asymptotic tail can be well reproduced if the basis states are well
chosen. With this method, a matrix representation of the Hamiltonian is obtained and the solutions are
computed by diagonalizing this matrix: $M$ upper bounds on the energies are determined with the corresponding
$M$ states, where $M$ is the order of the matrix. A very good accuracy is possible if $M$ is large enough.
Even for $M=1$, the accuracy can be better than the one provided by the AFM (see Sect.~\ref{sec:baryoncase}).
However, if one is interested in closed-form results, the matrix elements must have an analytical expression
and the number of computed states $M$ must be limited to 4. But even for $M=2$, the eigenvalues can have
very complicated expression, not usable in practice. So, the variational method can only provide, at best,
a very limited number of eigenvalues and eigenstates with an analytical form.

The WKB method is also a popular method to solve eigenvalue equations. \cite{QM2,WKB,J}\ In principle, it is
only valid for high values of the radial quantum number $n$, but it can sometimes yield very good results for
low-lying states. \cite{brau00}\ Wavefunctions are not necessary to compute some observables \cite{brau00} but
they can be determined for arbitrary value of $n$ with this method. An advantage is that their asymptotic
behavior can be correct but, unfortunately, the WKB method is mainly manageable for S-states. Indeed, for
$l\ne 0$, the interaction $V(r)$ must be supplemented by the centrifugal potential, which complicates
greatly the integrals to compute. Moreover, these wavefunctions are piecewise-defined whose different parts
must be connected properly at the turning points. So they are not very practical to use. 

\subsection{Improving the eigenenergies}
\label{sec:impeigene}

\subsubsection{Modifications of the principal quantum number}

We have \emph{a priori} no  idea of the dependence of the exact result
$\epsilon(\lambda;n,l)$ on the parameter $\lambda$ and on the quantum
numbers $(n,l)$. The AFM gives approximate answers: (\ref{eq:epsAFMr2}) or
(\ref{eq:epsAFMrm1}). 
Moreover, a simple glance at both formulae convinces us that they have the
same dependence in $\lambda$, but they differ in the dependence on $(n,l)$
only through a different expression for the principal quantum number $Q$
which is a remnant of the function $P(r)$ used in the AFM.

Owing to the fact that AFM gives the exact result for $\lambda = -1$ and
$\lambda=2$, a continuity argument suggests that the exact result should not
be too different from the previous formulae, but with a modified expression
for the principal quantum number. Thus we propose for the eigenvalues of the
Schr\"{o}dinger equation with power-law potential (\ref{eq:Hpowred2}) the
following prescription
\begin{equation}
\label{eq:epsAFMpwmod}
\epsilon_{\textrm{AFM}}(\lambda;n,l)= \frac{2+\lambda}{2 \lambda}
 |\lambda|^{\frac{2}{2+\lambda}}2^{-\frac{\lambda}{2+\lambda}}
Q(\lambda;n,l)^{\frac{2 \lambda}{2+\lambda}}.
\end{equation}
The choice of the function $Q(\lambda;n,l)$ is just a matter of guess but
we require that it is a continuous function of $\lambda$ and that it coincides
with $Q_{C}(n,l)$ for $\lambda=-1$ and with $Q_{HO}(n,l)$ for $\lambda= 2$. Besides,
there is some freedom. Asymptotically for large $l$, both $Q_{C}$ and $Q_{HO}$
are proportional to $l$ with a slope equal to unity. Owing to the fact that
$Q_{C}$ and $Q_{HO}$ lead to lower and upper bounds, a unity slope is
maintained asymptotically whatever the value of $\lambda$. Asymptotically for
large $n$, both $Q_{C}$ and $Q_{HO}$ are proportional to $n$ and we maintain
also this characteristic behavior for other values of $\lambda$.
In this work we propose two forms of $Q(\lambda;n,l)$ which improve largely
the results as compared to the $Q_{C}$ and $Q_{HO}$ values. Although in a
context slightly different, R.L. Hall has considered an improvement of the results
based on a function $P_{n,l}(\lambda)$ which was shown as monotone increasing.
His prescription to improve the results (see Ref.~\citen{hall2000}) was different from ours,
but the spirit is the same.

The first prescription depends on two free functions $b$, $g$ and is chosen as
\begin{equation}
\label{eq:formNs}
Q_s(\lambda;n,l)= b(\lambda)n+l+g(\lambda).
\end{equation}
The preceding conditions impose $b(2)=2$, $g(2)=3/2$ and $b(-1)=1$ = $g(-1)$.
The simple form $Q_s$ is linear (it is called sometimes the ``linear
approximation") both in $n$ and $l$ and, thus, looks as simple as $Q_{C}$ and
$Q_{HO}$ but allows much better results, as will be shown below.
The reader may use it if he needs a simple formulation giving an average
relative precision of order of $10^{-2}$.

The second prescription depends on five free functions $a$, $c$, $d$, $e$, $f$ and reads
\begin{equation}
\label{eq:formNq}
Q_q(\lambda;n,l)= \frac{a(\lambda)n^2+l^2+c(\lambda)nl+d(\lambda)n+e(\lambda)l
+f(\lambda)}{n+l+1}.
\end{equation}
The boundary conditions are in this case $(a,c,d,e,f)(\lambda=2) = (2,3,7/2,
5/2,3/2)$ and $(a,c,d,e,f)(\lambda=-1) = (1,2,2,2,1)$. The form $Q_q$ 
is a rational fraction with a quadratic numerator (it is called
sometimes the ``quadratic approximation"). This form is more sophisticated that
$Q_s$ but it gives results with a very high accuracy of order of $10^{-4}$. Some
examples are given in the next section. Other forms have been tested but 
formulae~(\ref{eq:formNs}) and (\ref{eq:formNq}) appeared very convenient.
Contrary to the case $Q=Q_{HO}$, for which $\epsilon_{\textrm{AFM}}$ is a
upper bound and the case $Q=Q_{C}$ for which it is a lower bound, we have no
certainty concerning the position of the AFM energy as compared to the exact
energy in the cases $Q=Q_s$ or $Q=Q_q$; but in any case, we will show that
 the approximate value is always very close to the exact value.

Except the very special case of the linear potential, which is studied in
detail above, no analytical expression for the exact energy $\epsilon(\lambda;n,l)$
exists (even in S-state) for potentials with $\lambda \neq -1$ or $2$. Therefore to
compare our AFM results to the exact ones, one needs very accurate eigenvalues
$\epsilon_{\textrm{num}}(\lambda;n,l)$ of Hamiltonian~(\ref{eq:Hpowred2}). For that
purpose, we rely on a very efficient method, called the Lagrange mesh method. \cite{lag}\
Without difficulty, one can get an accuracy of $10^{-8}$ much higher than what could
be expected from AFM. Thus, in the following, we identify fully the exact value
$\epsilon$ and the numerical value $\epsilon_{\textrm{num}}$.

The ultimate aim is to determine the approximate AFM energies $\epsilon_{\textrm{AFM}}$
that stick as close as possible to the exact values. Two kinds of approximations,
denoted generically $(a)$, are considered; in both of them $\epsilon^{(a)}_
{\textrm{AFM}}$ is given by (\ref{eq:epsAFMpwmod}) but in the linear approximation
($(a)=s$) the principal quantum number is given by (\ref{eq:formNs}) while in the
quadratic approximation ($(a)=q$) this number results from (\ref{eq:formNq}). The
free parameters of each approximation are denoted $\sigma(\lambda)$. There are
two of them $\sigma=(b,g)$ for the linear approximation and five of them $\sigma=
(a,c,d,e,f)$ for the quadratic approximation. 

\subsubsection{Comparisons to numerical results}
\label{compnumresdet}

We now search for an analytical formula giving $\sigma(\lambda)$. In order to do 
that, we proceed in the following way.
In a first step, we choose a sample $\{\lambda_1=-1,\lambda_2,\ldots,\lambda_p=2
\}$ of $p$ values of $\lambda$ in the considered domain $-1 \leq \lambda \leq 2$.
We found that $p=10$ is a good compromise between the researched quality and the
numerical effort.
For each value $\lambda_i$ belonging to the sample, we calculate, using the Lagrange mesh method
\cite{lag}, an array of ``exact" eigenvalues $\epsilon(\lambda_i;n,l)$
for $n_\textrm{max}+1$ values of the radial quantum number $n$ ($n =0,\ldots,n_\textrm{max}$) and for
$l_\textrm{max}+1$ values of the orbital quantum number $l$ ($l =0,\ldots,l_\textrm{max}$). For the
precise study that we want, it is enough to choose $n_\textrm{max}=8$ and $l_\textrm{max}=5$. The AFM
results depend on the free parameters $\sigma(\lambda_i) = \sigma_i$. The game
is to find the values of $\sigma_i$ which make the AFM energies the closest to
the exact ones.

A tool to do that consists in the minimization of the chi-square quantity
\begin{equation}
\label{chi2}
\chi(\sigma_i)=\frac{1}{(n_\textrm{max}+1)(l_\textrm{max}+1)}\sum_{n=0}^{n_\textrm{max}}
\sum_{l=0}^{l_\textrm{max}} \left(
\epsilon(\lambda_i;n,l) - \epsilon_{\textrm{AFM}}(\sigma_i;n,l)\right)^2.
\end{equation}
Other choices are possible but we find this one very convenient. In some very
specific situations, this function may lead to bad results and we find convenient
to introduce an alternative minimization based on the difference quantity
\begin{equation}
\label{delt2}
\Delta(\sigma_i)=\frac{1}{(n_\textrm{max}+1)(l_\textrm{max}+1)}\sum_{n=0}^{n_\textrm{max}}
\sum_{l=0}^{l_\textrm{max}} \left|
\epsilon(\lambda_i;n,l) -  \epsilon_{\textrm{AFM}}(\sigma_i;n,l) \right|.
\end{equation}
The minimization of one of these functions provides the values $\sigma_i$ which
will be denoted $\sigma_{\textrm{num}}(\lambda_i)$ because they rely on a numerical
prescription. Generally the chi-square and the difference procedures give very
similar results, but sometimes a glance at the results may lead us to favor one
against the other. Of course for $\lambda=-1$ and $\lambda=2$, we know that the
values $\sigma_{\textrm{num}}(\lambda_i)$ given by the values discussed earlier
lead to the exact results so that the minimization procedure is useless and both
$\chi(\sigma_i)$ and $\Delta(\sigma_i)$ vanish. Over the full range of $\lambda$
sample (10 values), we found that the maximal value of the chi-square is
$4.4\ 10^{-4}$ and $2.2\ 10^{-7}$ for the linear and quadratic approximations
respectively, while the accuracy given by $\Delta$ is $1.33\%$ and $0.07\%$
respectively. Thus, one sees that the very simple prescription $(a)=s$ gives
already a relative accuracy of the order of $10^{-2}$ over the whole spectrum (54 values),
while the more sophisticated prescription $(a)=q$ gives practically the exact results.
Several formulae exist in literature for approximate values of
$\epsilon(\lambda;n,l)$ resulting from various techniques. Among them, one of
the most precise is reported in Ref.~\citen{QR} with the use of a WKB approach. It
must be stressed that our method, even in its crudest version, looks much
simpler and gives much better results. Thus, we are confident that our formulae
are really an improvement compared to the previous existing ones. 

The second step of our study is now the guess of continuous functions $\sigma
(\lambda)$ which stick as much as possible to the values $\sigma_{\textrm{num}}
(\lambda_i)$ for the particular values $\lambda=\lambda_i$. The choice is huge
because we have freedom on the form of the function, and then on the parameters
entering it. Since we research simplicity above all, we find that an hyperbola
form for all the parameters is very well suited. The fit of the 3 parameters entering
the definition of an hyperbola is done
using again a chi-square function $\chi ' $ based on the numerical value $\sigma_i$
and the proposed value $\sigma(\lambda_i)$.

We summarize below the definite choice for the various parameters. For the
linear approximation (\ref{eq:formNs}), one gets
\begin{equation}
\label{paramlin}
b(\lambda)=\frac{41\lambda+86}{13\lambda+58}, \quad g(\lambda)=\frac{5\lambda+17}
{2\lambda+14}.
\end{equation}
For the quadratic approximation (\ref{eq:formNq}), the corresponding functions
look like
\begin{eqnarray}
\label{paramquad}
&&a(\lambda)=\frac{43\lambda+82}{15\lambda+54},\quad c(\lambda)=\frac{171\lambda+675}
{29\lambda+281},\quad d(\lambda)=\frac{136\lambda+330}{25\lambda+122}, \\  \nonumber
&&e(\lambda)=\frac{109\lambda+517}{30\lambda+234},\quad f(\lambda)=
\frac{225\lambda+729}{94\lambda+598}.
\end{eqnarray}
One verifies that for $\lambda=-1$, $2$ the exact values are recovered. We choose to use
integer numbers below 1000 in order to match at best the real numbers yielded by the 
minimization procedure.

As a matter of test, we present in detail the special case of the linear
potential ($\lambda=1$) studied in Sect.~\ref{sec:wfobserv}. $\epsilon(1;n,0)$ is then
just proportional to the opposite of the $(n+1)$th zero of the Airy function. An asymptotic
solution of  Hamiltonian~(\ref{eq:Hpowred2}) valid for large $n$ is given by (\ref{Enex}).
Equating (\ref{Enex}) and (\ref{eq:formNs}), we find $b(1)=
\pi/\sqrt{3} \approx 1.814$ and $g(1)=\sqrt{3}\pi/4 \approx 1.360$. These
values are the exact ones to describe the spectrum for $l=0$ and large
$n$, but not necessarily for $l \neq 0$ or low $n$ values. However the
values coming from our fit (\ref{paramlin}) are $b(1)=127/71 \approx 1.789$
and $g(1)=22/16 = 1.375$, very close to these new values. Assuming
that this procedure can be extended for any values of $l$, the hyperbola can be
determined completely by using the exact values $b(-1)=1$, $g(-1)=1$, $b(2)=2$,
and $g(2)=3/2$. The coefficients are then given by 
\begin{eqnarray}
\label{bc3}
z&=&\sqrt{3}\pi , \nonumber \\
b(\lambda)&=& \frac{(4z-18) \lambda+(18-2z)}
{(3z-15) \lambda+(21-3z)}, \nonumber \\
g(\lambda)&=& \frac{(7z-36) \lambda+(36-5z)}{(6z-32) \lambda+
(40-6z)}.
\end{eqnarray}
These relations could look quite complicated but it is worth noting the
symmetries existing between the absolute values of the numbers present in
the formula. When a number appears twice in a coefficient of the numerator
(denominator), it is the arithmetic mean of two other corresponding numbers
in the denominator (numerator). 
A number which appears twice in $g(\lambda)$ is the double than the corresponding
number in $b(\lambda)$. 

\begin{figure}
\centerline{
\includegraphics*[width=\halftext]{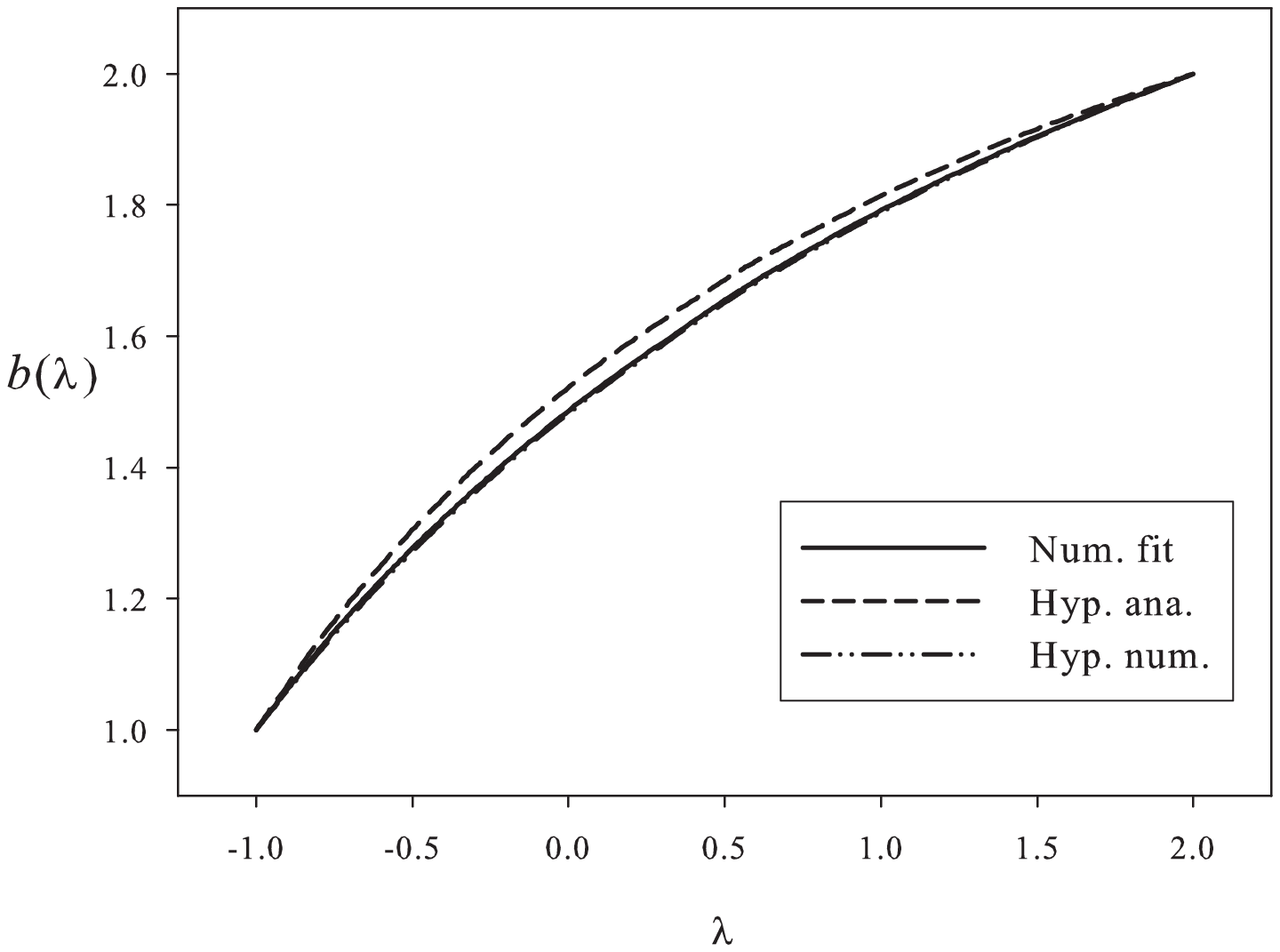}\quad
\includegraphics*[width=\halftext]{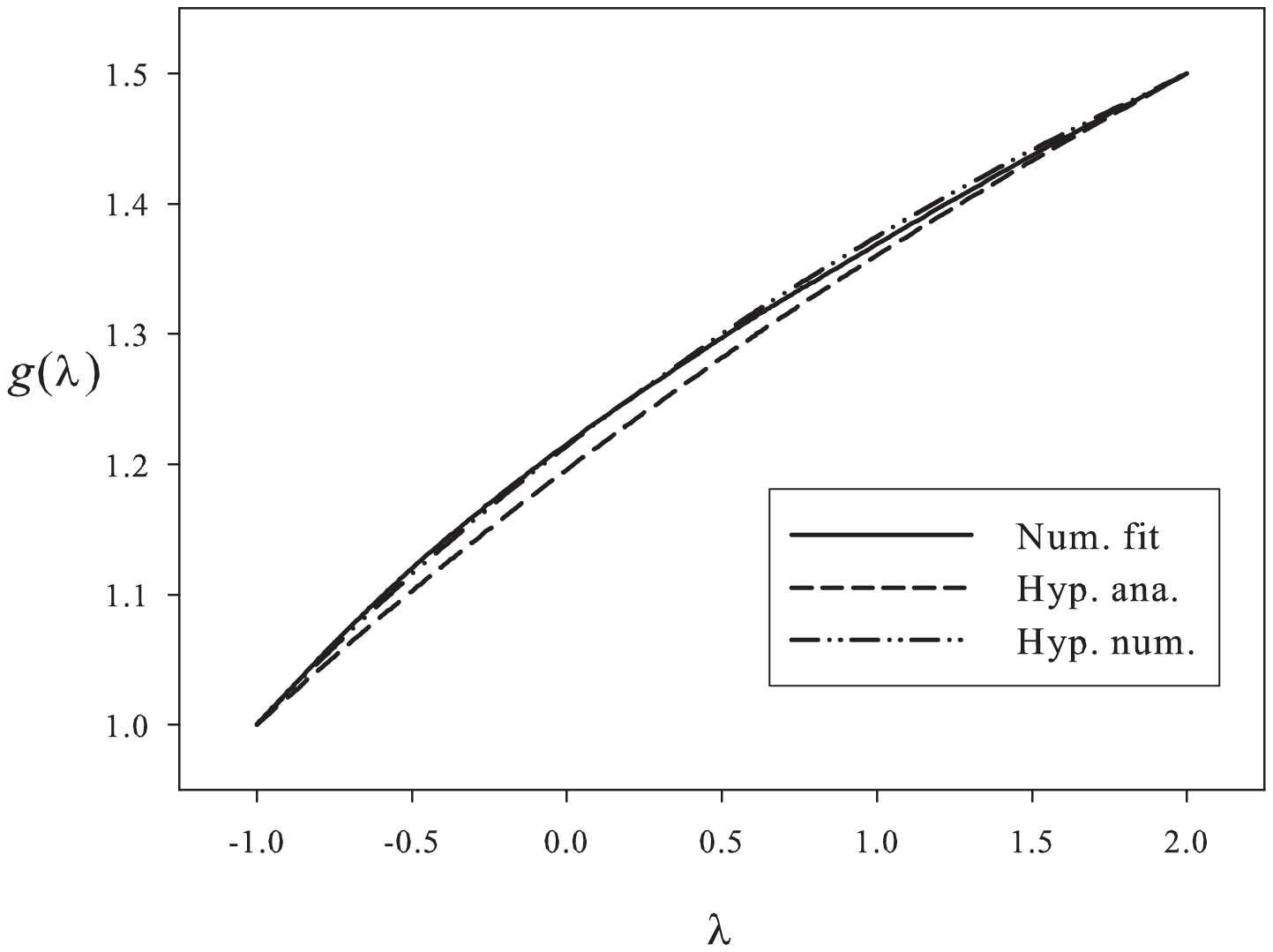}
}
\caption{Coefficients $b(\lambda)$ and $g(\lambda)$ from~(\ref{eq:formNs}) 
as a function of $\lambda$,
obtained from various approximations: numerical fit minimizing the function
$\chi$ (solid line), hyperbolae from~(\ref{bc3}) (short-dashed line),
hyperbolae from~(\ref{paramlin}) (dashed-dotted line). These last curves are
nearly indistinguishable from the solid lines.}
\label{bglamb}
\end{figure}

In Fig.~\ref{bglamb}, we give an idea of how good is the fit of
$\sigma(\lambda)$ with hyperbolae. We show only the plots for the functions
$b(\lambda)$ and $g(\lambda)$ for the linear approximation. The results look
similar for the functions $\sigma(\lambda)$ of the quadratic approximation.
The solid lines represent the numerical values $\sigma_{\textrm{num}}(\lambda_i)$
obtained by a chi-square procedure; in a sense they are the best that we can
find. The short-dashed line represent an hyperbola determined
analytically; the
corresponding functions are given by~(\ref{bc3}). The dashed-dotted line
represents the hyperbola which sticks as close as possible to the numerical
curve; the corresponding functions are given by~(\ref{paramlin}). This last
prescription appears to be very good.

\begin{table}[ht]
\caption{Values of $\epsilon(\lambda=1;n,l)$ for various approximations. For
each set $(n,l)$, the first line (bold case) is the exact result obtained by numerical
integration, the other lines are obtained with the AFM approximation~(\ref{eq:epsAFMpwmod}).
For the principal quantum number $Q(1;n,l)$, the second line uses the
prescription~(\ref{bc3}), the third line the prescription~(\ref{eq:formNs})
with~(\ref{paramlin}) (linear approximation), and the fourth line the
prescription~(\ref{eq:formNq}) with~(\ref{paramquad}) (quadratic approximation).}
\label{tab1}
\begin{center}
\begin{tabular}{ccccc}
\hline\hline\noalign{\smallskip}
 $l$ & $\epsilon(1;0,l)$ & $\epsilon(1;1,l)$ & $\epsilon(1;2,l)$ &
 $\epsilon(1;3,l)$ \\ [3pt]
\hline
 0 & \textbf{1.47292} &  \textbf{2.57525} &  \textbf{3.47773} &  \textbf{4.27536} \\
   & 1.46167 &  2.57138 &  3.47560 &  4.27394 \\
   & 1.47214 &  2.56575 &  3.45909 &  4.24854 \\
   & 1.47310 &  2.57544 &  3.47766 &  4.27508 \\
\noalign{\smallskip}
 1 & \textbf{2.11746} &  \textbf{3.07701} &  \textbf{3.91056} &  \textbf{4.66528} \\
   & 2.11057 &  3.08645 &  3.92585 &  4.68320 \\
   & 2.11929 &  3.08132 &  3.91032 &  4.65894 \\
   & 2.11743 &  3.07666 &  3.91021 &  4.66500 \\
\noalign{\smallskip}
 2 & \textbf{2.67619} &  \textbf{3.54649} &  \textbf{4.32712} &  \textbf{5.04580} \\
   & 2.67098 &  3.56156 &  4.35159 &  5.07529 \\
   & 2.67874 &  3.55678 &  4.33684 &  5.05198 \\
   & 2.67630 &  3.54615 &  4.32670 &  5.04544 \\
\noalign{\smallskip}
 3 & \textbf{3.18188} &  \textbf{3.98898} &  \textbf{4.72763} &  \textbf{5.41584} \\
   & 3.17757 &  4.00682 &  4.75742 &  5.45277 \\
   & 3.18469 &  4.00231 &  4.74332 &  5.43029 \\
   & 3.18209 &  3.98873 &  4.72722 &  5.41545 \\
\noalign{\smallskip}\hline 
\end{tabular}
\end{center}
\end{table}

In Table~\ref{tab1}, we present the lower part of the spectrum ($n_\textrm{max}=l_\textrm{max}=3$) obtained
for different types of approximations and compare the various results to the exact ones.
It is clear that all the considered approximations give very good results.
As expected, the approximation with~(\ref{bc3}) is very good for $l=0$
and large $n$ values, but it becomes worse and worse (although not so bad) as
$l$ increases. The linear approximation~(\ref{eq:formNs}) is very good
everywhere and should be considered as a good compromise between simplicity
and accuracy while the quadratic approximation is really excellent everywhere
and must be used if we want analytical accurate expression of the exact energies.

Other approximate forms for the eigenvalues of Hamiltonian~(\ref{eq:Hpowred2})
exist and give good results \cite{fab88,bose}, but they are not considered
here because they do not give the exact result for the cases $\lambda=-1$ and
$\lambda=2$.

Obviously the functions $\sigma(\lambda)$ depend on the points used for the
fit but also on the particular choice of the functions $\chi$ and
$\Delta$. Other definitions--relative error instead of absolute error
or different summations on quantum numbers--would have given other numbers.
Nevertheless, the quality of the results tends to prove that they are very close to the best possible ones.

\section{Schr\"odinger equation with an arbitrary potential}
\label{sec:schrodAFM}

\subsection{Scaling laws}
\label{sec:scalNRpowl}

Let $E(m,G,a)$ be the eigenvalues of a Schr\"{o}dinger equation corresponding
to a system of reduced mass $m$ subject to a potential of intensity $G$ and
characteristic inverse length $a$. The scaling law gives the relationship
between $E(m,G,a)$ and $E(m',G',a')$. 
Let us start from the corresponding Schr\"{o}dinger equations
\begin{eqnarray}
\label{eq:scheq1}
\left[ -\frac{1}{2m} \Delta_r + G V(ar) - E(m,G,a) \right] \Psi(\bm{r})=0, \\
\label{eq:scheq2}
\left[ -\frac{1}{2m'} \Delta_r + G' V(a'r) - E(m',G',a') \right]
\Psi'(\bm{r})=0.
\end{eqnarray}
The important point is that it is the \emph{same function} $V(x)$ which appears in
both equations.
In (\ref{eq:scheq2}), let us make the change of variables 
$\bm{r}=\alpha \bm{x}$ and multiply it by $\chi$. Now, 
we choose the arbitrary parameters $\alpha$ and $\chi$ in order to fulfill the
conditions $\chi/(m' \alpha^2)=1/m$ and $\alpha a'=a$. In other words, we
impose the following values
\begin{equation}
\label{eq:detparamscal}
\alpha = \frac{a}{a'}, \quad \chi = \frac{m'}{m} \left(\frac{a}{a'}
\right )^2.
\end{equation}
With these values, (\ref{eq:scheq2}) can be recast into the form
\begin{equation}
\label{eq:scheq2p}
\left[ -\frac{1}{2m} \Delta_x + G' \frac{m'}{m} \left(\frac{a}{a'} \right )^2
V(ax) - \frac{m'}{m} \left(\frac{a}{a'} \right )^2 E(m',G',a') \right]
\Psi'\left(\frac{a}{a'} \bm{x}\right)=0.
\end{equation}
Equation~(\ref{eq:scheq1}) can be recovered, provided one makes the
identification $G = G' (m'/m) (a/a')^2$ and a similar relation for the
energies.

The scaling law is thus expressed in its most general form as
\begin{equation}
\label{eq:scallawgenp}
E(m,G,a) = \frac{m'}{m} \left(\frac{a}{a'} \right )^2 E \left (
m', G'=G \frac{m}{m'} \left(\frac{a'}{a} \right )^2,a' \right ).
\end{equation}
In fact, it is always possible to define the function $V(x)$ so that
$a'=1$ (for example in defining a new position operator as $a'r$).
In what follows, and without loss of generality, we will apply
the scaling law for energies under the form
\begin{equation}
\label{eq:scallawgen}
E(m,G,a) = \frac{m'a^2}{m} E \left ( m', G'= \frac{mG}{m' a^2},1 \right ).
\end{equation}
This equality is very powerful since it is valid for the 
eigenvalues of a nonrelativistic Schr\"{o}dinger equation with an
\emph{arbitrary central potential}. It allows to express the energy in terms
of a dimensionless quantity and some dimensioned factors, as we will see below.
Moreover it is possible to give to the mass $m'$ any particular convenient
value (for example $m' = 1$) so that $E(m,G,a)$ is expressed through the
energies of a very simple reduced equation depending on a single parameter
$G'$. Instead, one can choose to impose a given value of the intensity $G'$
(for example $G'=1$) so that $E(m,G,a)$ is expressed through the
energies of a very simple reduced equation depending on a single parameter
$m'$.

To summarize the previous discussion, the scaling laws are very powerful
since they allow to express the most general form of the energy $E(m,G,a)$
in terms of the eigenvalues of a reduced equation in which two parameters
among the three available can be set at a given fixed value. The chosen
parameters as well as their values are determined in the most convenient
way.

\subsection{General formulae for energies and mean radius}
\label{sec:invformenerg}

In Sect.~\ref{sec:PowerLawpot}, we remarked that, using either $P(r)=r^2$ or $P(r)=-1/r$,
we obtain, for power-law potentials, AFM energies which have exactly the same
form but differ only by the expression of the principal quantum number $Q$.
In other words, this means that $\epsilon_{\textrm{AFM}}(\lambda;n,l)$ appears
always in the form $F(\lambda;Q(n,l))$ with a universal form of the function
$F$. The only difference is that, in $F$, we must use $Q_{HO}(n,l)$ for $P(r)=
r^2$ and $Q_{C}(n,l)$ for $P(r)=-1/r$. This is obvious from the comparison of
(\ref{eq:epsAFMr2}) and (\ref{eq:epsAFMrm1}) (or even from (\ref{eq:elogred}) and
(\ref{eq:elogred2})).

This property allowed us to modify the expression of this number $Q$ in order to
improve drastically the quality of the results.
This universal property seems also valid for the expression of the mean radius,
as you can see from (\ref{eq:minr0pw}) and (\ref{eq:minr0pw2}) (or even for
(\ref{eq:minr0log}) and (\ref{eq:minr0log2})).

We will show in this part that these invariance properties are even more general
and are valid not only for a power-law potential but whatever the potential
$V(r)$ to be used. \cite{bsb08b}\
This property is even stronger since it persists for any power-law type for
the starting function $P(r)$ and not only for a quadratic or Coulomb form.
Let us demonstrate this fundamental property.

For beginning, let us recall that the AFM energies for the Hamiltonian~(\ref{eq:fgplw})
are given by (see (\ref{eq:scallpw}) and (\ref{eq:epsAFMpwmod}))
\begin{equation}
\label{eq:eigenerplpot}
E_{\textrm{AFM}}(m,a,\lambda;n,l)=\frac{2+\lambda}{2 \lambda}
(a |\lambda|)^{\frac{2}{\lambda+2}} \left ( \frac{Q(\lambda;n,l)^2}{m} \right )
^{\frac{\lambda}{\lambda+2}}.
\end{equation}
If we ask that $E_{\textrm{AFM}}(m,a,\lambda;n,l)=E(m,a,\lambda;n,l)$, we can consider that this formula gives the definition of the principal quantum number 
$Q(\lambda;n,l)$ for all power-law potentials. The exact form is known only in a 
limited number of cases. Obviously, $Q(-1;n,l)=n+l+1$ and $Q(2;n,l)=2 n+l+3/2$.
Using the notations of Appendix~\ref{sec:obs_Ai}, we can write 
\begin{equation}
\label{eq:QAi}
Q(1;n,0)=2 \left( -\frac{\alpha_n}{3} \right)^{3/2} \approx \frac{\pi}{\sqrt{3}} n + \frac{\sqrt{3}}{4}\pi.
\end{equation}
In the other cases, we showed that (\ref{eq:eigenerplpot}) was able to give the results with a relative accuracy better than $10^{-4}$ if a good choice is made for the value of $Q$.
Following the philosophy of Sect.~\ref{sec:recproc}, we consider that
expression (\ref{eq:eigenerplpot}) represents the \emph{exact expression} of
our problem 
and pursue the general AFM method with the starting function $P(r) =
P^{(\lambda)}(r) = \textrm{sgn}(\lambda) r^\lambda$.

In this case, $K(r)$ = $V'(r)r^{1-\lambda}/|\lambda|$ and the extremization
condition (\ref{eq:trans2}) leads to the transcendental equation
\begin{equation}
\label{eq:r0pw1}
r_0^{\lambda+2} K(r_0) = \frac{Q(\lambda;n,l)^2}{m |\lambda|}.
\end{equation}
It seems that $r_0$ is really $\lambda$-dependent, but replacing $K(r)$ by
its expression in terms of the potential, this equation turns out to be
\begin{equation}
\label{eq:r0pw2}
r_0^3 V'(r_0) = \frac{Q(\lambda;n,l)^2}{m}.
\end{equation}
For a given potential, the inverse function ${\cal D}(x)$, 
\begin{equation}
\label{eq:defcalD}
\mathcal{D}(x^3 V'(x)) = x \quad \textrm{or} \quad \mathcal{D}(x)^3 V'(\mathcal{D}(x)) = x,
\end{equation}
is obviously universal and one has $r_0 = {\cal D}(Q^2/m)$.

The value of the AFM energy follows from (\ref{eq:Egen3c}) and reads
\begin{equation}
\label{eq:E0pw1}
E_{\textrm{AFM}} = \frac{Q(\lambda;n,l)^2}{2 m r_0^2}+V(r_0).
\end{equation}
Thus the function 
\begin{equation}
\label{eq:defcalF}
{\cal F}(x) = \frac{x}{2{\cal D}(x)^2} + V({\cal D}(x))
\end{equation}
is also universal and one has
\begin{equation}
\label{eq:E0pw1calF}
E_{\textrm{AFM}}={\cal F}(Q^2/m).
\end{equation}
Since it is always possible to choose the same mass for the genuine
Schr\"{o}dinger equation with $V(r)$ and the Schr\"{o}dinger equation with
the AFM potential $P^{(\lambda)}(r)$, the previous demonstration shows
that both the mean radius and the eigenenergy depends on $\lambda$
through the principal quantum number $Q(\lambda;n,l)$ only.

We end up with the very important conclusion that can be stated as a theorem
(expressed below for the energy but also valid for the mean radius)
\begin{quote}
\emph{If, in the expression $E(Q(\lambda;n,l))$ of the approximate energies
resulting from the AFM with $P^{(\lambda)}(r)$, one makes the substitution
$Q(\lambda;n,l) \rightarrow Q(\eta;n,l)$ (so that $E(Q(\lambda;n,l)) \rightarrow
E(Q(\eta;n,l))$ with the same functional form for $E$), one obtains the approximate
eigenenergies resulting from the AFM with $P^{(\eta)}(r)$}.
\end{quote}
In a sense, as long as we use a power-law potential $P^{(\lambda)}(r)$ 
as starting potential, there is a universality of the approximate AFM expression
of the eigenvalue, depending only on the potential $V(r)$. The only remainder
of the particular chosen potential $P^{(\lambda)}(r)$ is the expression of
$Q(\lambda;n,l)$, as given by (\ref{eq:formNs}) for instance. This result holds
whatever the form chosen for the potential $V(r)$, even if we are unable to
obtain analytical expressions for the AFM approximation.

This property was first remarked for the particular case of a power-law potential
$V(r)=r^\lambda$ switching from the harmonic oscillator ($\lambda=2$) to the
Coulomb potential ($\lambda=-1$). We proved here that it is in fact totally
general. It is probably related to a well-known property in classical mechanics:
one can pass from the motion of a harmonic oscillator to the Kepler motion by a
canonical transformation.
This universality property will be used extensively in the following applications;
it allows us to choose the form $P^{(\lambda)}(r)$ which is the most convenient
for our particular situation. Nevertheless, the existence of possible lower or upper
bounds can only be guaranteed for $\lambda=-1$ and $2$, since formula~(\ref{eq:eigenerplpot})
is exact only in these cases. 

By rewriting (\ref{eq:r0pw2}) and recalling that the kinetic energy is denoted by
$T(\bm p)$ with $T(x)=\frac{x^2}{2m}$, the AFM formula for the energies can be written
into the form
\begin{eqnarray}
\label{NRvir1}
E_{\textrm{AFM}} &=& T(p_0)+V(r_0), \\
\label{NRvir2}
p_0 &=& \frac{Q}{r_0}, \\
\label{NRvir3}
p_0 T'(p_0)&=& r_0 V'(r_0),
\end{eqnarray}
where $Q$ stands for $Q(\lambda;n,l)$ to lighten the notations and to stress that we
have some freedom in the choice of this quantity. Equation~(\ref{NRvir2})
defines a ``mean impulsion" $p_0$ from the mean radius $r_0$ resulting from~(\ref{eq:r0pw2}).
Looking at equation~(\ref{NRvir3}), one can recognize the general form of the virial theorem.
\cite{luch90}\ Finally, (\ref{NRvir1}) gives the energy as a sum of the kinetic energy
evaluated at the mean impulsion $p_0$ and the potential energy evaluated at the mean
radius $r_0$. This makes clearly appear the physical content of the AFM approximation.
Let us point out that $p_0$ is not a linear function of $Q$, as suggested by (\ref{NRvir2}),
since $r_0$ has a complicated dependence on $Q$ through (\ref{NRvir3}). The approximate eigenstate 
$| \nu_0 \rangle$ is a solution of $h(\nu_0)=\bm p^2/(2 m)+\textrm{sgn}(\lambda)\nu_0 r^\lambda$ and has a 
characteristic size depending on $m\nu_0=m K(r_0)$ 
(see (\ref{psiOH}), (\ref{psiHy}) and (\ref{psiAi})). This quantity
can be easily computed once (\ref{NRvir3}) is solved. 

The virial theorem applied to Hamiltonian $\tilde H(\nu_0)$, given by (\ref{eq:fgplw})
with $a\to \nu_0$, implies that
\begin{equation}
\label{NRvir4}
\left\langle \nu_0 \left| \frac{\bm p^2}{m} \right| \nu_0 \right\rangle = \left\langle
\nu_0 \left| \nu_0 |\lambda| r^\lambda \right| \nu_0 \right\rangle,
\end{equation}
with $\nu_0=K(r_0)$ given by~(\ref{eq:r0pw1}). The use of definition~(\ref{NRvir2})
with property (\ref{pr0}) gives
\begin{equation}
\label{NRvir5}
\left\langle \nu_0 \left| \frac{\bm p^2}{p_0^2} \right| \nu_0 \right\rangle =
\left\langle \nu_0 \left| \frac{r^\lambda}{r_0^\lambda} \right| \nu_0 \right\rangle = 1.
\end{equation}
The last equality is another way to write the fundamental property (\ref{pr0}). These equations,
as well as the boundary character of the solution, are only applicable when the exact 
form $Q(\lambda;n,l)$ is used. In practice, this occurs when $\lambda = -1$ or $2$, 
or $\lambda = 1$ for S-states only. Nevertheless, a better accuracy can be obtained by 
an appropriate choice of $Q$ as shown in Sect.~\ref{sec:impeigene}. 

In the rest of this section, several interactions will be studied with the AFM. Approximations
for all states in the spectrum will be given. When the potential allows only a finite number
of bound states, formulae for corresponding critical constants will be also presented. 

\subsection{Square root potentials}
\label{sec:sqrtpot}

The square root potential that we study in this section is fundamental for
the understanding of hybrid mesons, which are exotic mesons actively researched
nowadays. Indeed, there are two possible descriptions of hybrid mesons within constituent 
models:
\begin{itemize}
	\item First, a genuine three-body object made of a quark, an antiquark and a
	constituent gluon.
	\item Second, a two-body object made of a quark and an antiquark in the
	potential due to the gluon field in an excited state.
\end{itemize}
It has been shown that these two
pictures of the same object are, to a large extent, equivalent.
\cite{Bui06a,Bui06b,Bui07}\ In this part, we focus only on the second
aspect of the description.

In general, the string energy, and, therefore, the potential energy between the
static quark and antiquark in the excited gluon field is given by
\cite{Arv83,All98}
\begin{equation}
\label{eq:potex}
V(r)=\sqrt{a^2 r^2+b^2},
\end{equation}
where $a$ is the usual string tension while $b^2=2 \pi a K + C$ is a term
exhibiting the string excitation number $K$ and a constant $C$. These values
depend on the model adopted: a pure string theory \cite{Arv83} or a more
phenomenological approach. \cite{Bui07,All98}\
For the study of heavy hybrid mesons, it is therefore very interesting to
calculate the eigenenergies of the Schr\"{o}dinger equation governed by
the Hamiltonian
\begin{equation}
\label{eq:Hroot}
H=\frac{\bm{p}^2}{2m} + \sqrt{a^2 r^2+b^2},
\end{equation}
where $m$ is the reduced mass and where the parameters $a$ and $b$ depend
on the model adopted, which must be confronted to experimental results
ultimately.

In this section we give an analytical approximate expression for the 
eigenvalues of Hamiltonian~(\ref{eq:Hroot}). In Ref.~\citen{Sem09a}, those results
have been exploited to study the mass of a heavy hybrid meson in the 
picture of an excited color field. With
this mass formula, it was possible to predict the general behavior of
the mass spectrum as a function of the quantum numbers of the system and
search for possible towers of states. These results can then be used as
a guide for experimentalists. 

\subsubsection{Expression of the energies}
\label{sec:Esqrt}

Using scaling laws~(\ref{eq:scallawgen}), dimensionless variables
$\epsilon$ and $\beta$ can be defined so that the general energy $E(m,a,b)$
is easily calculated
\begin{equation}
\label{eq:eps}
E(m,a,b)=\left( \frac{2 a^2}{m} \right)^{1/3} \epsilon (\beta), \quad
\textrm{with} \quad \beta=b^2 \left( \frac{m}{2 a^2} \right)^{2/3}.
\end{equation}
$\epsilon (\beta)$ is an eigenvalue of the dimensionless reduced Hamiltonian
\begin{equation}
\label{eq:phroot}
h=\frac{\bm{p}^2}{4} + \sqrt{r^2+\beta}.
\end{equation}

Let us apply the AFM in order to find approximate expressions for the eigenvalues
of the Hamiltonian~(\ref{eq:phroot}). Equation~(\ref{eq:r0pw2}) gives immediately 
\begin{equation}
\label{eq:minsqr01}
4r_0^8 - Q^4 r_0^2 - \beta Q^4 = 0.
\end{equation}
A simple change of variable $x=2 r_0^2/Q^{4/3}$
plus the definition of the parameter
\begin{equation}
\label{eq:defYroot}
Y=\frac{16\beta}{3 Q^{4/3}}
\end{equation}
allows to transform the transcendental equation in the
fourth order reduced equation $4 x^4 - 8 x - 3Y = 0$ whose
solution is given in terms of the $G_{-}$ function discussed
in Sect.~\ref{sec:Fourtheq}.
This allows the determination of $\epsilon_{\textrm{AFM}}$ by (\ref{eq:E0pw1}). 
\begin{equation}
\label{eq:Enu0root}
\epsilon_{{\rm AFM}}(\beta;n,l)=2\sqrt{\frac{\beta}{3 Y}}
\left[G_{-}^2(Y)+\frac{1}{G_{-}(Y)} \right],
\end{equation}
with $Y=Y(\beta;n,l)$ given by~(\ref{eq:defYroot}).
The problem is entirely solved.

Equation~(\ref{eq:Enu0root}) is complicated but quite accurate. In order to
get a better insight into this formula, it is interesting to calculate
the limits when $Y \gg 1$ and $Y \ll 1$ because they allow an easier
comparison with experimental data. This technical point was treated
in Ref.~\citen{Sem09a}.

\subsubsection{Comparisons to numerical results}
\label{sec:compsqrt}

As we discussed in Sect.~\ref{sec:invformenerg}, it is
possible to improve drastically the results given by~(\ref{eq:Enu0root})
if, instead of using the natural forms $Q=Q_{HO}$ or $Q=Q_{C}$,
we switch to a more convenient expression. The only relevant dynamical
parameter in this case being $\beta$, it is natural to propose an expression
depending on this parameter: $Q(\beta;n,l)$. For power-law potentials we
proposed a ``linear" (in terms of $n,l$) approximation $Q_s$ and a
``quadratic" approximation $Q_q$. For our special potential, one can adopt
the same prescriptions, or choose a new better one, if one wishes. Since our
aim is not to obtain the most accurate possible analytic expressions at the
price of complicated formulae, but rather to show that the AFM is able to give
already very good results even with very simple expressions, we adopt here
a linear expression, so that we choose $Q(\beta;n,l)$ under the form
\begin{equation}
\label{eq:formNroot}
Q(\beta;n,l)=A(\beta)n+l+C(\beta).
\end{equation}
We allow the $\beta$ value to vary between $\beta = 0$ for which the potential
is purely linear and $\beta=\infty$ for which it is harmonic. We are again in
a situation for which $Q=Q_{HO}$ gives an upper bound while $Q=Q_{C}$ provides
a lower bound. 

\begin{table}[ht]
\caption{Comparison between the exact values
$\epsilon(\beta;n,l)$ (2nd line in bold) and analytical approximate expressions
$\epsilon_{{\rm AFM}}(\beta;n,l)$ for the eigenvalues of Hamiltonian~
(\ref{eq:phroot}) with $\beta=1$. For each set $(n,l)$, the exact
result is obtained by numerical integration. 3rd line: approximate results
are given by (\ref{eq:Enu0root}) with (\ref{eq:defYroot}),
(\ref{eq:formNroot}) and (\ref{eq:coefacd}); 1st line: upper bounds obtained
with $Q=Q_{HO}$; 4th line: lower bounds obtained with $Q=Q_{C}$.}
\label{tab:comproot} 
\begin{center}
\begin{tabular}{cccccc}
\hline\hline\noalign{\smallskip}
$l$ & $\epsilon(1;0,l)$ & $\epsilon(1;1,l)$ & $\epsilon(1;2,l)$ & $\epsilon(1;3,l)$ & $\epsilon(1;4,l)$\\ [3pt]
\hline
 0 & 1.94926 & 2.99541 & 3.90193 & 4.72059 & 5.47723 \\
   & \textbf{1.91247} & \textbf{2.89556} & \textbf{3.74112} & \textbf{4.50374} & \textbf{5.20859} \\
   & 1.89549 & 2.85420 & 3.69078 & 4.44883 & 5.15078 \\
   & 1.65395 & 2.22870 & 2.75000 & 3.23240 & 3.68492 \\
\noalign{\smallskip}
 1 & 2.49495 & 3.46197 & 4.32027 & 5.10556 & 5.83725 \\
   & \textbf{2.45074} & \textbf{3.34652} & \textbf{4.14232} & \textbf{4.87138} & \textbf{5.55148} \\
   & 2.44621 & 3.32970 & 4.11913 & 4.84403 & 5.52098 \\
   & 2.22870 & 2.75000 & 3.23240 & 3.68492 & 4.11355 \\
\noalign{\smallskip}
 2 & 2.99541 & 3.90193 & 4.72059 & 5.47723 & 6.18692 \\
   & \textbf{2.94841} & \textbf{3.77899} & \textbf{4.53310} & \textbf{5.23246} & \textbf{5.88996} \\
   & 2.95032 & 3.77678 & 4.52783 & 5.22459 & 5.87970 \\
   & 2.75000 & 3.23240 & 3.68492 & 4.11355 & 4.52250 \\
\noalign{\smallskip}
 3 & 3.46197 & 4.32027 & 5.10556 & 5.83725 & 6.52732 \\
   & \textbf{3.41419} & \textbf{4.19405} & \textbf{4.91307} & \textbf{5.58628} & \textbf{6.22329} \\
   & 3.41969 & 4.20097 & 4.91998 & 5.59242 & 6.22821 \\
   & 3.23240 & 3.68492 & 4.11355 & 4.52250 & 4.91485 \\
\noalign{\smallskip}
 4 & 3.90193 & 4.72059 & 5.47723 & 6.18692 & 6.85935 \\
   & \textbf{3.85430} & \textbf{4.59335} & \textbf{5.28251} & \textbf{5.93264} & \textbf{6.55111} \\
   & 3.86189 & 4.60620 & 5.29790 & 5.94903 & 6.56756 \\
   & 3.68492 & 4.11355 & 4.52250 & 4.91485 & 5.29295 \\
 \noalign{\smallskip}\hline
\end{tabular}
\end{center}
\end{table}

The procedure used is similar to the one presented in 
Sect.~\ref{sec:impeigene} and is described in detail in Ref.~\citen{Sem09a}. 
In order to obtain functions which are as simple as possible, continuous
in $\beta$, and which reproduce at best the exact values, we choose
hyperbolic forms for $A(\beta)$ and $C(\beta)$. Explicitly,
we find
\begin{equation}
\label{eq:coefacd}
A(\beta) = \frac{8 \beta + 102}{4 \beta + 57}, \quad
C(\beta) = \frac{30 \beta + 53}{20 \beta + 39}.
\end{equation}
The integers appearing in $A$ and $C$ are rounded numbers whose magnitude
is chosen in order to not exceed too much 100.
The $A$ and $C$ functions have been constrained to exhibit the right behavior
$A \to 2$ and $C \to 3/2$ for very large values of $\beta$. formulae~
(\ref{eq:coefacd}) give $A(0)=102/57\approx 1.789$ and $C(0)=53/39\approx 1.359$.
These values are such that $A(0)\approx \pi/\sqrt{3}\approx 1.814$ and $C(0)
\approx \sqrt{3}\pi/4\approx 1.360$, as expected from the results of a
nonrelativistic linear potential. 

Our results are exact for $\beta \to \infty$, 
and the error is maximal for small values of $\beta$. But, over the
whole range of $\beta$ values, the results given by our analytical expression
can be considered as excellent.
Just to exhibit a quantitative comparison, we report in Table~\ref{tab:comproot}
the exact $\epsilon(\beta;n,l)$ and approximate $\epsilon_{{\rm AFM}}(\beta;n,l)$
values obtained for $\beta=1$, a value for which the corresponding potential
is neither well approximated by a linear one nor a harmonic one. 
As can be seen, our approximate expressions are better than $1\%$ for any
value of $n$ and $l$ quantum numbers. Such a good description is general
and valid whatever the parameter $\beta$ chosen.

The upper bounds obtained with $P(r)=r^2$ are far better than the lower bounds
computed with $P(r)=-1/r$. This is expected since the potential
$\sqrt{a^2 r^2+b^2}$ is closer to a harmonic interaction than to a Coulomb one.
Better lower bounds could be obtained with $P(r)=r$. But, the exact form of $Q$
is not known for this potential, except for $l=0$ for which $Q$ is given by
(\ref{eq:QAi}). With the approximate form 
$Q=(\pi/\sqrt{3}) n+l+\sqrt{3}\pi/4$ \cite{bsb08a,bsb08b}, we have checked that
results obtained are good but the variational character cannot be guaranteed.

\subsection{Exponential-type potentials}
\label{sec:exppot}

In this section, we apply the AFM to find approximate closed
analytical formulae for central potentials of exponential form, that is $-\alpha\,
{\rm e}^{-(\beta\, r)^{\eta}}$ where $\alpha$, $\beta$, $\eta$
are positive real number.
We thus start with the following Hamiltonian
\begin{equation}\label{h0def}
  H=\frac{{\bm p}^2}{2m}-\alpha\, {\rm e}^{-(\beta\, r)^{\eta}}.
\end{equation}
Two particular cases are of current use in physics: 
\begin{itemize}
  \item The first one is the Gaussian potential ($\eta=2$) which was intensively used
in molecular and nuclear physics because it allows very often an exact analytical
expression of various matrix elements. 
  \item The second one is the pure exponential ($\eta=1$), used also in nuclear
physics and for which exact analytical expression for the energies in S-state are known.
\end{itemize}

\subsubsection{Expression of the energies}
\label{sec:Eexp}

Using the scaling laws, the eigenenergies $E(m,\alpha,\beta,\eta;n,l)$ can be written
\begin{equation}
\label{eq:scalexp}
E(m,\alpha,\beta,\eta;n,l)=\frac{\beta^2}{2m} \epsilon(\eta,g;n,l),
\end{equation}
where $\epsilon(\eta,g;n,l)$ is en eigenvalue of the dimensionless Hamiltonian 
\begin{equation}\label{hdefexp}
	h={\bm p}^2-g\, {\rm e}^{-r^{\eta}},
\end{equation}
 with
\begin{equation}\label{gdef}
	g=\frac{2m\alpha}{\beta^{2}}.
\end{equation}
An interesting feature of exponential-type potentials is that they all admit a
\emph{finite} number of bound states that depends on the dimensionless parameter
$g$, ruling the potential depth, and defined by (\ref{gdef}). There exists thus
``critical constants": Potential depths beyond which new bound states appear. We
refer the reader to Ref.~\citen{brau1} for detailed explanations about how to compute
critical constants in a given potential. The
definition and the properties of the Lambert function, that will frequently
appear in our calculations, are given in Appendix~\ref{sec:lambert}.
Some of these potentials have been
studied with the AFM in Ref.~\citen{bsb09a}.

The application of (\ref{eq:r0pw2}) gives  
\begin{equation}
\label{eq:r0exp}
r_0^{\eta+2} \exp(-r_0^\eta) = \frac{2 Q^2}{\eta\, g}.
\end{equation}
Introducing the parameter
\begin{equation}
\label{eq:defYexpl0}
Y_\eta = \frac{\eta}{\eta+2} \left( \frac{2 \,Q^2}{\eta\,g},
\right)^{\eta/(\eta+2)},
\end{equation}
the solution of~(\ref{eq:r0exp}) is 
obtainable using the properties of the Lambert function
\begin{equation}
\label{eq:x0expl0}
r_0=\left( -\frac{\eta+2}{\eta} W(-Y_\eta)\right)^{1/\eta}.
\end{equation}
Lastly the AFM energy is given by (\ref{eq:E0pw1}). A 
straightforward calculation gives the final result
\begin{eqnarray}
\label{eq:epsAFMexpl0}
\epsilon_{\rm AFM}(\eta,g;n,l) &=& -g \, \exp \left( \frac{\eta+2}{\eta}
W_0(-Y_\eta) \right) \left[1+\frac{\eta+2}{2} W_0(-Y_\eta) \right] \nonumber \\
&=& -g \left( \frac{-Y_\eta}{W_0(-Y_\eta)} \right)^{\frac{\eta+2}{\eta}}
 \left[1+\frac{\eta+2}{2} W_0(-Y_\eta) \right],
\end{eqnarray}
with $Y_\eta$ given by~(\ref{eq:defYexpl0}). 
The energy $\epsilon_{\rm AFM}$ must be negative, which implies that 
$W(-Y_\eta) \ge -2/(\eta+2)$. This is only possible for the branch $W_0$ with
a negative argument, which is the case since $-Y_\eta < 0$.
The quality of this formula for the Gaussian potential ($\eta=2$) is discussed
in Ref.~\citen{silv10} and for the pure exponential ($\eta=1$) in Refs.~\citen{AFMeigen,bsb09a}. 

Let us just mention that eigenenergies of Hamiltonian (\ref{hdefexp}) for the
pure exponential ($\eta=1$) can be analytically computed for $l=0$ only. But even
in this case, the expression of $\epsilon(g)$ is not very tractable since it is
formally defined by the relation \cite{flug}
\begin{equation}
\label{enexa}
J_{2\sqrt{-\epsilon(g)}}(2\sqrt g)=0,	
\end{equation}
where $J_\rho$ is a Bessel function of the first kind. Consequently, the AFM
result is interesting since it yields an analytical formula for the energy levels
of the pure exponential potential that is of simpler use than (\ref{enexa}) for
$l=0$ and, above all, that remains valid for arbitrary $n$ and $l$ quantum numbers.  

\subsubsection{Critical constants}
\label{sec:critconst}

By definition,
the critical constants $g_{\eta;nl}$ of the potential are such that
$\epsilon_{\rm AFM}(\eta,g=g_{\eta;nl};n,l)=0$, with $\epsilon$ given by
(\ref{eq:epsAFMexpl0}). Of course this is the AFM approximation of these
critical constants, but according to this last equation, one finds
that the energy vanishes for $Y_\eta=2/(\eta+2) \, {\rm e}^{-2/(\eta+2)}$,
a value which is lower than $1/{\rm e}$ as required by
(\ref{eq:x0expl0}). Equivalently (equating this particular value of
$Y_\eta$ with the general value~(\ref{eq:defYexpl0})), one can say that the
critical constants are given by
\begin{equation}\label{gc1}
g_{\eta;nl}=\left( \frac{\eta \, {\rm e}}{2} \right)^{2/\eta}\, Q(n,l)^2.	
\end{equation}
They are such that, if $g > g_{n_0l_0}$, the potential admits a bound state
with the given quantum numbers $n_0$, $l_0$. It is remarkable that our
approximation scheme, based on the potential $P(x)=x^\lambda$ for which an infinite
number of bound states is present, is able to predict that only a finite number
of bound states will be present in exponential potentials. This is another test
of the ability of the AFM to analytically reproduce the qualitative features of
a given eigenvalue problem. Let us note that the AFM does not give any
information about the optimal power-law potential to determine the form of
the number $Q(n,l)$. For that, we can rely on direct comparison with
numerical solution. It is also very remarkable that whatever the power $\eta$,
the critical constants are always proportional to the square of the principal
quantum number $Q$.

The critical constants for the Gaussian potential are discussed in Ref.~\citen{silv10}
and for the pure exponential in Ref.~\citen{bsb09a}. Let us just discuss a little bit this
last case. Equation~(\ref{enexa}) leads to a determination of the exact critical
constants $g_{n0}$ ($\eta=1$ is here always assumed and is no longer indicated);
let us denote them as $g^*_{n0}$. Indeed these critical constants are such that
they correspond to a energy level $\epsilon(g^*_{n0})=0$.  We are thus led to
the equation
\begin{equation}
	J_0\left(2\sqrt{ g^*_{n0}}\right)=0\Rightarrow  g^*_{n0}=\frac{j^2_n}{4},
\end{equation}
where $j_n$ is the $(n+1)^{{\rm th}}$ zero of the Bessel function $J_0$. At large
$n$, these are given by $j_n\approx \pi(n+3/4)$ \cite{abra}, leading to the
asymptotic expression
\begin{equation}
\label{gcbess}
	g^*_{n0}\approx\frac{\pi^2}{4}\, \left(n+\frac{3}{4}\right)^2.
\end{equation}
This result is qualitatively similar to (\ref{gc1}), stating that $g_{n0}\propto Q(n,0)^2$.  

We can try to use formula~(\ref{gcbess}) to improve the result~(\ref{gc1}). Assuming
that $Q_{nl}= b\, n+l+c$ with no constraint on parameters $b$ and $c$, it is easy to
see that formulae~(\ref{gc1}) and (\ref{gcbess}) coincide for $g_{n0}$ with $b=\pi/\rm e$
and $c=3\pi/(4\rm e)$. We can then try a new relation to compute the critical constants 
\begin{equation}
\label{gc1b}
g_{e;nl}=\left(\frac{\pi}{2} n+\frac{\rm e}{2} l+\frac{3\pi}{8}\right)^2.	
\end{equation}
This formula is in good agreement with exact results: For $n\in[0,5]$ and $l\in[0,5]$,
the minimal, maximal and mean relative errors are respectively 0.03\%, 8.7\% and 3.5\%. \cite{bsb09a}

\subsection{Yukawa potential}
\label{sec:Yukawapot}

Among all the central interactions, the Yukawa potential is widely used in atomic physics
(effective interaction), nuclear physics (long range behavior of the nucleon-nucleon
interaction, due to one pion exchange), hadronic physics (screened Coulomb force). It is
thus interesting to apply the AFM to this very important case. We start with the following
Hamiltonian
\begin{equation}\label{h0def2}
  H=\frac{{\bm p}^2}{2m}-\alpha \frac{{\rm e}^{- \beta\, r}}{r}.
\end{equation}

\subsubsection{Expression of the energies}
\label{sec:Eexp2}

Using the scaling laws, the eigenenergies $E(m,\alpha,\beta;n,l)$ can be written
\begin{equation}
\label{eq:scalexp2}
E(m,\alpha,\beta;n,l)=\frac{\beta^2}{2m} \epsilon(g;n,l),
\end{equation}
where $\epsilon(g;n,l)$ is en eigenvalue of the dimensionless Hamiltonian 
\begin{equation}\label{hdefyuk}
	h={\bm p}^2-g \frac{{\rm e}^{-r}}{r},
\end{equation}
 with
\begin{equation}\label{gdefyuk}
	g=\frac{2m\alpha}{\beta}.
\end{equation}
This potentials admits also a
finite number of bound states that depends on the dimensionless parameter
$g$, ruling the potential depth, and defined by (\ref{gdefyuk}). 

\begin{figure}
\centerline{
\includegraphics*[width=\halftext]{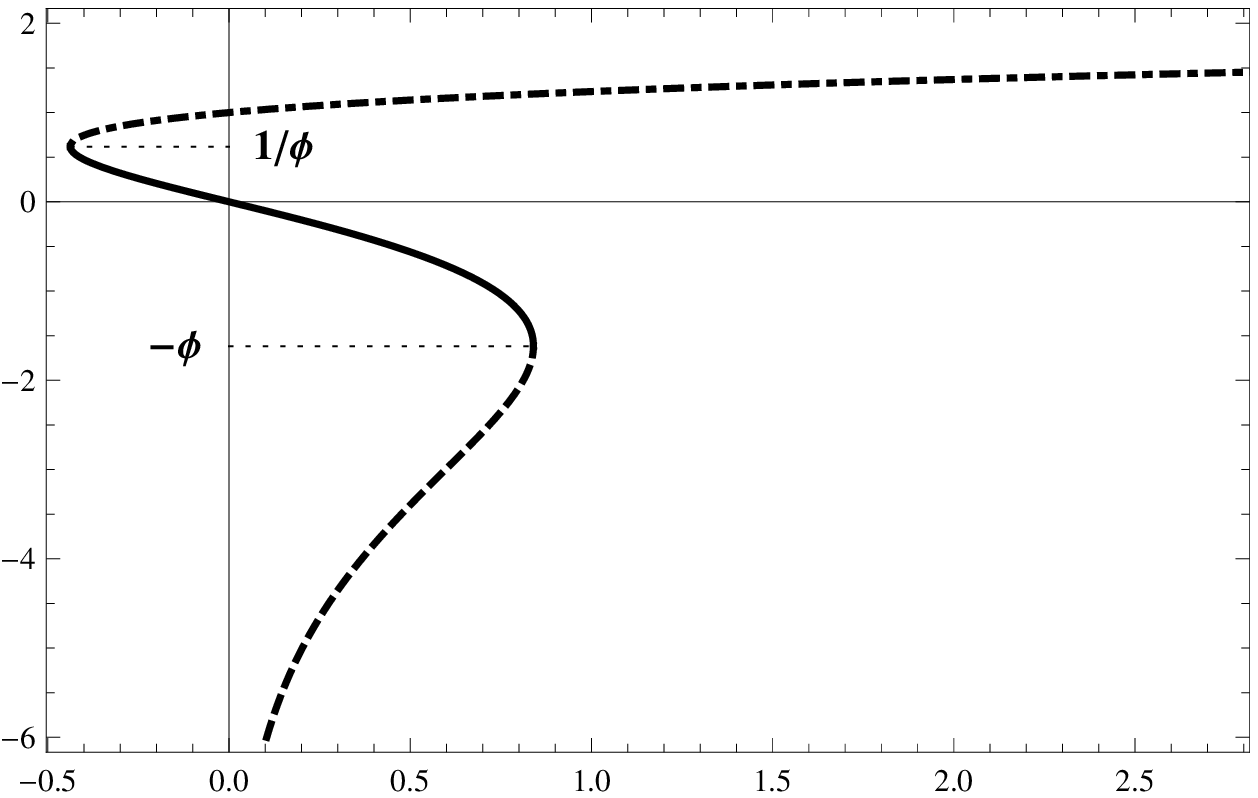}\quad
\includegraphics*[width=\halftext]{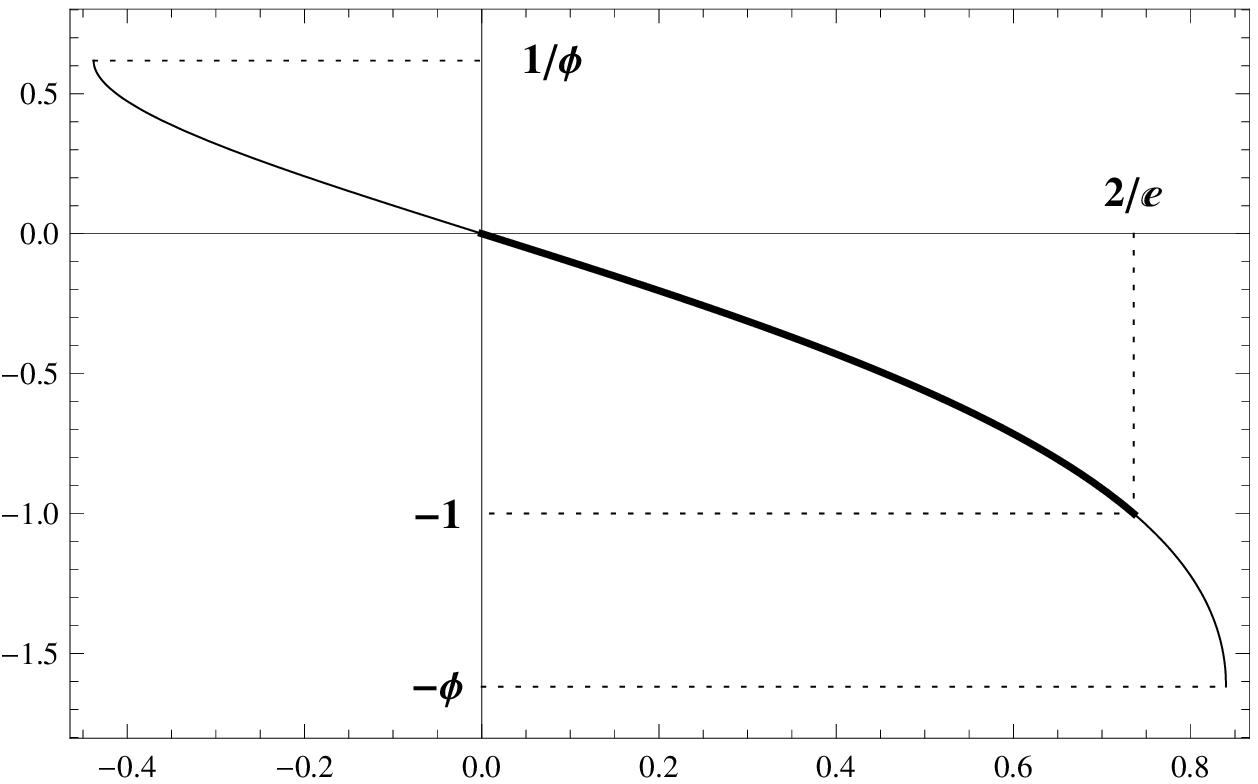}
}
\caption{Left: the three branches of the function $\Omega$. Right: in bold,
the function $\Omega(T)$ relevant to compute formula~(\ref{eq:EAFMexp1}). $\phi$
is the golden ratio.}
\label{fig:omega}
\end{figure}

The application of the AFM to this case leads to the transcendental equation
(see (\ref{eq:r0pw2}))
\begin{equation}
\label{eq:transexp1}
{\rm e}^{-r_0}r_0(1+r_0) = \frac{2Q^2}{g} = T.
\end{equation}
As far as we know, there does not exist an analytical expression giving
$r_0$ in terms of $T$.
Let us introduce formally the inverse function $\Omega=t^{-1}$ of the function
$t(u)={\rm e}^{u}u(u-1)$. By definition, we have the properties
\begin{equation}
\Omega[{\rm e}^{u}u(u-1)] = u = {\rm e}^{\Omega(u)}\Omega(u)(\Omega(u)-1).
\end{equation}
Thus one has $r_0 = -\Omega(T)$. A calculation giving the AFM energy through
(\ref{eq:E0pw1}) leads to the expression
\begin{equation}
\label{eq:EAFMexp1}
\epsilon_{\rm AFM}(g;n,l)=\frac{g}{2} \frac{T[1+\Omega(T)]}{\Omega(T)^2[\Omega(T)-1]}.
\end{equation}
$\Omega(u)$ is a multi-valued
function composed of 3 monotonic branches (see Fig.~\ref{fig:omega}). However,
we have the constraints that $r_0 > 0$ and $\epsilon_{\rm AFM} < 0$. This imposes the
condition $-1 \leq \Omega(T) < 0$, with $T > 0$. This condition automatically selects the
branches, passing through the origin, 
defined in $[-\exp(1/\phi)/\phi^3,\exp(-\phi)\phi^3]$ and whose image is in 
$[-\phi,1/\phi]$, where $\phi$ is the golden ratio. 
So, there is no ambiguity.
It is possible to obtain an approximate AFM energy formula in terms of usual simple functions, 
but it is then necessary to use an approximate form for the solution of the  transcendental
equation (\ref{eq:transexp1}). \cite{bsb09a}\ Since $\Omega(u)\approx -u$ for $u\ll 1$, we find
\begin{equation}
\label{eq:limEAFMexp1}
\lim_{g\to \infty}\epsilon_{\rm AFM}(g;n,l)=-\frac{g^2}{4 Q^2}.
\end{equation}
With the choice $Q=Q_C$, this corresponds to the Coulomb energy. It is expected since
$g\to \infty$ is equivalent to $\beta \to 0$.

Upper bounds can be obtained by choosing $Q=Q_C=n+l+1$ in (\ref{eq:EAFMexp1}). This seems
quite natural since a Yukawa potential is similar in form to a Coulomb one. It is possible
to improve the accuracy of the AFM formula by using an appropriate form $Q= A(g) n+l+ C(g)$
\cite{bsb09a} but the variational character of the approximation cannot then guaranteed.
An accurate energy formula has been found in Ref.~\citen{green82} from a fit of the
numerically computed energy levels of the Yukawa potential. In our notations, it reads 
\begin{equation}
\label{enap}
\epsilon_{nl}=-\frac{g}{4Q^2_{C}}(g-g_{G;nl}) \frac{g-2 A' (Q_{C}+\sigma)^2+
2 B' Q^2_{C}}{g-g_{G;nl}+2 B' Q^2_{C}}
\end{equation}
with
\begin{equation}
\label{eq:gnlemp}
	g_{G;nl}=2\left(\sqrt{Z_l}+\frac{n}{S_l}\right)^2,
\end{equation}
where $A'=1.9875$, $B'=1.2464$ and $\sigma=0.003951$, and where  
\begin{eqnarray}
	&&Z_l=Z_0(1+\rho l+\tau l^2),\quad S_l=S_0(1+\gamma l+\delta l^2),
	\nonumber \\
	&&Z_0=0.839908,\quad \rho=2.7359,\quad \tau=1.6242, \nonumber \\
	&&S_0=1.1335,\quad \gamma=0.019102,\quad \delta=-0.001684. 
\end{eqnarray}
This formula
is rather different from ours. However, they coincide
at the limit $g \to \infty$ when $Q=Q_c$ (see (\ref{eq:limEAFMexp1})). 
Equation~(\ref{eq:gnlemp}) gives the critical constants of the Yukawa
potential with a relative accuracy around 0.4\%.

\begin{table}[htb]
\caption{Eigenvalues $\epsilon(g;n,l)$ for a Yukawa potential with $g=30$
as a function of $(n,l)$ sets. First line: exact value; Second line:
formula~(\ref{eq:EAFMexp1}) with $Q=Q_C=n+l+1$;
Third line: formula~(\ref{enap}).
A * indicates a non real or a non negative value.}
\label{tab:yuk3} 
\begin{center}
\begin{tabular}{ccccc}
\hline\hline\noalign{\smallskip}
$l$ & $\epsilon(30;0,l)$ & $\epsilon(30;1,l)$ & $\epsilon(30;2,l)$ & $\epsilon(30;3,l)$ \\ [3pt]
\hline
0 & $\bm{-196.44}$ & $\bm{-31.51}$ & $\bm{-5.47}$ & $\bm{-0.22}$ \\
  & $-195.98$ & $-29.97$ & $-2.92$ & * \\
  & $-196.36$ & $-31.62$ & $-5.67$ & $-0.30$ \\
\noalign{\smallskip}
1 & $\bm{-30.74}$ & $\bm{-4.94}$ & $\bm{-0.029}$ & - \\
  & $-29.97$ & $-2.92$ & * & - \\
  & $-30.52$ & $-4.85$ & $-0.019$ & - \\
\noalign{\smallskip}
2 & $\bm{-3.81}$ & - & - & - \\
  & $-2.92$ & - & - & - \\
  & $-3.70$ & - & - & - \\ 
\noalign{\smallskip}\hline
\end{tabular}
\end{center}
\end{table}

The quality of the two approximations presented above can be appraised in Tables~\ref{tab:yuk3}
for a particular value of $g$. 
Not all bound states can be found with (\ref{eq:EAFMexp1}). Let us note that formula~(\ref{enap}) 
gives generally the correct number of bound states, but not in all cases. \cite{bsb09a}\  
Formula~(\ref{eq:EAFMexp1}) with $Q=Q_C$ gives quite good results for the lowest eigenvalues. The
quality of the fit for formula~(\ref{enap}) is better but comparable
to the quality of AFM formula~(\ref{eq:EAFMexp1}) with a more sophisticated parameterization
of $Q$. \cite{bsb09a}\ Nevertheless, formula~(\ref{enap}) is empirical, while AFM result
(\ref{eq:EAFMexp1}) is obtained from an explicit analytical resolution of the
Schr\"{o}dinger equation.

\subsubsection{Critical constants}
\label{sec:critconst2}

Let us discuss below more deeply the critical constants relative to the Yukawa
potential. It is easy to find from~(\ref{eq:EAFMexp1}) that the $T$ value which cancels
the AFM energy is equal to $T=2/{\rm e}$. Using the definition~(\ref{eq:transexp1}),
one obtains the AFM critical constants for the Yukawa potential
\begin{equation}
\label{eq:critconstyuk}
g_{yuk;nl}= {\rm e} Q^2.
\end{equation}
It is again remarkable that the critical constants for the Yukawa potential are
also proportional to the square of the principal quantum number, a property
that seems universal for exponential-like potentials.

Due to the smallness of $\gamma$ and $\delta$ in (\ref{eq:gnlemp}), we can replace $S_l$ by $S_0$ in
(\ref{eq:gnlemp}). It clearly appears that $g_{G;nl} \propto \left(S_0\sqrt{Z_0\beta}\,
l+n\right)^2$ asymptotically. This is precisely the quadratic behavior predicted by
our analytical results. It is important to stress that formula~(\ref{eq:gnlemp})
was specially designed for fitting the Yukawa critical constants and that it
needs several free parameters, while our AFM result~(\ref{eq:critconstyuk})
follows a general algorithm valid in any circumstance. The fact that both give
the same asymptotic behavior is again a very strong argument in favor of the relevance of the
AFM. Introducing a constant which allows to obtain the exact result for $n=l=0$
and considering the asymptotic expansion for large values of $l$, 
(\ref{eq:gnlemp}) is approximately given by
\begin{equation}
\label{eq:gnlemp2}
	g_{G;nl}\approx \left( 1.248\, n+1.652\, l+1.296\right)^2.
\end{equation}
This is not too different from our result~(\ref{eq:critconstyuk}) with $Q=Q_C=n+l+1$
\begin{equation}
\label{eq:critconstyuk2}
g_{yuk;nl}(Q_c)= (\sqrt{\rm e}\,n+\sqrt{\rm e}\,l+\sqrt{\rm e})^2,
\end{equation}
since $\sqrt{\rm e}\approx 1.649$.
This formula is unfortunately not very accurate:
For $n\in[0,4]$ and $l\in[0,4]$, the minimal, maximal and mean relative errors
are respectively 6\%, 42\% and 26\%. 
The minimal, maximal and mean relative errors are respectively 0.01\%, 0.9\% and
0.3\% with formula~(\ref{eq:gnlemp}). The very good quality of these last results
is due to the use of a complicated formula fitted on exact results. Our formula
is simpler and the general behavior is predicted by the AFM.
Let us note that it is possible to improve the values of AFM critical constants for
the Yukawa potential using an appropriate form $Q=A n+B l +C$, with $A$, $B$ and $C$
fixed, and keeping the simplicity of the original formula~(\ref{eq:critconstyuk}). \cite{bsb09a}

\subsection{Sum of power-law potentials}
\label{sec:sumplpot}

In Sect.~\ref{sec:ext} we suggested that potentials of type $aP(r)+V(r)$
could be treated with the AFM in some occasions. Since a power-law potential is
very often a good starting point, it is interesting to see whether an analytical
AFM solution exists for a second contribution which is itself a power-law
potential. The resulting total potential is thus the sum of two power-law
potentials. Those types of potential are frequently used in various domains of
quantum mechanics. This is the subject of this section, whose calculations are based on 
(\ref{eq:minr0mod}) and (\ref{eq:energr0mod}).

\subsubsection{Expression of the energies}
\label{sec:Esumpw}
It is very instructive to start with the potential 
$P(r)=\textrm{sgn}(\eta)r^\eta = P^{(\eta)}(r)$. In this case, the extremization
condition (\ref{eq:minr0mod}) is written (see also (\ref{eq:r0pw1}))
\begin{equation}
\label{eq:spwmin2}
a |\eta| r_0^{\eta+2}+ r_0^3 V'(r_0) = \frac{Q_\eta^2}{m},
\end{equation}
where $Q_\eta$ is \emph{a priori} the optimal principal quantum number for $P^{(\eta)}(r)$. 
The AFM energy (\ref{eq:energr0mod}) is given by
\begin{equation}
\label{eq:Epws1}
E_{\textrm{AFM}} = \frac{Q_\eta^2}{2mr_0^2}+\textrm{sgn}(\eta) a r_0^\eta +V(r_0).
\end{equation}
With this in mind, let us introduce the potential $V(r)$ in the form of a
power-law potential $\textrm{sgn}(\lambda) b r^{\lambda}$ so that the total
potential for our system is given by
\begin{equation}
\label{eq:sumpowlaw}
W(r)=\textrm{sgn}(\eta) a r^{\eta} + \textrm{sgn}(\lambda) b r^{\lambda}.
\end{equation}
The above considerations show that the extremization condition and the AFM energy look like
\begin{eqnarray}
\label{eq:sumpwmin}
&&a |\eta| r_0^{\eta+2} + b |\lambda| r_0^{\lambda+2} = \frac{Q_\eta^2}{m} \\
\label{eq:sumpwE1}
&&E_{\textrm{AFM}} = \frac{Q_\eta^2}{2mr_0^2}+\textrm{sgn}(\eta) a r_0^\eta +
+\textrm{sgn}(\lambda) b r_0^\lambda.
\end{eqnarray}

Should have we taken instead $P(r)=P^{(\lambda)}(r)$ and
$V(r)=P^{(\eta)}(r)$, we would have recovered the same physical problem with the potential
$W(r)$ given by (\ref{eq:sumpowlaw}). Due to the form of basic equations
(\ref{eq:minr0mod}) and (\ref{eq:energr0mod}) the only change would have been
$Q_\eta \to Q_\lambda$. Because of the symmetry of the problem there is no
reason to prefer the choice $Q_\eta$, as in the previous equations, instead
of the choice $Q_\lambda$; moreover in any application, we already saw that
it is convenient to change the value of the principal quantum number. In
consequence, in the following, we will use the more neutral notation $Q$.
This result can also be directly obtained by using (\ref{NRvir1})-(\ref{NRvir3}), 
but it is interesting to illustrate that the AFM depends only on the potential
$P(r)$ through the principal quantum number $Q$. 

In general, one does not have an analytical expression for the root of the
extremization equation (\ref{eq:sumpwmin}). This is possible only for very
specific values of the powers $\eta$ and $\lambda$.

In this section, we will study such a problem for the most favorable
cases, where $\eta$ is chosen to give an exact expression for the eigenvalues
(in practice $\eta=2$ and $\eta=-1$) and where $\lambda$ is chosen in order
to have an analytical result. In the following, it is assumed that $a > 0$,
$b > 0$ and $-2 \leq \lambda \leq 2$.

\subsubsection{Solvable potentials}
\label{sec:SolvablePotentials}

Let us examine first the case $P(r)=r^2$. The extremization equation is
a transcendental equation for which an analytical solution does not 
exist automatically. The only cases for which we are sure that
an analytical solution exists is when it can be transformed into a polynomial
of degree less than or equal to 4. In order to investigate this condition,
let us put $(\lambda+2)/4 = p/q$ ($p$ and $q$ are relatively prime integers),
and define the new variable $X = r_0^{4/q}$. The corresponding extremization
condition is transformed into
\begin{equation}
\label{eq:sumpwmin2}
2 a X^q + b |\lambda| X^p = \frac{Q^2}{m}.
\end{equation}

All the solvable potentials should verify the conditions $0 \leq p \leq 4$,
$1 \leq q \leq 4$. An exhaustive research of all the solvable potentials
in our domain for $\lambda$ leads to the following, non trivial, values of the power
$\lambda$:
\begin{equation}
\label{vallambr2}
\lambda = -2,\: -1,\: -\frac{2}{3},\: \frac{2}{3},  1.
\end{equation}
We will study in detail the cases:
\begin{itemize}
\item $\lambda = -2$ because it corresponds to a centrifugal term with a
real parameter instead of the usual $l(l+1)$ term;
\item $\lambda = - 1$ because it corresponds to a simplified potential
for hadronic systems with a short range Coulomb potential and a quadratic
confinement;
\item $\lambda = 1$ because this anharmonic potential is sometimes used
in molecular physics.
\end{itemize}

Now we investigate the case $P(r)= -1/r$. The same type of argument with
$\lambda+2 = p/q$ and $X=r_0^{1/q}$ leads to the following polynomial
equation
\begin{equation}
\label{eq:sumpwmin3}
a X^q + b |\lambda| X^p = \frac{Q^2}{m}.
\end{equation}
An exhaustive list of the non trivial solvable potentials is given below
\begin{equation}
\label{vallamb1sr}
\lambda = -2,\: -\frac{7}{4}, \:-\frac{5}{3},\: -\frac{3}{2},\:
-\frac{4}{3}, \: -\frac{5}{4},\: -\frac{2}{3},\:-\frac{1}{2}, \: 1, \: 2.
\end{equation}
Among them we will study:
\begin{itemize}
\item $\lambda = 2$ because it corresponds to a potential $-a/r + b r^2$ which
is already studied with $P(r)=r^2$. This is the only potential that can be
described with either $P(r)=r^2$ or $P(r)=-1/r$ and this property allows very
fruitful comparisons.
\item $\lambda = 1$ because it corresponds to the funnel potential (Coulomb +
linear) which is widely used in hadron spectroscopy. \cite{sema04}\ Finding approximate
analytical values for the energies corresponding to this potential is thus
a very interesting question. To our knowledge, such formulae have not been 
proposed in the literature.

\end{itemize}

It could be also interesting to introduce potentials with $\lambda \leq -2$ but with
the restriction that it is repulsive at the origin (for instance, van der Waals
forces or Lennard-Jones type of potentials). In this case, the transcendental
equation is modified a little bit since $\lambda+2 \leq 0$ implying a negative
fraction $p/q$. An exhaustive research of the different values of $\lambda$ gives
the following results:

\begin{equation}
\label{vallamb1srp}
\lambda = -5,\: -4, \: -3, \:-\frac{5}{2}, \:-\frac{7}{3},\: -2.
\end{equation}
In this list, we will just consider the case:
\begin{itemize}
\item $\lambda = -2$ because the corresponding potential, known as the Kratzer
potential, exhibits its spectrum under an analytical form for all values of radial
quantum number $n$ and orbital quantum number $l$.
\end{itemize}

\subsubsection{Kratzer potential}
\label{sec:kratzer}

The Kratzer potential \cite{flug} is defined in its simpler form as
\begin{equation}
\label{eq:Kratz}
V(r)=\frac{a^2}{r^2}- \frac{2a}{r}.
\end{equation}
It can mimic the interaction of two atoms in a diatomic molecule. Moreover,
it presents some interest as a benchmark since it is one of the rare potentials
for which ones knows an exact analytical expression of the energies valid for
any $n$ and $l$ quantum numbers.
In fact, one can define the Kratzer potential in a more general formulation
(see Ref.~\citen{chad}) as
\begin{equation}
\label{eq:Kratz2}
V(r)=G \left( \frac{1}{r^2}- \frac{f}{r} \right).
\end{equation}
One recovers the simplest form putting $G=a^2$ and $f=2/a$.

It can be shown (see for example the hint given in Ref.~\citen{flug}) that the exact
eigenenergies read
\begin{equation}
\label{eq:energKratex}
E(n,l) = - \frac{mG^2f^2}{2 \left [n+1/2 + \sqrt{(l+1/2)^2 + 2mG} \right ]^2}.
\end{equation}

Applying the previous AFM expressions to this potential leads
to the following mean radius ($\eta=-1$, $\lambda=-2$, $a=Gf$ and $b=-G$
in (\ref{eq:sumpowlaw}))
\begin{equation}
\label{eq:r0Krat}
r_0=\frac{2mG+Q^2}{mGf}
\end{equation}
which, inserted in (\ref{eq:sumpwE1}), gives the desired result
\begin{equation}
\label{eq:energKrata}
E^{(\textrm{K})}(n,l)= - \frac{mG^2f^2}{2 \left [ 2mG + Q^2 \right ]}.
\end{equation}
It is natural to set $Q=Q_C=n+l+1$, considering that $P(r)=-1/r$.
We thus remark that the approximate
value $E^{(\textrm{K})}$ presents the correct asymptotic behavior for large
$n$ and for large $l$.
Just to have an idea of the quality of this approximation, let us calculate
the difference $\delta$ between the terms in brackets appearing in the denominators of $E$ and 
$E^{(\textrm{K})}$. It is just a matter of simple algebra to find
\begin{equation}
\label{eq:difKrat}
\delta=(2n+1)(l+1/2) \left [ \sqrt{1+\frac{2mG}{(l+1/2)^2}} -1 \right].
\end{equation}
One always has $\delta > 0$ so that $E(n,l) > E^{(\textrm{K})}(n,l)$ meaning
that the AFM approximation is a lower bound. Moreover, for small intensity
and/or mass, $mG \ll 1$, or for large angular momentum, $l \gg 1$, the
approximate value tends to the exact one and we have more explicitly
\begin{equation}
\label{eq:difKrata}
\delta \to (2n+1) \frac{mG}{(l+1/2)} \ll 1.
\end{equation}
For these limit conditions, the AFM approximation tends towards the exact result.
This behavior is easily understandable because, under those conditions, the
contribution due to $1/r$ is predominant as compared to the contribution of
$1/r^2$, and both expressions tend towards the same exact Coulomb result.

\subsubsection{Quadratic + centrifugal potential}
\label{sec:quadcent}

We consider now the potential (for an attractive centrifugal potential, not
all values of $b$ are relevant \cite{case50})
\begin{equation}
\label{eq:quadcentpot}
V(r)= a r^2 \pm \frac{b}{r^2}.
\end{equation}
The incorporation of the term $\pm b/r^2$ into the $l(l+1)/r^2$ term already present in
$\bm{p}^2$ allows to get the exact eigenvalue using the same kind of arguments
than those developed in the harmonic oscillator case. \cite{flug}\ Explicitly, we
obtain
\begin{equation}
\label{eq:Eqpcent0}
E(n,l)=\sqrt{\frac{a}{2m}} \left[ 2(2n+1)+\sqrt{(2l+1)^2 \pm 8 m b}\right].
\end{equation}

Using the AFM with $P(r)=r^2$ and setting $Y = 2mb/Q^2$, the value $r_0$
that extremizes this energy comes from a first degree equation and reads
\begin{equation}
\label{eq:r0qpcent}
r_0 = \left( \frac{b(1 \pm Y)}{aY} \right)^{1/4}.
\end{equation}
Substituting this value into the energy (\ref{eq:sumpwE1}), 
one obtains a very simple expression for the approximate energy
$E=2\sqrt{ab(1 \pm Y)/Y}$ or, alternatively
\begin{equation}
\label{eq:Eqpcent}
E^{(\textrm{qc})}(m,a,b;n,l)=2 \sqrt{\frac{a}{2m}} \sqrt{Q^2 \pm 2mb}.
\end{equation}

This quantity and the corresponding exact one depend on three parameters $m$,
$a$, $b$ but we know that the general scaling law properties allow us
to write them in a more pleasant form 
\begin{equation}
\label{eq:newfm1}
E(m,a,b;n,l) = \sqrt{\frac{a}{2m}} \; \epsilon(\beta=2mb;n,l),
\end{equation}
where $\epsilon(\beta;n,l)$ is an eigenvalue of the reduced
Schr\"{o}dinger equation for which the Hamiltonian depends now on a single
dimensionless parameter $\beta$
\begin{equation}
\label{eq:redschqpcent}
h=\bm{p}^2 + r^2 \pm \frac{\beta}{r^2}.
\end{equation}
The exact eigenvalues of this Hamiltonian are given by
\begin{equation}
\label{eq:engexqpcent}
\epsilon(\beta;n,l)=2(2n+1)+\sqrt{(2l+1)^2 \pm 4 \beta}.
\end{equation}
The approximate values immediately come from (\ref{eq:Eqpcent})
\begin{equation}
\label{eq:epsqpcent}
\epsilon^{(\textrm{qc})}(\beta;n,l) = 2 \sqrt{Q^2 \pm \beta}.
\end{equation}
It seems natural to choose $Q=Q_{HO}$, considering that $P(r)=r^2$. In this case, one can check that
the relative error between $\epsilon$ and $\epsilon^{(\textrm{qc})}$
decreases as $l^{-3}$ for a fixed value of $n$ and large $l$; it
decreases as $n^{-1}$ for a fixed value of $l$ and large $n$.
This behavior is easily understandable because, for large values of the quantum numbers,
the contribution due to $r^2$ is predominant as compared to the contribution of
$1/r^2$ and both expressions tends towards the same exact harmonic oscillator result.

Let us assume that $\beta \ll 1$ and let us choose $Q=Q_{HO}$; a Taylor expansion truncated to first order
leads to
\begin{equation}
\label{eq:qpcentfo}
\epsilon^{(\textrm{qc})}(\beta;n,l) \approx 2Q_{HO} \pm \frac{\beta}{Q_{HO}}.
\end{equation}
It is easy to check that this expression can also be
obtained by perturbation theory.
In particular for $\beta=0$, one recovers the exact value $2 Q_{HO}$, as expected.

\subsubsection{Anharmonic potential}
\label{sec:anarpot}

The potential under consideration reads
\begin{equation}
\label{eq:anharm}
V(r)= a r^2 + 2 b r.
\end{equation}
The extremization condition writes
\begin{equation}
\label{eq:anarr0}
2ar_0^4 + 2b r_0^3 = \frac{Q^2}{m}.
\end{equation}
Let us introduce the parameter
\begin{equation}
\label{eq:Yanar}
Y = \frac{8}{3} a \left ( \frac{Q^2}{m b^4} \right )^{1/3}
\end{equation}
and the new variable $x = \left( \frac{Q^2}{mb} \right)^{1/3} \frac{1}{r_0}$.
The previous equation (\ref{eq:anarr0}) is put in the simplest form
$4 x^4 - 8 x - 3Y = 0$.
This is the reduced quartic equation presented in 
Sect.~\ref{sec:Fourtheq} whose
solution is given by $x(Y)=G_{-}(Y)$ (see (\ref{eq:rootquarteq})).
Substituting this value into the energy (\ref{eq:sumpwE1}) and making
a little algebra leads to the desired approximate energy
\begin{equation}
\label{eq:Eanar}
E^{(\textrm{an})}(m,a,b;n,l)=\frac{3 b^2}{8a} Y \left ( G_{-}^2(Y) +
\frac{1}{G_{-}(Y)} \right ),
\end{equation}
with $Y$ given by (\ref{eq:Yanar}). It is natural to use the value $Q=Q_{HO}$,
considering that $P(r)=r^2$.

As in the previous case, this quantity and the corresponding exact one
depend on three parameters $m$, $a$, $b$ but the general scaling law allows us to
write them in terms of a reduced quantity depending on a single parameter $\beta$
\begin{equation}
\label{eq:newfm2}
E(m,a,b;n,l) = \sqrt{\frac{2a}{3m}} \; \epsilon \left( \beta=\frac{3b^2}{16}
\sqrt{\frac{3m}{2a^3}};n,l \right ),
\end{equation}
where $\epsilon(\beta;n,l)$ is an eigenvalue of the reduced
Schr\"{o}dinger equation for the Hamiltonian
\begin{equation}
\label{eq:redschanar}
h= \frac{\bm{p}^2}{4} + 3 r^2 + 8 \sqrt{\beta} r.
\end{equation}
The approximate value corresponding to this reduced equation follows from
(\ref{eq:Eanar}) 
\begin{equation}
\label{eq:epsanar}
\epsilon^{(\textrm{an})}(\beta;n,l) = 2 \, \beta \, Y \left ( G_{-}^2(Y) + 
\frac{1}{G_{-}(Y)} \right ), \quad Y = \left ( \frac{Q}{\beta} \right )^{2/3}.
\end{equation}
The parameter $\beta$ could also be associated with the quadratic potential,
but this less interesting case is not considered here. 

Let us assume that $\beta \ll 1$ and let us choose $Q=Q_{HO}$; a Taylor expansion truncated to first order
leads to
\begin{equation}
\label{eq:anarfo}
\epsilon^{(\textrm{an})}(\beta;n,l) \approx \sqrt 3 Q_{HO} + 4 \sqrt{\frac{2 \beta \, Q_{HO}}
{\sqrt 3}}.
\end{equation}
In particular for $\beta=0$, one recovers the exact value $\sqrt 3 Q_{HO}$, as
it should be. This result comes also from the perturbation theory.

The limit $\beta\to\infty$ is not physically relevant, but it is interesting to
consider it in order to check the formula. In this limit, we find
\begin{equation}
\label{eq:anbetinf}
\epsilon^{(\textrm{an})}(\beta;n,l) = 3(4 \beta Q^2)^{1/3}+ O\left(
\beta^{-1/3} \right).
\end{equation}
The dominant term is the result expected for a pure linear potential, as given by
(\ref{eq:eigenerplpot}).

\subsubsection{Quadratic + Coulomb potential}
\label{sec:quadcoul}

We study now the quadratic + Coulomb potential defined as
\begin{equation}
\label{eq:quadcoulpot}
V(r)= a r^2 - \frac{b}{r}.
\end{equation}
The equation giving the mean radius
looks like
\begin{equation}
\label{eq:qpcoulr0}
2ar_0^4 + b r_0 = \frac{Q^2}{m}.
\end{equation}
Let us introduce the parameter
\begin{equation}
\label{eq:Yqpcoul}
Y = \frac{8 Q^2}{3m} \left ( \frac{4a}{b^4} \right )^{1/3}
\end{equation}
and the new variable $x = \left ( \frac{4a}{b} \right )^{1/3} r_0$.
The equation that leads to the extremization of the energy is then
$4 x^4 + 8 x - 3Y = 0$.
This reduced quartic equation is studied in Sect.~\ref{sec:Fourtheq}.
Its solution is given by $x(Y)=G_{+}(Y)$ (see (\ref{eq:rootquarteq})).
Substituting this value into the energy (\ref{eq:sumpwE1}), one is led,
after some manipulations, to the desired approximate energy
\begin{equation}
\label{eq:Eqpcoul}
E^{(\textrm{qC})}(m,a,b;n,l)=\frac{3}{4} \left ( \frac{a b^2}{2} \right )^{1/3}
\left [ \frac{Y}{G_{+}^2(Y)} - \frac{4}{G_{+}(Y)} \right ],
\end{equation}
with $Y$ given by (\ref{eq:Yqpcoul}). As usual with the AFM, the quality of the approximation
depends on the choice of $Q$. For instance, the use of $Q=Q_C$
and $Q=Q_{HO}$ in the expression for $Y$ gives respectively a lower bound and
an upper bound on the true energy.
 
The quantity (\ref{eq:Eqpcoul}) and the corresponding exact one depend again on
three parameters $m$, $a$, $b$ but the general scaling law allows us to write
them in terms of a reduced quantity depending on a single dimensionless parameter
$\beta$. One can imagine two formulations depending on whether $\beta$ is part
of the quadratic contribution or of the Coulomb contribution:
\begin{eqnarray}
\label{eq:newfm3}
E(m,a,b;n,l)&=&4 \sqrt{\frac{2a}{3m}} \; \epsilon \left ( \beta=\frac{1}{4}
\left ( \frac{54 m^3 b^4}{a} \right )^{1/6};n,l \right ), \\
\label{eq:newfm4}
E(m,a,b;n,l)&=&\frac{3 m b^2}{16} \; \eta \left ( \beta'=4 \left (
\frac{a}{54 m^3 b^4} \right )^{1/6};n,l \right ).
\end{eqnarray}
The $\epsilon$ and $\eta$ energies are the eigenvalues of the reduced
Schr\"{o}dinger equations for the respective Hamiltonians $h_\epsilon$ et
$h_\eta$:
\begin{eqnarray}
\label{eq:redschqpcoul1}
h_\epsilon&=&\frac{3 \bm{p}^2}{16} +\frac{r^2}{4} - \frac{\beta^{3/2}}{r}, \\
\label{eq:redschqpcoul2}
h_\eta&=&\frac{3 \bm{p}^2}{16} - \frac{\sqrt 2}{r} + {\beta '}^{6} r^2.
\end{eqnarray}
The approximate values corresponding to these reduced Hamiltonians follow from
(\ref{eq:Eqpcoul}):
\begin{eqnarray}
\label{eq:epsqpcoul}
\epsilon^{(\textrm{qC})}(\beta;n,l)&=&\frac{3 \beta}{8} 
\left [ \frac{Y}{G_{+}^2(Y)} - \frac{4}{G_{+}(Y)} \right ], \quad Y = 
\left ( \frac{Q}{\beta} \right )^2, \\
\label{eq:etaqpcoul}
\eta^{(\textrm{qC})}(\beta ';n,l)&=&\frac{3 {\beta '}^2}{4} 
\left [ \frac{Y'}{G_{+}^2(Y')} - \frac{4}{G_{+}(Y')} \right ], \quad Y' = 
\left ( Q \beta' \right )^2.
\end{eqnarray}

Let us assume that $\beta \ll 1$ or $\beta' \ll 1$ (it is equivalent to the formulation $V(r) \ll P(r)$):
\begin{itemize}
\item A Taylor expansion truncated to first order for the formulation based
on the $\epsilon$ form
leads to
\begin{equation}
\label{eq:qpcoulfo1}
\epsilon^{(\textrm{qC})}(\beta;n,l) \approx \frac{\sqrt 3}{4} Q - \sqrt{\frac{2 \beta^3}
{Q \sqrt 3}}.
\end{equation}
In particular for $\beta=0$, one recovers the exact value $\sqrt 3 Q/4$
if $Q=Q_{HO}$. 
\item The Taylor expansion for the formulation based on the $\eta$ form gives
\begin{equation}
\label{eq:qpcoulfo2}
\eta^{(\textrm{qC})}(\beta ';n,l) \approx - \frac{8}{3 Q^2} +
\frac{9 {\beta '}^6 Q^4}{128}.
\end{equation}
In particular for $\beta '=0$, one recovers the exact value $- 8/(3 Q^2)$ if
$Q=Q_{C}$. 
\end{itemize}
In all these cases, the approximate formulae resulting from AFM agree with the result
of perturbation theory.

The limits $\beta\to\infty$ and $\beta'\to\infty$ are interesting to consider in order to check the formulae: 
\begin{itemize}
\item The $\epsilon$-based formulation gives 
\begin{equation}
\label{eq:qcbetinf1}
\epsilon^{(\textrm{qC})}(\beta;n,l) = -\frac{4\beta^3}{3 Q^2}+O\left(
\beta^{-2} \right).
\end{equation}
If $Q=Q_{C}$, the dominant term is the exact result for a Coulomb
potential.
\item The $\eta$-based formulation gives
\begin{equation}
\label{eq:qcbetinf2}
\eta^{(\textrm{qC})}(\beta';n,l) = \frac{\sqrt{3}}{2}{\beta'}^3 Q+
O\left( {\beta'}^{3/2} \right).
\end{equation}
If $Q=Q_{HO}$, the dominant term is the exact result for a
quadratic potential.
\end{itemize}

Let us emphasize the point that both (\ref{eq:redschqpcoul1}) and
(\ref{eq:redschqpcoul2}) can be related with the scaling laws. 
As a consequence, it can be shown that both the
exact eigenvalues and the AFM approximate ones fulfill the relation
\begin{equation}
\label{eq:linkepseta}
\epsilon^{(\textrm{qC})}(\beta;n,l)=\frac{\beta^3}{2}
\eta^{(\textrm{qC})}(1/\beta;n,l). 
\end{equation}

It is worth mentioning that analytical solutions of the Schr\"{o}dinger
equation with quadratic plus Coulomb potentials are presented in Ref.~\citen{kand05},
within the framework of a two-dimensional system of two-interacting electrons
($-1/r$) in a confining magnetic field ($r^2$). Closed-form solutions are found 
for particular values of
magnetic field and spatial confinement length. It is argued that 
a generalization to a three-dimensional space is possible. But, as the corresponding
formulae are not given explicitly, a comparison with our results is not
available. More generally, the
Schr\"{o}dinger equation with a $-a/r+b\, r+c\, r^2$ potential (particular
cases are studied in Sect.~\ref{sec:quadcoul} and \ref{sec:funnel})
is directly linked with the biconfluent Heun's equation. \cite{ronv95} 

\subsubsection{Funnel potential}
\label{sec:funnel}

In this section, we study the funnel potential defined as
\begin{equation}
\label{eq:funpot}
V(r)= a r - \frac{b}{r}.
\end{equation}
This potential is particularly important and its interest is discussed 
in Sect.~\ref{sec:SolvablePotentials}. The extremization
condition writes in this case
\begin{equation}
\label{eq:minfunnu}
br_0 + a r_0^3 =\frac{Q^2}{m}.
\end{equation}
Let us introduce the parameter
\begin{equation}
\label{eq:Yfun}
Y = \frac{3Q^2}{2mb}  \sqrt{ \frac{3a}{b}}
\end{equation}
and the new variable $x = \sqrt{\frac{3a}{b}}r_0$.
The extremization condition for the energy is the solution of the equation
$x^3 + 3 x - 2Y = 0$. This reduced cubic equation is studied in Sect.~\ref{sec:Thirdeq}.
Its solution is given by $x(Y)=F_{+}(Y)$.
Inserting this value into the energy (\ref{eq:sumpwE1}), one finds the
expression of the approximate AFM energy
\begin{equation}
\label{eq:Efun}
E^{(\textrm{f})}(m,a,b;n,l) 
= \sqrt{3 a b} \left [ \frac{Y}{F_{+}(Y)^2} - \frac{2}{F_{+}(Y)} \right ],
\end{equation}
with $Y$ given by (\ref{eq:Yfun}).
Let us mention that another form of this equation can be found thanks to
the following relation
\begin{equation}
\label{eq:Efun2}
\frac{Y}{F_{+}(Y)^2} - \frac{2}{F_{+}(Y)} 
= \sinh \theta -\frac{1}{4 \sinh \theta},
\end{equation}
with the change of variables $Y=\sinh (3 \theta)$.

The quantity (\ref{eq:Efun}) and the corresponding exact one depend again on
three parameters $m$, $a$, $b$ but the general scaling law allows us to write
them in terms of a reduced quantity depending on a single dimensionless parameter
$\beta$. In hadronic physics, the dominant interaction between a quark and an
antiquark is a confining linear potential. \cite{sema04}\ So, although the linear
potential has no analytical exact solution for all values of the quantum numbers,
it is also interesting to consider two formulations depending on whether $\beta$
is part of the linear contribution or of the Coulomb contribution:
\begin{eqnarray}
\label{eq:newfm5}
E(m,a,b;n,l)&=&3 \left ( \frac{a^2}{2m} \right )^{1/3} \epsilon \left ( \beta=
\left ( \frac{4 m^2 b^3}{27 a} \right )^{1/4};n,l \right ), \\
\label{eq:newfm6}
E(m,a,b;n,l)&=&\frac{2 m b^2}{3^{5/3}} \; \eta \left( \beta'=\left (\frac{27 a}
{4 m^2 b^3} \right )^{1/4};n,l \right ).
\end{eqnarray}
The $\epsilon$ and $\eta$ energies are the eigenvalues of the reduced
Schr\"{o}dinger equations for the respective Hamiltonians $h_\epsilon$ et
$h_\eta$:
\begin{eqnarray}
\label{eq:redschfun1}
h_\epsilon&=&\frac{\bm{p}^2}{3} +\frac{r}{3} - \frac{\beta^{4/3}}{r}, \\
\label{eq:redschfun2}
h_\eta&=&\frac{\bm{p}^2}{3} - \frac{3^{1/3}}{r} + {\beta '}^{4} r.
\end{eqnarray}
The approximate values corresponding to these reduced Hamiltonian follow from
(\ref{eq:Efun}):
\begin{eqnarray}
\label{eq:epsfun}
\epsilon^{(\textrm{f})}(\beta;n,l)&=&\beta^{2/3} 
\left [ \frac{Y}{F_{+}(Y)^2} - \frac{2}{F_{+}(Y)} \right ], 
\quad Y = \left ( \frac{Q}{\beta} \right )^2, \\
\label{eq:etafun}
\eta^{(\textrm{f})}(\beta ';n,l)&=&3^{2/3} {\beta '}^2 
\left [ \frac{Y'}{F_{+}(Y')^2} - \frac{2}{F_{+}(Y')} \right ], \quad Y' = \left ( Q
\beta ' \right )^2.
\end{eqnarray}

Let us assume that $\beta \ll 1$ or $\beta' \ll 1$:
\begin{itemize}
\item The Taylor expansion for the formulation based on the $\epsilon$ form gives
\begin{equation}
\label{eq:funfo1}
\epsilon^{(\textrm{f})}(\beta;n,l) \approx \frac{Q^{2/3}}{2^{2/3}}  - \left(
\frac{\beta^4}{2 Q^2} \right)^{1/3}.
\end{equation}
In particular for $\beta=0$, one recovers the value expected for a pure linear
potential, as given by (\ref{eq:eigenerplpot}).
\item The Taylor expansion for the formulation based on the $\eta$ form gives
\begin{equation}
\label{eq:funfo2}
\eta^{(\textrm{f})}(\beta ';n,l) \approx - \frac{3^{5/3}}{4 Q^2} + \frac{2 Q^2
{\beta '}^4} {3^{4/3}}.
\end{equation}
In particular for $\beta'=0$, one recovers the exact value $- 3^{5/3}/(4 Q^2)$
for the Coulomb potential if $Q=Q_C$.
\end{itemize}
In all these cases, the approximate formulae resulting from AFM agree with the result
of perturbation theory.

The limits $\beta\to\infty$ and $\beta'\to\infty$ are interesting to consider
in order to check the formulae: 
\begin{itemize}
\item The $\epsilon$-based formulation gives 
\begin{equation}
\label{eq:funbetinf1}
\epsilon^{(\textrm{f})}(\beta;n,l) = -\frac{3\beta^{8/3}}{4 Q^2}+O\left( 
\beta^{-4/3} \right).
\end{equation}
The dominant term is the exact result for a Coulomb potential if $Q=Q_C$.
\item The $\eta$-based formulation gives 
\begin{equation}
\label{eq:funbetinf2}
\eta^{(\textrm{f})}(\beta';n,l) = \left(\frac{3}{2}{\beta'}^4 Q\right)^{2/3} +
O\left( {\beta'}^{4/3} \right).
\end{equation}
The dominant term is the result expected for a pure linear potential, as given by
(\ref{eq:eigenerplpot}).
\end{itemize}

Here again, both (\ref{eq:redschfun1}) and (\ref{eq:redschfun2}) can
be related with the scaling laws. As a consequence, it can be shown that both
exact eigenvalues and AFM approximate ones fulfill the relation
\begin{equation}
\label{eq:funfo3}
\epsilon^{(\textrm{f})}(\beta;n,l) = \frac{\beta^{8/3}}{3^{2/3}} 
\eta^{(\textrm{f})} (1/\beta;n,l).
\end{equation}

\subsubsection{Comparison with numerical results}
\label{sec:compnumfun}

The potentials studied in this section are rather sophisticated and, presumably,
the AFM should lead to less good results than power-law potentials. Thus, it is
interesting to test this method in these less favorable cases. A detailed
discussion can be found in Ref.~\citen{bsb08b}. Here we focus our analysis on the
funnel potential only, because of its physical importance, and rely on the
reduced Hamiltonian $h_\epsilon$ (\ref{eq:redschfun1}). The AFM energies
$\epsilon^{(\textrm{f})}(\beta;n,l)$ are given by~(\ref{eq:epsfun}). For most of
the hadronic systems, the physical values of $\beta$ vary from 0 to about 1.5.

As usual, the simplest prescription to improve harmonic oscillator or Coulomb-like
results is to adopt a
principal quantum number of the form
\begin{equation}
\label{eq:Nbetfun}
Q(\beta)=b(\beta)n+l+c(\beta).
\end{equation}
The protocol to determine the best values for the $b$ and $c$ coefficients
follows essentially the same steps than those
explained in Sect.~\ref{compnumresdet} and is not repeated here. The
only difference is the form of the functions $b(\beta)$ and $c(\beta)$ which
are better fitted with Gaussian functions than hyperbolae. Explicitly, the
parameterization retained is 
\begin{equation}
\label{eq:bcfun}
b(\beta)= 1+p_1 \exp\left( -p_2^2 \beta^2\right), \quad
c(\beta)= 1+q_1 \exp\left( -q_2^2 \beta^2\right).
\end{equation}
With this choice, $b(\infty)=1=c(\infty)$ as it should be since in this case we
have a pure coulomb potential. For $\beta=0$ we are in presence of a pure linear
potential and for $l=0$ we do know the exact result. With this in mind, we
choose $b(0)=\pi/\sqrt{3}$ and $c(0)=\sqrt{3}\pi/4$. The only free parameter is
$p_2$ for $b(\beta)$ and $q_2$ for $c(\beta)$. The parameters retained for our study are:
$p_1=\frac{\pi}{\sqrt{3}}-1$, $p_2=0.416$, $q_1=\frac{\sqrt{3}\pi}{4}-1$, $q_2=1.245$.

\begin{table}[htb]
\caption{Eigenvalues $\epsilon(\beta;n,l)$ of
Hamiltonian~(\ref{eq:redschfun1}) with $\beta=0.5$, for some sets $(n,l)$. 
First line: $\epsilon_{\textrm{num}}(\beta;n,l)$ from numerical
integration considered as the exact ones; 
second line: $\epsilon^{(\textrm{f})}(\beta;n,l)$ given by
(\ref{eq:epsfun}) with $Q(\beta)$ defined by (\ref{eq:Nbetfun}) and (\ref{eq:bcfun});
third line: $\epsilon^{(\textrm{f})}(\beta;n,l)$ given by (\ref{eq:epsfun})
with $Q(\beta)=Q_{C}$.}
\label{tab:fun3}
\begin{center}
\begin{tabular}{ccccc}
\hline\hline\noalign{\smallskip}
$l$ & $\epsilon(0.5;0,l)$ & $\epsilon(0.5;1,l)$ &
$\epsilon(0.5;2,l)$ & $\epsilon(0.5;3,l)$ \\ [3pt]
\hline
0 & \textbf{0.39711} & \textbf{1.11714}  & \textbf{1.64558}  & \textbf{2.09628} \\
 & 0.42779 & 1.16223 & 1.68099 & 2.12205 \\
 & 0.26827 & 0.79105 & 1.15440 & 1.45987 \\
\noalign{\smallskip}
1 & \textbf{0.90598}  & \textbf{1.45955}  & \textbf{1.92580}  & \textbf{2.34167} \\
 & 0.88794 & 1.46673 & 1.93564 & 2.34911 \\
 & 0.79105 & 1.15440 & 1.45987 & 1.73269 \\
\noalign{\smallskip}
2 & \textbf{1.25749}  & \textbf{1.74247}  & \textbf{2.17133}  & \textbf{2.56288} \\
 & 1.23307 & 1.73892 & 2.17323 & 2.56506 \\
 & 1.15440 & 1.45987 & 1.73269 & 1.98358 \\
\noalign{\smallskip}
3 & \textbf{1.55457} & \textbf{1.99727}  & \textbf{2.39917}  & \textbf{2.77168} \\
 & 1.52908 & 1.98937 & 2.39764 & 2.77183 \\
 & 1.45987 & 1.73269 & 1.98358 & 2.21833 \\
\noalign{\smallskip}\hline
\end{tabular}
\end{center}
\end{table}

Allowing a $\beta$-dependence for the coefficients $b$ and $c$ improves
greatly the approximate eigenvalues since one gains a factor $10^{-2}$ to
$10^{-3}$ on the chi-square values. \cite{bsb08b}\  
The quality of the fits can be appraised for $\beta=0.5$ by examining the values of
approximate results compared with exact ones presented 
in Table~\ref{tab:fun3}. The values $Q=Q_{HO}$ and $Q=Q_C$ give respectively
an upper and a lower bound on the exact energy. However, as can be seen from
Table~\ref{tab:fun3} for the case $Q=Q_C$, these bounds are not very close
to the exact value and, thus not very interesting 
(the error is of the order of $10\%$ or more). The value $Q=Q(\beta)$
gives more accurate results (the error is less than $5\%$ and is often of order of
$2$-$3\%$), but we have no certainty concerning the
(anti)variational character of the approximation. The global accuracy is less good than 
for power-law potentials but remains quite acceptable.
 
\section{Two-body spinless Salpeter equation}
\label{sec:tbsalp}

In this section, we will apply the AFM to the spinless Salpeter Hamiltonian 
for which the kinetic energy has a semirelativistic form
\begin{equation}
\label{eq:Htsr}
H=\sigma \sqrt{\bm{p}^2+m^2}+V(r).
\end{equation}
The term ``relativistic" is generally not used because such a Hamiltonian is not
manifestly covariant.
The parameter $\sigma$ is equal to 1 for one-body systems ($r$ is then the distance of 
the particle from the origin of the force field) and to 2 
for two-body systems with equal masses ($r$ is then the distance between the two particles).
However, it is very convenient to
choose it as an arbitrary positive real parameter. We will prefer the
notation $M$ instead of $E$ for the eigenvalue of this kind of Hamiltonian
because it does include the rest mass of the particles.

Obviously, there is much less material concerning the properties of eigenvalues for
semirelativistic equations as compared to the huge bulk of results for nonrelativistic treatments.
Nevertheless the quadratic potential has been studied taking the Fourier transform of
of the potential. \cite{Lucha2001}\ Moreover, working in momentum space, Boukraa and Basdevant
were able to obtain interesting properties concerning a class of potentials. \cite{book88}
 
For two-body systems, the presence of two different masses cannot be dealt with a reduced
mass, so that we need the introduction of two different auxiliary fields to
treat the kinetic energy term. This situation makes the problem much more complex. 
It is briefly studied in Sect.~\ref{sec:unqmass}. Thus, in the following,
we focus on Hamiltonian~(\ref{eq:Htsr}). 

\subsection{Scaling laws}
\label{sec:scalsalpeq}

Let us consider the spinless Salpeter equation
\begin{equation}
\label{sl1}
\left(\sigma\, \sqrt{\bm p^2+m^2}+G\, V(a\, \bm r)-M(m,G,a,\sigma)\right)\,
 \Psi(\bm r\,)=0,
\end{equation}
where $G$ and $a$ are two real numbers defining the potential, $a$ being
related to its extension and $G$ to its intensity. The eigenvalues
$M$ depend \emph{a priori} on the four parameters $m,G,a,\sigma$. Scaling laws allow
to obtain the general form of $M(m,G,a,\sigma)$ in terms of the energies of
another system depending on less free parameters. Let us examine this point.

Defining $\bm \rho=a\bm r$, $\bm\pi=\bm p/a$, one has
\begin{equation}
\label{sl2}
\left(\sqrt{\bm \pi^2+(m/a)^2}+\frac{G}{a\sigma}\, V(\bm \rho)-\frac{1}
{a\sigma} M(m,G,a,\sigma)\right)\, \Psi(\bm \rho/a)=0.
\end{equation}
This is just the spinless Salpeter equation 
\begin{equation}
\label{sl3}
\left(\sqrt{\bm \pi^2+m'^2}+G'\, V(\bm \rho)-M'(m',G',1,1)\right)\,
\varphi(\bm \rho)=0,
\end{equation}
with 
\begin{equation}
	m'=\frac{m}{a},\quad G'=\frac{G}{a\sigma},\quad M'(m',G',1,1)=
\frac{M(m,G,a,\sigma)}{a\sigma}.
\end{equation}
We are thus led to the following scaling law for the mass spectrum
\begin{equation}\label{scalsr}
	M(m,G,a,\sigma)=a\, \sigma\ M\left(\frac{m}{a},\frac{G}{a\sigma},1,1\right).
\end{equation}
It means that both one parameter of the potential (here the extension of the
potential $a$) and the number of constituents in the system $\sigma$ can always be
set equal to 1 in order to simplify the computations without loss of generality.
The general energy formula will then be recovered thanks to~(\ref{scalsr}).

\subsection{Generalities}
\label{sec:salpgen}

To deal with the worrying square root, it is interesting to introduce one
auxiliary field $\mu$ in order to build the AFM Hamiltonian corresponding to
(\ref{eq:Htsr})
\begin{equation}
\label{eq:AFMH5}
\tilde{H}(\mu) = \frac{\bm{p}^2+m^2}{\mu}+\frac{\sigma^2}{4} \mu +V(r).
\end{equation}
The semirelativistic operator is then replaced by a nonrelativistic counterpart with 
a parameter $\mu$ to be determined. This technique to get rid of the square root
operator is not new and has been used previously to study hadronic physics. 
\cite{simo89a,simo89b,morgunov99,kalash00,sema04}\ 
Let us remark that, as $\tilde{H}(\mu)$ and $H$ have not the same kinetic part,
most of the results about lower and upper bounds developed in Sect.~\ref{sec:ulb}
are not applicable. 
At this stage, one can imagine two cases which are discussed below.

\subsubsection{Potential $V(r)$ is solvable}

Let us assume that the analytical solution of the Schr\"{o}dinger
equation with the potential $V(r)$ is known. 
In particular, this approach is interesting if $V(r)=r^2$
or $V(r)=-1/r$. 
One can then write the eigenvalues of $\tilde{H}(\mu)$ as 
\begin{equation}\label{eq:enu1}
	M(\mu)=\frac{m^2}{\mu}+\frac{\sigma^2}{4}\mu+e(\mu),
\end{equation}
where $e(\mu)$ is the eigenenergy of the Hamiltonian $h(\mu)=\bm{p}^2/\mu +
V(r)$. The final spectrum is given by $M(\mu_0)$, where $\mu_0$ is such
that $\left. \partial_\mu M(\mu)\right|_{\mu=\mu_0}=0$ which implies that
\begin{equation}\label{eq:enu2}
\frac{\sigma^2}{4}+ e'(\mu_0)=\frac{m^2}{\mu^2_0}.
\end{equation}
The prime denotes a derivation with respect to $\mu$. Provided that $e(\mu)$ is
analytically known, the only difficulty in this procedure is the analytical
resolution of this last transcendental equation, especially when $m\neq 0$.
Reporting the value of $\mu_0$ given by (\ref{eq:enu2}) in the expression of
the energy (\ref{eq:enu1}) allows to write
\begin{equation}\label{enu4}
M(\mu_0) = \frac{\sigma^2}{2}\mu_0+\left.\left(\mu\, e(\mu)\right)'
\right|_{\mu=\mu_0}.
\end{equation}
From Ref.~\citen{luchabd}, we know that $M(\mu) \ge M$ for any state,
and that $M(\mu_0) \ge M$, since it is also true for the particular value $\mu_0$
of the auxiliary field. The AFM yields in this case an upper bound. This can be
also demonstrated using a simpler version of the AFM which can be used only when 
it is interesting to replace an operator by its square. \cite{bsb09c}

Using the Hellmann-Feynman theorem \cite{feyn}, it has been shown that 
the value of $\mu_0$ is given by \cite{sema04,bsb09c}
\begin{equation}\label{eq:enu3}
\mu_0=\frac{2}{\sigma}\sqrt{\left\langle \bm p^2+m^2 \right\rangle},
\end{equation}
where the mean value is computed with an eigenstate of $\tilde{H}(\mu_0)$.
An estimation of the difference $\delta$ between the exact and approximate eigenenergies
is given by \cite{bsb09c} 
\begin{equation}
\delta=\sigma \sqrt{\left\langle \bm p^2+m^2 \right\rangle}- \sigma \left\langle
\sqrt{\bm p^2+m^2} \right\rangle.
\end{equation}

At the limit of low mass, (\ref{eq:Htsr}) can be written 
\begin{equation}\label{sshlowm}
H\approx H^{\textrm{ur}}+\frac{\sigma\, m^2}{2 \sqrt{\bm p^2}} \quad
\textrm{with} \quad H^{\textrm{ur}}=\sigma\, \sqrt{\bm p^2}+V(r).
\end{equation}
The contribution $\Delta(m)$ of the mass $m$ to an eigenenergy of the
ultrarelativistic Hamiltonian $H^{\textrm{ur}}$ ($m=0$) appears as a small contribution
that can be computed as a perturbation. The mean value being taken with an
eigenfunction of $H^{\textrm{ur}}$, we have \cite{bsb09c} 
\begin{equation}\label{mcontur}
\Delta(m)=\left\langle \frac{\sigma\, m^2}{2 \sqrt{\bm p^2}}\right\rangle
\approx\frac{\sigma\, m^2}{2 \sqrt{\left\langle\bm p^2\right\rangle}} =
\frac{m^2}{\bar \mu_0},
\end{equation}
where $\bar \mu_0$ is the auxiliary field obtained with the AFM applied to
the Hamiltonian $H^{\textrm{ur}}$.

\subsubsection{Potential $V(r)$ is not solvable}

If the eigenvalues of $h=\bm p^2/\mu+V(r)$ are not known,
it is tempting to treat the kinetic energy term and the potential term on equal
footing, in the spirit of what is done in Sect.~\ref{sec:AFM}, introducing the
auxiliary field $\mu$ to deal with $T$ and the auxiliary field $\nu$ to deal
with $V$. The case of a general function $P(r)$ cannot be solved analytically,
but restricting to the important power-law potential $P(r)= \textrm{sgn}
(\lambda) r^\lambda$, very interesting expressions can be obtained.

The eigenenergies $e(\mu,\nu)$ of the nonrelativistic Hamiltonian $h(\mu,\nu)=
\bm p^2/\mu+\nu P(r)$ are given by (\ref{eq:scallpw}) and (\ref{eq:epsAFMpwmod})
and the AFM mass by (the $K$ and $J$ functions are defined as usual)
\begin{equation}
\label{eq:massmunu}
M(\mu,\nu)=e(\mu,\nu)+\frac{\sigma^2}{4} \mu+\frac{m^2}{\mu}+V(J(\nu))-
\textrm{sgn}(\lambda) \nu J(\nu)^\lambda.
\end{equation}
The extremization of $M$ needs to solve two coupled equations with respect to
$\mu$ and $\nu$ which provide the values $\mu_0$ and $\nu_0$. Introducing the new variable
\begin{equation}
\label{eq:defsrx0}
x_0 = \left( \frac{|\lambda|}{2}\mu_0 \nu_0 \right )^{2/(\lambda+2)},
\end{equation}
it can be shown that both $\mu_0$ and $\nu_0$ can be expressed in terms of $x_0$
only. Namely
\begin{eqnarray}
\label{eq:mu0x0}
\mu_0(x_0)&=&\frac{2}{\sigma}\sqrt{m^2+Q_\lambda^{2\lambda/(\lambda+2)}x_0}, \\
\label{eq:nu0x0}
\nu_0(x_0)&=&K(Q_\lambda^{2/(\lambda+2)}/\sqrt{x_0}).
\end{eqnarray}
Reporting these values into the definition (\ref{eq:defsrx0}) allows the
determination of $x_0$ through the transcendental equation
\begin{equation}
\label{eq:detx0rel}
\sigma x_0^{(\lambda+2)/2} = |\lambda| \sqrt{m^2+Q_\lambda^{2\lambda/
(\lambda+2)}x_0}\;K(Q_\lambda^{2/(\lambda+2)}/\sqrt{x_0}).
\end{equation}
One deduces easily the expression of the AFM mass
\begin{equation}
\label{eq:massx0rel}
M_{\textrm{AFM}}=\sigma \sqrt{m^2+Q_\lambda^{2\lambda/(\lambda+2)}x_0}
+V(Q_\lambda^{2/(\lambda+2)}/\sqrt{x_0}).
\end{equation}
It is possible to go further in the simplification using the expression of the
$K$ function and defining the new mean radius
\begin{equation}
\label{eq:defr0rel}
r_0 = \frac{Q_\lambda^{2/(\lambda+2)}}{\sqrt{x_0}}.
\end{equation}
The transcendental equation (\ref{eq:detx0rel}) and the value of the AFM mass
(\ref{eq:massx0rel}) are given now by simpler expressions:
\begin{eqnarray}
\label{eq:detr0rel}
\sigma Q_\lambda &=&r_0^2V'(r_0)\sqrt{1+\left(\frac{m r_0}{Q_\lambda}\right)^2}, \\
M_{\textrm{AFM}}
 & = & \frac{\sigma Q_\lambda}{r_0} \sqrt{1+\left(\frac{m r_0}{Q_\lambda}\right)^2} + V(r_0).
\label{eq:massr0rel}
\end{eqnarray}
Despite the fact that two auxiliary fields were introduced, only one transcendental
equation is necessary to solve the problem. This great simplification is due to
the use of a power-law expression for the basic function $P(r)$.

It is also clear from (\ref{eq:detr0rel}) and (\ref{eq:massr0rel})
that, very much in the same way than the nonrelativistic case, the expression of
the mean radius and of the mass depend on the $\lambda$ variable only through the
value of the principal quantum number $Q(\lambda;n,l)$, which can thus be
modified at will in order to improve the results. So, in the following, this
quantum number will be simply denoted $Q$.

As in the nonrelativistic case, the AFM also yields approximations for the eigenstates.
The approximant
states, which are eigenstates of $\bm p^2/\mu_0 + \nu_0 \textrm{sgn}(\lambda) r^\lambda$,
are characterized by a size parameter depending on the product 
$\mu_0 \nu_0$ (see (\ref{psiOH}), (\ref{psiHy}) and (\ref{psiAi})). This quantity is given by
\begin{equation}
\label{eq:prodmu0nu0}
\mu_0 \nu_0 = \frac{2 Q^2}{|\lambda| r_0^{\lambda+2}}.
\end{equation}
A lot of observables can then be computed in terms of $r_0$ only.

Another procedure to solve the problem is to start from formula~(\ref{eq:enu1}) and
replace the exact value $e(\mu)$ by a corresponding approximation given by an AFM
solution obtained from the Hamiltonian $h(\mu,\nu_0)= \bm p^2/\mu+\nu_0 P(r)$ with
$P(r)= \textrm{sgn} (\lambda) r^\lambda$. After this first step, the optimal value
$\nu_0$ depends not only on $V(r)$ but also on $\mu$. One can then show that the
subsequent extremization on $\mu$ gives exactly the results (\ref{eq:detr0rel}) and
(\ref{eq:massr0rel}) presented above. 

This procedure in two steps is nevertheless interesting to obtain informations about
the bounds on exact energies. The replacement of the square root operator by a
nonrelativistic counterpart without modifying the potential, $H\to \tilde{H}(\mu)$
(see (\ref{eq:Htsr}) and (\ref{eq:AFMH5})), yields upper bounds. If the exact potential
$V(r)$ is then replaced by an AFM approximant $\tilde V(r) \ge V(r)$, the upper bounds
will be further increased. On the contrary, if the new potential is such that $\tilde V(r)
\le V(r)$, the upper bounds will be decreased. As a consequence, the energies given by
(\ref{eq:detr0rel}) are upper bounds on the exact energies if $\tilde V(r) \ge V(r)$, or
equivalently if the function $g(x)=V(P^{-1}(x))$ is concave. When $\tilde V(r) \le V(r)$
or $g(x)$ is convex, the antivariational character of the approximation cannot be guaranteed.

The previous equations take a particularly simple form for massless particles.
Indeed, for $m=0$, they reduce to
\begin{eqnarray}
\label{eq:detr0relur}
\sigma Q &=&r_0^2V'(r_0), \\
\label{eq:massr0relur}
M_{\textrm{AFM}}&=& \frac{\sigma Q}{r_0}  + V(r_0).
\end{eqnarray}
Let us define the function $D(x)$ as the inverse function of $x^2 V'(x)$
\begin{equation}
\label{eq:deffuncD}
D(x^2 V'(x)) = x \quad \textrm{or} \quad (D(x))^2 V'(D(x)) = x
\end{equation}
and the related function $F(x)$ defined by
\begin{equation}
\label{eq:deffuncF}
F(x) = \frac{x}{D(x)}+V(D(x)).
\end{equation}
Once the potential $V$ is given, these functions are universal in the sense that
they do not depend on the mass $m$ (except if the potential is mass-dependent)
and can be computed once and for all.
The mass for a system in the ultrarelativistic limit is thus written in a
very simple form, namely
\begin{equation}
\label{eq:massurfuncF}
M_{\textrm{AFM}} = F(\sigma Q).
\end{equation}

By rewriting (\ref{eq:detr0rel})-(\ref{eq:massr0rel}) and denoting the kinetic
energy by $T(|\bm p|)$ with $T(x)=\sigma\sqrt{x^2+m^2}$, the AFM formula for the
energies can be written into the form
\begin{eqnarray}
\label{SRvir1}
M_{\textrm{AFM}} &=& T(p_0)+V(r_0), \\
\label{SRvir2}
p_0 &=& \frac{Q}{r_0}, \\
\label{SRvir3}
p_0 T'(p_0)&=& r_0 V'(r_0).
\end{eqnarray}
As in the nonrelativistic case, (\ref{SRvir2}) defines the mean impulsion $p_0$ from
the mean radius $r_0$. This radius is defined by (\ref{eq:detr0relur}), and one can
recognize again in (\ref{SRvir3}) the general form of the virial theorem. \cite{luch90}\ 
Finally, (\ref{SRvir1}) gives the energy as a sum of the kinetic energy evaluated at
the mean impulsion $p_0$ and the potential energy evaluated at the mean radius $r_0$. These
equations simplify greatly for $m=0$ and reduce to (\ref{NRvir1})-(\ref{NRvir3}),
with $M_{\textrm{AFM}}=2 m+E_{\textrm{AFM}}$, for $m\to \infty$ and $\sigma=2$ ($m$
is then the common mass of the two identical particles and not the reduced mass).

The parameter $\mu_0$ can be considered as an effective mass for the relativistic particle.
\cite{simo89a,simo89b,morgunov99,kalash00,sema04}\ Using (\ref{eq:mu0x0}),
(\ref{eq:defr0rel}) and (\ref{SRvir2}), one finds
\begin{equation}
\label{eq:mu0p0}
\mu_0=\frac{2}{\sigma}\sqrt{p_0^2+m^2},
\end{equation}
which is a kind of ``total energy of a free particle". 

After some algebra, one can check that the virial theorem applied to Hamiltonian
$\tilde H(\mu_0,\nu_0)$, whose solutions are $M(\mu_0,\nu_0)$ given by (\ref{eq:massmunu}),
implies that
\begin{equation}
\label{SRvir5}
\left\langle \mu_0,\nu_0 \left| \frac{\bm p^2}{p_0^2} \right| \mu_0,\nu_0 \right\rangle =
\left\langle \mu_0,\nu_0 \left| \frac{r^\lambda}{r_0^\lambda} \right| \mu_0,\nu_0 \right
\rangle = 1,
\end{equation}
where $\left| \mu_0,\nu_0 \right\rangle$ is an eigenstate of $\tilde H(\mu_0,\nu_0)$. 
These equations,
as well as the boundary character of the solution, are only applicable when the exact 
form $Q(\lambda;n,l)$ is used. In practice, this occurs when $\lambda = -1$ or $2$, 
or $\lambda = 1$ for S-states only. Nevertheless, a better accuracy can be obtained by 
an appropriate choice of $Q$ as we will see below. 

\subsection{Power-law potentials}
\label{sec:salppw}

\subsubsection{General case}
\label{sec:Esalppw}

The power-law potential was considered as a prototype for testing the AFM in
the framework of the Schr\"{o}dinger equation. It was shown that an analytical
expression for the eigenenergy is obtained whatever the power $\lambda$ under
consideration. We apply in this section the AFM to find approximate analytical
energy formulae for the spinless Salpeter Hamiltonian (\ref{eq:Htsr}) with the
potential $V(r)=\textrm{sgn}(\lambda) a\, r^\lambda$ where $a$ is
positive. We consider that $\lambda>-2$, as needed in the nonrelativistic case
to have bound states. In this cases, (\ref{eq:detr0rel}) and (\ref{eq:massr0rel}) become
\begin{eqnarray}
\label{eq:epl3bis}
\sigma Q^2 &=&a |\lambda| r_0^{\lambda+1}\sqrt{Q^2+ m^2 r_0^2}, \\
\label{eq:epl3ter}
M(\lambda)
 & = & \sigma \sqrt{\frac{Q^2}{r_0^2}+m^2} + \textrm{sgn}(\lambda) a\, r_0^\lambda.
\label{enfin1}
\end{eqnarray}
With the change of variable
\begin{equation}
\frac{1}{r_0^2}=\left(\frac{a|\lambda|}{2}\right)^{\frac{2}{\lambda+2}}Q^{-\frac{4}{\lambda+2}}x_0,
\end{equation}
(\ref{eq:epl3bis}) can be recast into  
\begin{equation}\label{eq:epl3}
\frac{\sigma^2}{4}x_0^{\lambda+2}-A_\lambda\, x_0-m^2=0 \quad {\rm with} \quad A_\lambda=
\left( \frac{a |\lambda|}{2}\right)^{\frac{2}{\lambda+2}}\, Q^{\frac{2\lambda}
{\lambda+2}},
\end{equation}
a form which will be very convenient to use. With this convention, the mass~(\ref{eq:epl3ter})
is also written in a simple form
\begin{equation}\label{eq:MlaAl}
M(\lambda)=\frac{\sigma}{\lambda}
\frac{\lambda m^2+(\lambda+1)A_\lambda x_0}{\sqrt{m^2+A_\lambda x_0}}.
\end{equation}

In order to obtain analytical energy formulae, one should be able to
analytically solve (\ref{eq:epl3}). Let us now examine the cases for which
this is possible. First, we set $\lambda+2=p/q$ and $X_0=x_0^{1/q}$, where
$p$ and $q$ are co-prime integers. Then, (\ref{eq:epl3}) becomes
\begin{equation}
	\frac{\sigma^2}{4}X_0^{p}-A_\lambda\, X_0^q-m^2=0.
\end{equation}
A polynomial possesses analytical roots if its order is less or equal to 4;
therefore all the solvable potentials should verify the conditions $0 \leq p
\leq 4,\, 1 \leq q \leq 4$. An exhaustive research of all the solvable
potentials leads to the following values for the power $\lambda$:
\begin{equation}
\label{vallambr3}
\lambda = -2,\: -\frac{7}{4},\: -\frac{5}{3},\:- \frac{3}{2}, \: -\frac{4}{3},
\: -\frac{5}{4},\: -1,\: -\frac{2}{3}, \: -\frac{1}{2},\: 0,\: 1,\: 2.
\end{equation}
Among these allowed values, three are of particular interest: $\lambda=-1$ and
$2$ because the Coulomb problem and the harmonic oscillator play a central role
in theoretical physics, and $\lambda=1$ since a linearly rising potential is
generally considered to be a relevant approximation of the confining potential
in QCD. \cite{lucha}\ These three cases are explicitly solved in the following
sections. Thus, in contrast to the nonrelativistic case where an analytical
expression exists for an arbitrary value of $\lambda$, the spinless Salpeter
equation leads to a restricted list of favorable values (\ref{vallambr3}).
Fortunately, the most interesting potentials belong to this list.

Let us briefly discuss the existence of physical solutions, as function of
the parameters. Starting from (\ref{eq:epl3}) and (\ref{eq:MlaAl}), it is
possible to prove the following properties:
\begin{itemize}
  \item For $\lambda > 0$, there exists always a physical solution with
  $M(\lambda) > 0$.
  \item For $-1 \leq \lambda < 0$ there exists a physical solution with
  $M(\lambda) > 0$ as long as the parameter $a$ is less than a critical
  value $a_c(\lambda)$ given by
\begin{equation}
\label{aclambda}
a_c(\lambda) = \sigma \left ( \frac{Q}{\sqrt{|\lambda|}} \right )^{|\lambda|}
\left ( \frac{m^2}{1+\lambda} \right )^\frac{1+\lambda}{2}.
\end{equation}
Let us remark that $a_c(\lambda)$ depends on $m$, except $a_c(-1)=\sigma\,Q$.
  \item For $-2 \leq \lambda < -1$ either there is no solution for
(\ref{eq:epl3}), or there exists two solutions but, in the latter case, these
solutions are not compatible with the nonrelativistic expressions when $m$
tends toward infinity. In both cases, the corresponding solutions are not
physical so that these types of potential must be discarded from our study.
\end{itemize}
In the following, we will denote $Q_\lambda=Q(\lambda;n,l)$ the best possible form of the
principal quantum number for the power-law potential $V(r)=\textrm{sgn}(\lambda)
a\, r^\lambda$ in a Schr\"odinger equation. We have then $Q_2=Q_{HO}$,
$Q_{-1}=Q_C$ and $Q_{1}$ given by (\ref{eq:QAi}). 
For other values of $\lambda$, a formula of type~(\ref{eq:formNs})
can be chosen for instance. 

\subsubsection{Harmonic potential}
\label{sec:salpharm}

If $\lambda=2$, (\ref{eq:epl3}) becomes a quartic equation of the form
\begin{equation}
\label{quartic1}
	\frac{\sigma^2}{4}x^4_0-A_2\, x_0-m^2=0 \quad {\rm with}\quad A_2=
	\sqrt a\, Q.
\end{equation}
Defining
\begin{equation}
x_0=\left(\frac{2A_2}{\sigma^2}\right)^{1/3}\, X\quad {\rm and}\quad
Y_2=\frac{m^2}{3} \left(\frac{16\sigma}{a Q^2}\right)^{2/3},
\end{equation}
(\ref{quartic1}) can be rewritten as $4X^4-8X-3Y_2=0$,
which is precisely of the form (\ref{eq:redcubeq2}). Following~(\ref{eq:rootquarteq}), 
the solution is given by
$X(Y_2)=G_{-}(Y_2)$, and, after a rearrangement of~(\ref{eq:MlaAl}), the
mass spectrum reads (under one of the equivalent forms)
\begin{eqnarray}
\label{E2_3f}
	M(\lambda=2) &= &\sigma\, m \sqrt{\frac{3}{Y_2}}  \frac{Y_2+4G_-(Y_2)}
	{\sqrt{8G_-(Y_2)+3Y_2}} \\
	& = & \sigma\, m \sqrt{\frac{3}{Y_2}} \left( \frac{2}{G_-(Y_2)} +
	\frac{Y_2}{2 G_-^2(Y_2)} \right)\\
	& = & \frac{2 \sigma\, m}{\sqrt{3 Y_2}} \left(G_-^2(Y_2) +
	\frac{1}{G_-(Y_2)}\right).
\end{eqnarray}
With $Q=Q_2$ these formulae yield an upper bound on the energy, since
this setting gives the exact solution for the corresponding nonrelativistic
Hamiltonian.

In the nonrelativistic limit ($Y \to \infty$), the last equations reduce to
\begin{equation}\label{hoylarge}
	M - \sigma\, m \approx \sqrt{\frac{2 \sigma a}{m}} Q,
\end{equation}
as expected for a nonrelativistic harmonic oscillator. For large value of $m$, the choice $Q=Q_2$ is clearly
optimal. For small values of $m$, another choice could give better results.

\subsubsection{Linear potential}
\label{sec:salplin}

The resolution of the case $\lambda=1$ is rather similar to the one of
the harmonic oscillator. Defining
\begin{equation}
	x_0=2\left(\frac{A_1}{3\sigma^2}\right)^{1/2} X\quad{\rm and}\quad
	Y_1=\frac{3^{3/2}m^2\sigma}{4A_1^{3/2}},
\end{equation}
(\ref{eq:epl3}) simply becomes a cubic equation of the form
(\ref{eq:redcubeq1}), that is $X^3-3X-2Y_1=0$.
Notice that $A_1=\left(a Q/2\right)^{2/3}$ so that
\begin{equation}
Y_1=\frac{3}{2} \frac{\sqrt{3} \sigma m^2}{a Q}.
\end{equation}
Following~(\ref{eq:rootcubeq}), the solution is given by
$X(Y_1)=F_-(Y_1)$. After a rearrangement of~(\ref{eq:MlaAl}), the mass
spectrum reads (under one of the equivalent forms)
\begin{eqnarray}
\label{E1_3f}
	M(\lambda=1) & = & \sigma\, m \sqrt{\frac{2}{Y_1}} \frac{Y_1+3F_-(Y_1)}
	{\sqrt{3F_-(Y_1)+2Y_1}} \\
	& = &\sigma\, m \sqrt{\frac{2}{Y_1 F_-(Y_1)}} \left( 3 + \frac{Y_1}
	{F_-(Y_1)} \right) \\
	& = & \frac{\sigma\, m }{\sqrt{2 Y_1 F_-(Y_1)}} \left( 3 + F_-^2(Y_1)\right). 
\end{eqnarray}

In the nonrelativistic limit ($Y \to \infty$), the last equations reduce to
\begin{equation}\label{hoylarge2}
	M \approx \sigma\, m + \frac{3}{2} \left( \frac{\sigma a^2 Q^2}{m} \right)^{1/3} ,
\end{equation}
as expected for a nonrelativistic Hamiltonian with a linear potential. 
For large value of $m$, the choice $Q=Q_1$ is
a good one. For small values of $m$, another choice could give better results.

\subsubsection{Coulomb potential}
\label{sec:salpcoul}

Equation~(\ref{eq:epl3}) is considerably simplified when $\lambda=-1$; the value
of $x_0$ can directly be extracted in this case and reads
\begin{equation}
	x_0=\frac{4m^2}{\sigma^2-4A_{-1}} \quad  {\rm with}\quad A_{-1}=
	\left(\frac{a}{2Q}\right)^2.
\end{equation}
The energy spectrum (\ref{eq:MlaAl}) is finally given by 
\begin{equation}\label{encoulsr}
	M(\lambda=-1)=\sigma m \sqrt{1-\frac{a^2}{\sigma^2Q^2}}.
\end{equation}
It is obvious from this last equation that bound states exist only when $a <
\sigma\, Q$, in agreement with (\ref{aclambda}). The most stringent upper bound
for $a$ is actually found for the ground state, that is $a<\sigma \left.Q
\right|_{n=l=0}$. With $Q=Q_{-1}$, (\ref{encoulsr}) yields an upper bound on the
energy since this setting gives the exact solution for the corresponding
nonrelativistic Hamiltonian. This is
confirmed by the results of Refs.~\citen{luchabd,srcoul3} in which formula~(\ref{encoulsr})
was already given.

As the coupling constant $a$ is dimensionless in this case, the only mass scale
of the Hamiltonian is $m$. So the nonrelativistic limit cannot be obtained by
setting $m \to \infty$, as usual. It is well known that the mean speed of the
particle is independent of $m$ and proportional to $a$ in the Coulomb case. 
So, the nonrelativistic
limit is achieved for $a \to 0$. When $a \ll 1$, (\ref{encoulsr}) reduces to
\begin{equation}\label{encoulsrasmall}
	M \approx \sigma m -\frac{a^2 m}{2\sigma Q^2},
\end{equation}
which corresponds to the rest energy plus the Coulomb binding energy, as expected. 
For large values of $m$, the choice
$Q=Q_{-1}$ is clearly optimal. For small values of $m$, another choice could give
better results.

Several works have been devoted to the spinless Salpeter equation with a Coulomb
potential \cite{luchabd,srcoul3,srcoul1,srcoul2,srcoul4}, and it is
interesting to compare our results with previously found ones. First of all, it
has been shown in Ref.~\citen{srcoul1} that the energy spectrum of the
semirelativistic Coulomb problem is unbounded from below if 
\begin{equation}
	a> a_c = \frac{2\sigma}{\pi}.
\end{equation}
Moreover, a lower bound on the ground-state energy $M_g$ is \cite{srcoul1}
\begin{equation}
	M_g\geq \sigma m\sqrt{1-\frac{a^2}{a_c^2}}.
\end{equation}

A last result of interest is the analytical determination of the ground-state
energy performed in Ref.~\citen{srcoul4}. It is found that 
\begin{equation}\label{ground}
	M_g |_{a=a_c} \lesssim \sigma\,m\times 0.4842564\dots
\end{equation}
Our formula~(\ref{encoulsr}) leads in this case to 
\begin{equation}
M|_{a=a_c} = \sigma m\sqrt{1-\frac{4}{\pi^2\left.Q\right|_{n=l=0}^2}} .
\end{equation}
This last expression is equal to 0.77$\, \sigma m$ if $Q = Q_{-1}$, but agrees with
(\ref{ground}) if $\left.Q\right|_{n=l=0}\approx 0.73$ is taken.

\subsubsection{Ultrarelativistic limit}
\label{sec:ultrarel}

In the ultrarelativistic limit, that is to say $m=0$, the various elimination
equations for the auxiliary field and the energy formulae become simpler. 
Equation~(\ref{eq:epl3}) is then written as
\begin{equation}
\label{x0urcoul}
x_0 \, \left(\frac{\sigma^2}{4}\, x_0^{\lambda+1}-A_\lambda\right)=0,	
\end{equation}
and the final energy spectrum is given by (using the non trivial solution of (\ref{x0urcoul}))
\begin{equation}\label{ezerom}	
M^{{\rm ur}}_\lambda(a,\sigma,N)=\frac{\lambda+1}{\lambda} \left(a
|\lambda|\right)^{\frac{1}{\lambda+1}}\, \left(\sigma Q\right)^{\frac{\lambda}
{\lambda+1}}.
\end{equation}
Let us stress that, in this special case, an analytical expression is obtained
whatever the power $\lambda$ of the potential, in contrast to the particular
values (\ref{vallambr3}) resulting from the general case. One can check that
(\ref{E2_3f}) and (\ref{E1_3f}) reduce to (\ref{ezerom}) when $m\to 0$.

The quantity $M^{{\rm ur}}_\lambda$ is physically a mass. It appears that
$M^{{\rm ur}}_\lambda <0$ when $-1 < \lambda < 0$ and $M^{{\rm ur}}_{-1}=0$.
No bound state of massless constituent particles can be found in these cases.
Although (\ref{ezerom}) is positive for $-2<\lambda<-1$, this range of values for
$\lambda$ has been proved to be unphysical.

For the particular case $\lambda=1$, this last formula becomes $M_1^{{\rm ur}}
=2\sqrt{\sigma\, a\, Q}$, and one has consequently $(M^{{\rm ur}}_1)^2\propto Q$.
Such a linear behavior between the squared masses and the quantum numbers
of the different states (with $Q=Q_1$ for instance) is a well-known property of
the spinless Salpeter Hamiltonian with a linear potential and massless
constituents. Such a Hamiltonian is one of the simplest ways to describe a light
meson in a potential approach of QCD, and it is experimentally checked that the
square masses of the light mesons mainly grow linearly with their angular
momentum (Regge trajectories). See for example Refs.~\citen{lucha,goity} for a discussion
of that point.  

It is worth looking at the ultrarelativistic harmonic oscillator. 
The energy spectrum of the Hamiltonian 
\begin{equation}\label{hamho}
  h_{{\rm ho}}=2\sqrt{\bm p^{\,2}}+a\, r^2
\end{equation}
can be analytically computed for $l=0$ and reads \cite{luchaho}
\begin{eqnarray}\label{srhowf}
M_{{\rm ho}}&=&-(4a)^{1/3}\alpha_n,
\end{eqnarray}
where $\alpha_n<0$ are the zeros of the regular Airy function (see Appendix~\ref{sec:obs_Ai}).
Using (\ref{eq:QAi}), (\ref{srhowf}) reads
\begin{equation}
	M_{{\rm ho}}=3\left(\sqrt a\,Q(1;n,0)\right)^{2/3}.
\end{equation}
This last expression can be compared to the AFM result (\ref{ezerom}) for
$\lambda=\sigma=2$, \textit{i.e.}
\begin{equation}	
M^{{\rm ur}}_2=3\left(\sqrt a\, Q\right)^{2/3}.
\end{equation}
With (\ref{hamho}) as a starting point, a natural choice for $Q$ is $Q_2$, but 
it is clear that both $M_{{\rm ho}}$ and $M^{{\rm ur}}_2$ are identical
provided that $Q=Q(1;n,0)$, the principal quantum number giving the exact 
energy formula in the case of a nonrelativistic
kinetic energy with linear potential. The explanation of this fact is that
the Fourier transform of Hamiltonian (\ref{hamho}) is a nonrelativistic
Hamiltonian with linear potential. This is a supplementary
indication that $Q$ should be different in the semirelativistic and
nonrelativistic cases for the same potential. 

\subsubsection{Improved formulae for ultrarelativistic power-law potentials}

Exactly as it was the case in the nonrelativistic framework, it is possible to
improve drastically the energy formulae of the spinless Salpeter simply by
changing the value of the principal quantum number $Q$. 
To study this problem, we employ the scaling law properties and consider the
dimensionless Hamiltonian valid for massless particles
\begin{equation}
\label{hpowdimless}
H=2\sqrt{\bm p^2}+ r^\lambda
\end{equation}
with $\lambda>0$. Very accurate numerical values for the eigensolutions of 
this Hamiltonian can be obtained with the Lagrange mesh method. \cite{lag}

The approximate energy spectrum is given by (\ref{ezerom})
with $a=1$ and $\sigma=2$. With the choice $Q=Q_2=2\, n+l+3/2$, upper bounds
are obtained. As mentioned before, another choice for the $n$- and
$l$-dependences of $Q$ can greatly improve the results. By using the form
\begin{equation}
\label{gnlambdapow}
Q=b(\lambda)\, n + d(\lambda) \, l + c(\lambda),
\end{equation}
we find smooth variations for coefficients $b$, $c$ and $d$ for $\lambda \in
]0,2]$ ($l \le 3$ and $n \le 3$). In particular, $d(\lambda)\approx 1$ with
relative variations less than 2\%. Coefficients $b(\lambda)$ and $c(\lambda)$
can be fitted with various functions and similar agreement. Finally, we choose
\begin{equation}
\label{bcdlambdapow}
b(\lambda)=\frac{3.00\, \lambda+3.67}{\lambda+3.40}, \quad
c(\lambda)=\frac{2.69\, \lambda+8.69}{\lambda+8.27}, \quad
d(\lambda)=1.
\end{equation}
Agreement with exact results is very good but the variational character of the
approximation is no longer guaranteed.
With the choice (\ref{bcdlambdapow}), the maximal relative error for $l \le 3$
and $n \le 3$ and for $\lambda \in [0.1,2]$ is located between 0.3 and 1.1\%.
With the choice $Q=Q_2$, the corresponding error is located between 4.5 and
12.7\%. Just to give a quantitative idea of the quality of the approximation,
results for $\lambda =1$ are presented in Table~\ref{tab:powla1}.

\begin{table}[htb]
\caption{Eigenvalues $\epsilon(n,l)$ of the Hamiltonian~(\ref{hpowdimless}) with
$\lambda=1$, for some sets $(n,l)$. 
First line: value from numerical integration (exact values); 
second line: approximate result (\ref{ezerom}) with $Q$ defined by (\ref{gnlambdapow}) and
(\ref{bcdlambdapow});
third line: approximate result (\ref{ezerom}) with $Q=2\, n+l+3/2$ (upper bounds).}
\label{tab:powla1}
\begin{center}
\begin{tabular}{ccccc}
\hline\hline\noalign{\smallskip}
$l$ & $\epsilon(0,l)$ & $\epsilon(1,l)$ & $\epsilon(2,l)$ & $\epsilon(3,l)$ \\ [3pt]
\hline
0 & \textbf{3.1577} & \textbf{4.7109} & \textbf{5.8913} & \textbf{6.8742} \\
  & 3.1338 & 4.6849 & 5.8374 & 6.7973 \\
  & 3.4641 & 5.2915 & 6.6333 & 7.7460 \\
\noalign{\smallskip}
1 & \textbf{4.2248} & \textbf{5.4575} & \textbf{6.4837} & \textbf{7.3767} \\
  & 4.2215 & 5.4725 & 6.4866 & 7.3623 \\
  & 4.4721 & 6.0000 & 7.2111 & 8.2462 \\
\noalign{\smallskip}
2 & \textbf{5.0789} & \textbf{6.1304} & \textbf{7.0470} & \textbf{7.8671} \\
  & 5.0814 & 6.1602 & 7.0764 & 7.8869 \\
  & 5.2915 & 6.6333 & 7.7460 & 8.7178 \\
\noalign{\smallskip}
3 & \textbf{5.8108} & \textbf{6.7425} & \textbf{7.5775} & \textbf{8.3387} \\
  & 5.8156 & 6.7785 & 7.6207 & 8.3787 \\
  & 6.0000 & 7.2111 & 8.2462 & 9.1652 \\
\noalign{\smallskip}\hline
\end{tabular}
\end{center}
\end{table}

It was shown that $\left. Q\right|_{l=0}$ must be given by (\ref{eq:QAi}) for
a quadratic potential. Formulae~(\ref{bcdlambdapow}) give $b(2)=1.79$ close
to $\pi/\sqrt{3}\approx 1.81$ and $c(2)=1.37$ close to $\pi\sqrt{3}/4\approx
1.36$. This is also in agreement with formulae~(71) in Ref.~\citen{bsb08a} which
predict, in the case of a nonrelativistic Hamiltonian with a linear potential,
$b(1)=1.79$ and $c(1)=1.38$.

\subsubsection{Improved formulae for relativistic Coulomb potential}
\label{sec:rcoulp}

In this section, we study the relativistic Coulomb potential and use the
scaling properties to set $m=1$. With dimensionless variables, the
semirelativistic Coulomb Hamiltonian is written in this case
\begin{equation}
\label{hcouldimless}
h=2\sqrt{\bm p^2+1}-\frac{a}{r}
\end{equation}
with $0 \le a < a_c$. Again, very accurate numerical values for the eigensolutions of 
this Hamiltonian can be obtained with the Lagrange mesh method. \cite{lag}

\begin{table}[htb]
\caption{Eigenvalues $\epsilon(n,l)$ of the Hamiltonian~(\ref{hcouldimless}) with
$a=1$, for some sets $(n,l)$. 
First line: value from numerical integration (exact values); 
second line: approximate result (\ref{encoulsr}) with $Q$ defined by (\ref{gnacoul}) and (\ref{bcdacoul});
third line: approximate result (\ref{encoulsr}) with $Q=n+l+1$ (upper bounds).}
\label{tab:coula6}
\begin{center}
\begin{tabular}{ccccc}
\hline\hline\noalign{\smallskip}
$l$ & $\epsilon(0,l)$ & $\epsilon(1,l)$ & $\epsilon(2,l)$ & $\epsilon(3,l)$ \\ [3pt]
\hline
0 & \textbf{1.65817} & \textbf{1.92184} & \textbf{1.96739} & \textbf{1.98231} \\
  & 1.65982 & 1.92356 & 1.96680 & 1.98151 \\
  & 1.73205 & 1.93649 & 1.97203	& 1.98431 \\
\noalign{\smallskip}
1 & \textbf{1.93515} & \textbf{1.97122} & \textbf{1.98389} & \textbf{1.98973} \\
  & 1.93476 & 1.97012 & 1.98291 & 1.98895 \\
  & 1.93649 & 1.97203 & 1.98431 & 1.98997 \\
\noalign{\smallskip}
2 & \textbf{1.97187} & \textbf{1.98416} & \textbf{1.98987} & \textbf{1.99297} \\
  & 1.97296 & 1.98416 & 1.98961 & 1.99266 \\
  & 1.97203 & 1.98431 & 1.98997 & 1.99304 \\
\noalign{\smallskip}
3 & \textbf{1.98428} & \textbf{1.98993} & \textbf{1.99301} & \textbf{1.99487} \\
  & 1.98528 & 1.99021 & 1.99302 & 1.99477 \\
  & 1.98431 & 1.98997 & 1.99304 & 1.99489 \\
\noalign{\smallskip}\hline
\end{tabular}
\end{center}
\end{table}

The approximate energy spectrum is given by
(\ref{encoulsr}) with $m=1$ and $\sigma=2$. With the choice $Q=Q_{-1}=n+l+1$,
upper bounds are obtained and the results are exact in the limit $a \to 0$.
As mentioned before, another choice for the $n$- and $l$-dependences of
$Q$ can greatly improve the results. By using the form
\begin{equation}
\label{gnacoul}
Q=b(a)\, n + d(a) \, l + c(a)
\end{equation}
and imposing $b(0)=c(0)=d(0)=1$, we find smooth variations for coefficients $b$,
$c$ and $d$ for $a \in ]0,a_c]$ ($l \le 3$ and $n \le 3$). These coefficients
can be fitted with various functions and similar agreement. Finally, we choose
\begin{equation}
\label{bcdacoul}
b(a)=\frac{1.03\, a-1.48}{a-1.48}, \quad
c(a)=\frac{1.07\, a-1.64}{a-1.64}, \quad
d(a)=\frac{0.96\, a-1.56}{a-1.56}.
\end{equation}
Agreement with exact results is very good for $a\lesssim 1.2$ but the variational
character of the approximation is no longer guaranteed. With the choice
(\ref{bcdacoul}), the maximal relative error for $l \le 3$ and $n \le 3$ and for
$a \in [0.2,1.2]$ is located between 0.005 and 0.3\%. With the choice $Q=Q_{-1}$,
the corresponding error is located between 0.004 and 17.3\%. 
To obtain a good accuracy in the domain $a\approx a_c=4/\pi\approx 1.273$, special
method must be used as the one presented in Ref.~\citen{srcoul4}. To recover the value
obtained for the ground state in this paper, it is necessary to have $c(a_c)=0.73$.
Our formula gives $c(a_c)=0.77$. Results for $a=1$ are presented in
Table~\ref{tab:coula6}. 

One can see that $Q=Q_{-1}=n+l+1$ is a better choice for large values of $n$ or $l$.
This can be understood as a kind of nonrelativistic behavior since the limits
$a \to 0$ and $Q \to \infty$ have similar effects. 

\subsection{Square root potential}
\label{sec:salpsqrt}

In Sect.~\ref{sec:sqrtpot}, we introduced the square root potential
$V(r)=\sqrt{a^2 r^2+b^2}$ and commented its important role in the theoretical
description of hybrid mesons in a nonrelativistic framework. Nevertheless, such
a type of potential allied with a relativistic kinetic energy operator is 
important to describe hybrid mesons with low mass quarks as well. The general formula
with a non vanishing quark mass looks rather difficult and we prefer here
to present the two extreme limits and show the close connection with the previous
section through a Fourier transform.

\subsubsection{Nonrelativistic kinematics}

The nonrelativistic limit follows from the Hamiltonian $H=\frac{\bm p^2}{2 m}+
\sqrt{a^2 r^2+b^2}$ and as been studied in Sect.~\ref{sec:sqrtpot}. It is
easy to show that the Fourier transform of this Hamiltonian, denoted as
$H^{\textrm{FT}}$, reads
\begin{equation}
	H^{\textrm{FT}}=\sigma \sqrt{\bm p^2+M^2}+\kappa\,r^2, 
\end{equation}
with 
\begin{equation}
	\sigma=\left( \frac{4 a}{m^2} \right)^{1/3},\quad M=\frac{b}{\sigma},\quad
	\kappa = \sigma\frac{m a}{8}.
\end{equation}
This last Hamiltonian is nothing else than a spinless Salpeter one~(\ref{eq:Htsr})
with a harmonic potential. Provided that the proper substitution rules are taken
into account, the energy spectrum in this case is the same than the one computed
in Sect.~\ref{sec:salpharm}. We have thus 
\begin{eqnarray}
	E^{{\rm nr}}_{{\rm sq}}&= &b \sqrt{\frac{3}{Y_2}}  \frac{Y_2+4G_-(Y_2)}
	{\sqrt{8G_-(Y_2)+3Y_2}} \\
	& = & b \sqrt{\frac{3}{Y_2}} \left( \frac{2}{G_-(Y_2)} + \frac{Y_2}
	{2 G_-^2(Y_2)} \right)\\
	& = & \frac{2 b}{\sqrt{3 Y_2}} \left(G_-^2(Y_2) + \frac{1}{G_-(Y_2)}\right),
\end{eqnarray}
with
\begin{equation}
Y_2=\frac{b^2}{3}\left(\frac{32 \,m}{a^2Q^2}\right)^{2/3}.
\end{equation}
This potential was studied in more detail in Sect.~\ref{sec:sqrtpot}, and
we recovered here the results presented there (see (\ref{eq:Enu0root})). This
remarkable result proves that AFM is an approximation consistent with the
Fourier transform.

\subsubsection{Ultrarelativistic limit}

In the ultrarelativistic limit ($m=0$), the corresponding Salpeter Hamiltonian
is $H=\sigma\sqrt{\bm p^2}+\sqrt{a^2 r^2+b^2}$. To get the eigenenergies, it is
not necessary to start a new calculation. In fact, the Fourier transform of $H$
is simply
\begin{equation}
	H^{\textrm{FT}}=\sigma \sqrt{\bm p^2+M^2}+a\, r,
\end{equation}
with $M=b/\sigma$.
But this last Hamiltonian appears to be a spinless Salpeter one~(\ref{eq:Htsr})
with a linear potential. Provided that the proper substitution rules are taken
into account, the energy spectrum in this case is the same than the one computed
in Sect.~\ref{sec:salplin}. We have thus 
\begin{eqnarray}
	M^{{\rm ur}}_{{\rm sq}} & = & b \sqrt{\frac{2}{Y_1}} \frac{Y_1+3F_-(Y_1)}
	{\sqrt{3F_-(Y_1)+2Y_1}} \\
	& = &b \sqrt{\frac{2}{Y_1 F_-(Y_1)}} \left( 3 + \frac{Y_1}{F_-(Y_1)} \right) \\
	& = & \frac{b}{\sqrt{2 Y_1 F_-(Y_1)}} \left( 3 + F_-^2(Y_1)\right), 
\end{eqnarray}
with
\begin{equation}
Y_1=\frac{3^{3/2}\, b^2}{2a\, \sigma Q}.
\end{equation}
When $b\to 0$, these last equations logically reduce to (\ref{ezerom})
with $\lambda=1$.

\subsection{Funnel potential}
\label{sec:salpfun}

As we stressed in Sect.~\ref{sec:funnel}, the funnel potential
$V(r)=ar-b/r$, with $a$ and $b$ both positive, is of crucial importance in
hadronic physics. It was shown that an analytical expression is available in a
nonrelativistic framework. However the glueball resonances or the spectroscopy
of very light mesons need a semirelativistic description and a spinless Salpeter
equation governed with a funnel potential is the most economic way to consider
this situation.

\subsubsection{Massless particle}
\label{sec:salpfunm0}

The ultrarelativistic limit ($m=0$), is particularly important for the
description of a glueball composed of two massless gluons or a meson composed of two massless quarks. 
Then, the extremization condition (\ref{eq:detr0rel}) becomes $a r_0^2 + b = \sigma Q$
whose solution is
\begin{equation}
\label{eq:r0m0fun}
r_0=\sqrt{\frac{\sigma Q - b}{a}}.
\end{equation}
Inserting this value in the expression of the mass (\ref{eq:massr0rel})
leads to the mass formula
\begin{equation}\label{funnelur}
M^{{\rm ur}}_{{\rm f}}=	2 \sqrt{a \left( \sigma Q \ - b \right)}.
\end{equation}
It is astonishing that such a complicated potential used in conjunction with
a semirelativistic kinetic energy admits analytical approximate eigenenergies
of such simple form. For $b=0$, (\ref{funnelur}) reduces to $2\sqrt{a\sigma Q}$,
that is the expected expression from (\ref{ezerom}) in the case $\lambda=1$.
For $a=0$, $M^{{\rm ur}} \to 0$ as shown in Sect.~\ref{sec:ultrarel}.
If the prescription $Q=Q_2$ is chosen, then (\ref{funnelur}) is an upper bound
on the exact result. As already mentioned, other choice for $Q$ could improve
the accuracy of the formula but without guarantee about the variational
character of the result. This point is studied in the next section.

\subsubsection{Improved formulae for the ultrarelativistic limit}
\label{sec:impformrelatfun}

Written with dimensionless variables, the ultrarelativistic Hamiltonian
for two massless particles interacting via the funnel potential is given by
\begin{equation}
\label{hfundimless}
h=2\sqrt{\bm p^2}+r-\frac{\beta}{r}
\end{equation}
with $\beta \ge 0$. The approximate energy spectrum is given by (\ref{funnelur})
with $\sigma=2$, $a=1$ and $b$ replaced by $\beta$ (to avoid confusion with the
coefficient of $n$). This kind of Hamiltonian is often used in hadronic physics
with typical values for $\beta \approx 0.4$. Very accurate numerical values for the eigensolutions of 
this Hamiltonian can be obtained with the Lagrange mesh method. \cite{lag}

With the choice $Q=Q_2=2\,n+l+3/2$,
upper bounds are obtained. As shown in previous cases, another choice for
the $n$- and $l$-dependences of $Q$ can greatly improve the results.
By using the form
\begin{equation}
\label{gnbetfun}
Q=b(\beta)\, n + d(\beta) \, l + c(\beta),
\end{equation}
we find smooth variations for coefficients $b$, $c$ and $d$ for $\beta \in [0,1]$
($l \le 3$ and $n \le 3$). 
These coefficients can be fitted with various functions
and similar agreement. Finally, we choose
\begin{equation}
\label{bcdbetfun}
b(\beta)=\frac{1.88\, \beta-5.34}{\beta-3.51}, \quad
c(\beta)=\frac{1.99\, \beta-4.40}{\beta-3.49}, \quad
d(\beta)=\frac{0.76\, \beta-2.46}{\beta-2.54}.
\end{equation}
Agreement with exact results is very good but the variational character of the
approximation is no longer guaranteed. With the choice (\ref{bcdbetfun}), the
maximal relative error for $l \le 3$ and $n \le 3$ and for $\beta \in [0,1]$ is
located between 0.6 and 4.9\%. With the choice $Q=Q_2$, the corresponding error
is located between 12.7 and 42.2\%. 
Results for $\beta =0.4$ are presented in Table~\ref{tab:funbet04}.

For $\beta=0$, one obtains $b=1.52$, $c=1.26$, and $d=1.09$. From
(\ref{bcdlambdapow}) with $\lambda=1$, one obtains $b=1.52$, $c=1.23$, and
$d=1$. These values are close to each other as expected.

\begin{table}[htb]
\caption{Eigenvalues $\epsilon(n,l)$ of the Hamiltonian~(\ref{hfundimless}) with
$\beta=0.4$, for some sets $(n,l)$. 
First line: value from numerical integration (exact values); 
second line: approximate result (\ref{funnelur}) with $Q$ defined by (\ref{gnbetfun}) and (\ref{bcdbetfun});
third line: approximate result (\ref{funnelur}) with $Q=2\, n+l+3/2$ (upper bounds).}
\label{tab:funbet04}
\begin{center}
\begin{tabular}{ccccc}
\hline\hline\noalign{\smallskip}
$l$ & $\epsilon(0,l)$ & $\epsilon(1,l)$ & $\epsilon(2,l)$ & $\epsilon(3,l)$ \\ [3pt]
\hline
0 & \textbf{2.7821} & \textbf{4.3709} & \textbf{5.5874} & \textbf{6.5938} \\
  & 2.7804 & 4.4196 & 5.5977 & 6.5678 \\
  & 3.2249 & 5.1381 & 6.5115 & 7.6420 \\
\noalign{\smallskip}
1 & \textbf{3.9944} & \textbf{5.2365} & \textbf{6.2744} & \textbf{7.1772} \\
  & 3.9737 & 5.2529 & 6.2765 & 7.1552 \\
  & 4.2895 & 5.8652 & 7.0993 & 8.1486 \\
\noalign{\smallskip}
2 & \textbf{4.8993} & \textbf{5.9549} & \textbf{6.8772} & \textbf{7.7028} \\
  & 4.8837 & 5.9710 & 6.8887 & 7.6978 \\
  & 5.1381 & 6.5115 & 7.6420 & 8.6255 \\
\noalign{\smallskip}
3 & \textbf{5.6588} & \textbf{6.5927} & \textbf{7.4311} & \textbf{8.1957} \\
  & 5.6489 & 6.6115 & 7.4508 & 8.2046 \\
  & 5.8652 & 7.0993 & 8.1486 & 9.0774 \\
\noalign{\smallskip}\hline
\end{tabular}
\end{center}
\end{table}

\subsubsection{General case}
\label{sec:salpfungen}

In order to treat the general case
\begin{equation}
\label{eq:hamilfunwdim}
H =\sigma \sqrt{\bm p^2 + m^2} + a\,r - \frac{b}{r},
\end{equation}
we use the dimensionless Hamiltonian $h=H/\sqrt{\sigma\, a}$ given by
\begin{equation}
\label{eq:hamilfunsimp}
h =\sqrt{\bm p^2 + \chi^2} + r - \frac{\beta}{r},
\end{equation}
with $\chi=m\sqrt{\sigma/a}$ and $\beta=b/\sigma$.
The transcendental equation (\ref{eq:detr0rel}) writes in this case
\begin{equation}
Q = (r_0^2+\beta) \sqrt{1+(\chi r_0/Q)^2}.
\end{equation}
Defining a new variable $x$ and a new parameter $\alpha$ by
\begin{equation}
\label{eq:defxalpha}
x = \frac{\chi^2}{Q^2}r_0^2 \quad; \quad \alpha = \frac{\chi^2}{Q^2} \beta,
\end{equation}
the previous equation can be recast as a third order polynomial
equation:
\begin{equation}
x^3+(1+2\alpha)x^2+(\alpha^2+2\alpha)x+\alpha^2(1-Q^2/\beta^2) = 0.
\end{equation}
In consequence, it is analytically solvable and can be put under the
reduced form (\ref{eq:redcubeq1}).
Reporting the obtained $x$ value in the expression (\ref{eq:massr0rel}) for the
mass leads to the final expression.

Explicitly, the procedure to get the analytical AFM expression for the
spinless Salpeter equation based on the Hamiltonian (\ref{eq:hamilfunsimp})
needs the following steps.
Calculate the $\alpha$ value from (\ref{eq:defxalpha}), then introduce
the parameter
\begin{equation}
\label{eq:defYfunrel}
Y = \textrm{sgn}(\alpha-1)+\frac{27 \alpha^2 Q^2}{2 \beta^2 |\alpha - 1|^3}.
\end{equation}
The $x$ value is obtained from the expression
\begin{equation}
\label{eq:defxrelfun}
x=\frac{1}{3} \left [ |\alpha - 1| F_{-}(Y) - (1+2\alpha) \right ],
\end{equation}
while the AFM mass is given by
\begin{equation}
\label{eq:defmassrelfun}
M_{\textrm{AFM}}=\frac{1}{\chi\sqrt{x}} \left [ \chi^2 \sqrt{1+x} + (x - \alpha)Q 
\right ].
\end{equation}
Once again, one sees that the AFM is very powerful to get analytical expressions,
even for quite sophisticated problems. We have checked that (\ref{funnelur})
[with $\sigma=1$, $a=1$ and $b=\beta$] is recovered in the limit $\chi\to 0$. It
can also be checked that (\ref{eq:defmassrelfun}) and (\ref{eq:Efun})
[with $\sigma=1$, $m=\chi$, $a=1$ and $b=\beta$]  tend towards the same limit when
$\chi\to \infty$.

\subsubsection{Low mass expression}
\label{sec:salpfunlm}


By a ``small mass", we mean the condition $m \ll \sqrt{a}$. For $m=0$, we saw
already that $r_0 = \sqrt{(\sigma Q - b)/a}$. The principle of the method is based
on a limited expansion of the equations at first order in $m^2$. Thus, we set
\begin{equation}
r_0^2=\frac{\sigma Q - b}{a} (1+\epsilon).
\end{equation}
From the transcendental equation, it is easy to show that
\begin{equation}
\epsilon=- \frac{\sigma m^2}{2 a Q}.
\end{equation}
Inserting these values in the expression (\ref{eq:massr0rel}) for the mass, we
obtain the final result
\begin{equation}
\label{eq:enerfunlm}
M^{{\rm lm}}_{{\rm f}}=	\sqrt{\frac{\sigma Q\ - b}{a}}
	\left(2a+\frac{\sigma m^2}{2Q}\right).
\end{equation}
Here again, it is amazing that the approximate eigenvalues of a so sophisticated
Hamiltonian take such a simple form.

\subsection{Unequal masses}
\label{sec:unqmass}

Some particular problems require to deal with a system of
two particles with unequal masses. In this case, a general spinless Salpeter
Hamiltonian is given in the rest frame by
\begin{equation}
\label{Hm1m2}
	H=\sqrt{\bm p^2+m^2_1}+\sqrt{\bm p^2+m^2_2}+V(r).
\end{equation}

A general result can be obtained for Hamiltonians with two different
masses. Let us consider the following two-body Hamiltonians $H=T_1+T_2+V$,
$H_1=2 T_1+V$, and $H_2=2 T_2+V$ whose ground-state energies are respectively
$M=\langle\phi|H|\phi\rangle$, $M_1=\langle\phi_1|H_1|\phi_1\rangle$, and
$M_2=\langle\phi_2|H_2|\phi_2\rangle$. Since $H=(H_1+H_2)/2$, we can write
\begin{equation}
\langle\phi|H|\phi\rangle= \frac{1}{2} \left( \langle\phi|H_1|\phi\rangle +
\langle\phi|H_2|\phi\rangle \right). 
\end{equation}
The Ritz theorem implies that
\begin{equation}
M \ge \frac{1}{2} \left( M_1 + M_2 \right). 
\end{equation}
For particular cases, this approximation can be quite good. In Ref.~\citen{sema94},
it is shown that $M \approx ( M_1 + M_2 )/2$ for a relativistic Hamiltonian of
kind (\ref{Hm1m2}) with $V(r)=a\, r$ and $m_i \ll \sqrt{a}$.

The square roots appearing in the kinetic terms can still be treated by resorting
to the AFM. But this time, two auxiliary fields, $\mu_1$ and $\mu_2$, have to
be introduced. One is led to the Hamiltonian
\begin{eqnarray}\label{eq:hamuneq}
	\tilde{H}(\mu_1,\mu_2)&=&\frac{\mu_1+\mu_2}{2}+\frac{m^2_1}{2\mu_1}+
\frac{m^2_2}{2\mu_2}+\frac{\bm p^2}{2\bar{M}(\mu_1,\mu_2)}\nonumber \\
&&+\textrm{sgn}(\lambda) \nu r^\lambda+V(J(\nu))-\textrm{sgn}(\lambda) \nu J(\nu)^\lambda,
\end{eqnarray}
with 
\begin{equation}
	\bar{M}(\mu_1,\mu_2)=\frac{\mu_1\mu_2}{\mu_1+\mu_2}
\end{equation}
playing the role of a reduced mass. It has been shown in Ref.~\citen{sema12} that the AFM
leads to the system~(\ref{SRvir1})-(\ref{SRvir3}), but with $T(x)=\sqrt{x^2+m_1^2}+\sqrt{x^2+m_2^2}$.

It is then easy to obtain numerical upper bounds, but one can wonder if it is possible to obtain analytical solutions? Due to the presence of the square roots, it is only conceivable with very special conditions: one mass is vanishing ($m_1=0$ and $m_2=m$) and the potential is well chosen. Analytical solutions can be obtained for the funnel potential, often used in hadronic physics \cite{sema12}. Let us note that an analytical upper bound for a logarithmic potential (also relevant in hadronic physics) can also be computed. 

The solution for the general funnel potential is very complicated. As an illustration, the case of the Coulomb interaction, $-a/r$, is only presented here. The solution for the associated system~(\ref{SRvir1})-(\ref{SRvir3}) gives
\begin{equation}
\label{cpr0}
r_0 = \frac{Q}{m}\frac{\sqrt{a (2 Q-a)}}{a-Q}
\end{equation}
and
\begin{equation}
\label{eq:Ecoul0}
M=2 m \sqrt{\frac{a}{2Q} \left(1 - \frac{a}{2Q} \right)}.
\end{equation}
This formula is an upper bound if $Q=n+l+1$, but, as we pointed out
several times, it is justified to take a more sophisticated expression in order
to get a better accuracy.
It can be seen from (\ref{eq:Ecoul0}) that $\lim_{m\to 0} M =0$, as expected. 

From equations above, a solution $M$ exists only when the following condition is verified:
\begin{equation}
\label{cpcond}
\frac{a}{2} < Q < a.
\end{equation}
The right part of this inequality corresponds to the cancellation of the binding ($M\to m$ when $Q\to a$), and the left part corresponds to a collapse  ($M\to 0$ when $Q\to a/2$). Indeed, the mean radius $r_0 \to \infty$ when $Q\to a$, and $r_0 \to 0$ when $Q\to a/2$. A sufficiently strong interaction must exist to bind the system, but unphysical values of the mass could appear if it is too strong. With the Coulomb potential, the nonrelativistic limit is not defined by $m\to\infty$ but by $a\to 0$. So, (\ref{cpcond}) implies that this limit is irrelevant for the solution (\ref{eq:Ecoul0}).

\subsection{Semiclassical interpretation}
\label{sec:semiclass}

\begin{figure}[ht]
\begin{center}
\includegraphics*[width=6cm]{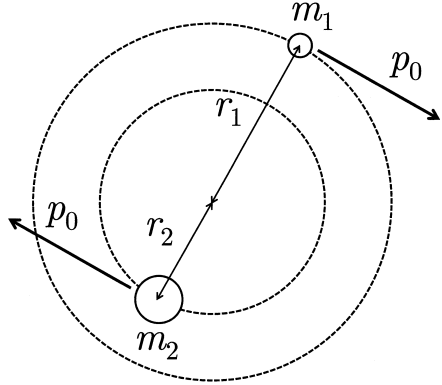}
\caption{Classical circular motion of the two relativistic particles. \label{fig:class}}
\end{center}
\end{figure}

A semiclassical interpretation of the system~(\ref{SRvir1})-(\ref{SRvir3}) is also possible. Let us assume a classical circular motion for the two particles, as illustrated in Fig.~\ref{fig:class}. In this case, the force $F_i$ acting on particle $i$ is given by
\begin{equation}
\label{Fi}
F_i = m_i\, \gamma(v_i)\, \frac{d v_i}{dt} \quad \textrm{with} \quad \frac{d v_i}{dt} = \frac{v_i^2}{r_i}.
\end{equation}
Taking into account that both particles are characterized by the same momentum $p_0$, This equation becomes
\begin{equation}
\label{Fi2}
F_i =\frac{p_0^2}{\sqrt{p_0^2+m_i^2}}\frac{1}{r_i}.
\end{equation}
The rigid rotation constraint, $v_1/r_1 = v_2/r_2$, implies that 
\begin{equation}
\label{r0r1r2}
r_0 = r_1 + r_2 = r_i \frac{\sqrt{p_0^2+m_1^2}+\sqrt{p_0^2+m_2^2}}{\sqrt{p_0^2+m_j^2}},
\end{equation}
with $i\ne j$. If the force acting on $i$ comes from the potential $V(r)$ generated by $j$, then $F_1=F_2=V'(r_0)$. (\ref{Fi2}) and (\ref{r0r1r2}) can be recast onto the form (\ref{SRvir3}), and it is obvious than (\ref{SRvir1}) gives the mass of the system. The total orbital angular momentum is $r_1\, p_0+r_2\, p_0=r_0\, p_0$. A semiclassical quantification gives thus $r_0\, p_0 = L+1/2$, and we obtain a system very similar to (\ref{SRvir1})-(\ref{SRvir3}). Nevertheless, the AFM produces more general results.

\section{$N$-body problems}
\label{sec:Nbodprob}

Up to now, the AFM was applied essentially to one- and two-body systems with
either a nonrelativistic or a relativistic kinetic energy operator. It was
shown that, in many situations, the method was able to provide an accurate
analytical expression for the corresponding eigenenergies. In this section we
address the important point of whether AFM is also able to give informations
concerning the eigenenergies of a system composed of $N$ particles.
Except the very special case of particles whose dynamics is governed by
quadratic interactions, which is treated in detail in Appendix~\ref{sec:nboh},
no analytical solution is known for $N$-body systems. Nevertheless
a number of very interesting physical problems need to solve a $N$-body
equation. As in the 2-body case, the AFM cannot pretend to give a very high
accuracy in the results, but it exhibits the interesting virtue to clearly
show the dependence of the results in terms of the physical parameters and
the various quantum numbers. 

Indeed, the envelope theory was used in the $N$-body problem to find bounds
for the nonrelativistic binding energies if the two-body potential fulfills
certain restrictive conditions. \cite{hall1980}\ In this section, we apply AFM
to the $N$-body problem in a more general context
\begin{itemize}
  \item we consider the possibility of dealing with a semirelativistic kinetic
  energy potential;
  \item we allow the presence of one-body terms in addition to two-body
  interactions.
\end{itemize}

Moreover, we are not interested only by the existence of bounds which can sometimes give
rather poor values, but the purpose of the present section is to show that the
AFM can be successfully applied to find approximate analytical mass formulae for
general relativistic $N$-body Hamiltonians of the form
\begin{equation}\label{hamgene}
H=\sum^{N}_{i=1}\sqrt{\bm p^2_i+m^2_i}+\sum^{N}_{i=1}
U_i(|\bm r_i-\bm R|)+\sum^{N}_{i<j=1}V_{ij}(|\bm r_i-\bm r_j|),
\end{equation}
with $\sum^{N}_{i=1} \bm p_i=\bm P=\bm 0$, where $\bm r_i$ and $\bm p_i$
are respectively the position and momentum of particle $i$ with a mass $m_i$,
and where $\bm R$ is a global variable defined in Appendix~\ref{sec:Jacobcoord}.
As stated, we consider a relativistic kinematics and allow for both pairwise $V$
and one-body $U$ interactions. Particular systems exhibiting one-body potentials
of the previous form will be studied below. Most of the material of this section
can be found in Ref.~\citen{silv10}.

A quantity which appears very often in this section is the number of interacting
particles. In order to simplify the notations, we use a special symbol to
designate this number and set
\begin{equation}
C_N = \frac{N(N-1)}{2}.
\end{equation}
  
\subsection{Main mass formulae}
\label{sec:geneqNbod}

\subsubsection{General case}

As stated, the AFM can be straightforwardly generalized to the case of
$N$-body Hamiltonians of the form~(\ref{hamgene}). The basic idea is
to introduce auxiliary fields so that this Hamiltonian is formally replaced by a
Hamiltonian for which an analytical solution can eventually be found. The
auxiliary fields are denoted as $\mu_i$, $\nu_i$, and $\rho_{ij}$, and are
introduced as follows   
\begin{eqnarray}\label{hamgene2}
H(\mu_k,\nu_k,\rho_{kl})&=&\sum^{N}_{i=1}\left[\frac{\mu_i}{2}+
\frac{m^2_i}{2\mu_i}+\frac{\bm p^2_i}{2\mu_i}\right]+\sum^{N}_{i=1}
\left[\nu_i\, P(r_i)+U_i(I_i(\nu_i))-\nu_i\, P(I_i(\nu_i))\right]\nonumber\\
&&+\sum^{N}_{i<j=1} \left[\rho_{ij}\, S(r_{ij})+
V_{ij}(J_{ij}(\rho_{ij}))-\rho_{ij}\, S(J_{ij}(\rho_{ij}))\right]	,
\end{eqnarray}
where (notice the definition of $r_i \ne |\bm r_i|$)
\begin{equation}
r_i=|\bm r_i-\bm R|,\quad r_{ij}=|\bm r_i-\bm r_j|,	
\end{equation}
 and where
\begin{equation}\label{idef}
I_i(x)=K^{-1}_i(x),	\quad  K_i(x)=\frac{U'_i(x)}{P'(x)},\quad
J_{ij}(x)=L^{-1}_{ij}(x),\quad L_{ij}(x)= \frac{V'_{ij}(x)}{S'(x)},
\end{equation}
where the prime denotes the derivative with respect to the argument.
It is sufficient to set $\mu_k=m_k$ $\forall k$ to treat a nonrelativistic
kinematics. We stress that, in order for the AFM to apply, it is assumed that
the mass spectrum of~(\ref{hamgene2}) is analytically computable. In practice,
this is only possible for the choice $P(x)=S(x)=x^2$ under some conditions
(see Appendix~\ref{sec:nboh}).

The reason of such a definition is that Hamiltonians~(\ref{hamgene}) and
(\ref{hamgene2}) are equivalent if the auxiliary fields are properly eliminated.
It can indeed be checked that 
\begin{eqnarray}
\label{elim}
\left.\delta_{\mu_i} H(\mu_k,\nu_k,\rho_{kl})\right|_{\mu_i=\hat\mu_i}=0
&\Rightarrow& \hat\mu_i=\sqrt{\bm p^2_i+m^2_i}, \\
\label{elim2}
\left.\delta_{\nu_i} H(\mu_k,\nu_k,\rho_{kl})\right|_{\nu_i=\hat\nu_i}=0
&\Rightarrow& \hat\nu_i=K_i(r_i), \\
\label{elim3}
\left.\delta_{\rho_{ij}} H(\mu_k,\nu_k,\rho_{kl})\right|_{\rho_{ij}=
\hat{\rho}_{ij}}=0&\Rightarrow& \hat{\rho}_{ij}=L_{ij}(r_{ij}),
\end{eqnarray}
and finally that $H(\hat\mu_k,\hat\nu_k,\hat{\rho}_{kl})=H$. 

The approximation underlying the AFM is now that, as we have done in previous
sections, the auxiliary fields will be seen as real variational parameters. Let
us note 
\begin{equation}
H(\mu_{k,0},\nu_{k,0},\rho_{kl,0}) |\varphi_0\rangle = M_0 \,|\varphi_0\rangle,  	
\end{equation}
where $\mu_{k,0}$, $\nu_{k,0}$, $\rho_{kl,0}$ are the real values of the
auxiliary fields such that $M_0$ is extremal. $|\varphi_0\rangle$ is an
eigenstate with fixed quantum numbers and a given symmetry. Using the same
mathematical techniques than in Sect.~\ref{sec:genAFM}, one can show that
\begin{eqnarray}
\label{muk0}
\mu_{k,0}&=&\sqrt{\langle\varphi_0|\bm p_i^2+m_i^2|\varphi_0\rangle}
\quad \textrm{with} \quad \bm P=\bm 0, \\
\langle\varphi_0|P(r_i)|\varphi_0\rangle&=&P(r_{i,0}) \quad \textrm{with}
\quad r_{i,0}=I_i(\nu_{i,0}), \\
\label{Sr0}
\langle\varphi_0|S(r_{ij})|\varphi_0\rangle&=&S(r_{ij,0}) \quad \textrm{with}
\quad r_{ij,0}=J_{ij}(\rho_{ij,0}).
\end{eqnarray}
Using the Hellmann-Feynman theorem \cite{feyn}, it can be shown that $\langle\varphi_0|\hat \mu_i^2|\varphi_0\rangle
= \mu_{i,0}^2$, $\langle\varphi_0|P(I_i(\hat \nu_i))|\varphi_0\rangle =
P(I_i(\nu_{i,0}))$ and $\langle\varphi_0|S(J_{ij}(\hat{\rho}_{ij}))|\varphi_0\rangle =
S(J_{ij}(\rho_{ij,0}))$. 
Let us rewrite the Hamiltonian~(\ref{hamgene2}) under the form 
\begin{equation}
H(\mu_k,\nu_k,\rho_{kl}) = T(\mu_k) + \sum^{N}_{i=1}
\tilde{U}_i(\nu_i,r_i) + \sum^{N}_{i<j=1} \tilde{V}_{ij}(\rho_i,r_{ij}),
\end{equation}
where $T(\mu_k)$ stands for the kinetic part. One can also check that
\begin{align}
&\tilde{U}_i(\nu_{i,0},r_{i,0}) = U(r_{i,0}), \quad \tilde{U}'_i(\nu_{i,0},r_{i,0})
= U'(r_{i,0}), \\
&\tilde{V}_{ij}(\rho_{ij,0},r_{ij,0}) = V(r_{ij,0}), \quad \tilde{V}'_{ij}
(\rho_{ij,0},r_{ij,0}) = V'(r_{ij,0}).
\end{align}
This means that the approximate potential $\tilde{U}_i$ ($\tilde{V}_{ij}$) and
the corresponding genuine potential $U_i$ ($V_{ij}$) are tangent at, at least,
one point. If the following conditions $\tilde{U}_i(\nu_{i,0},r_i) \ge U(r_i)$
and $\tilde{V}_{ij}(\rho_{ij,0},r_{ij}) \ge V(r_{ij})$ are fulfilled $\forall i$
and $\forall j$ and for all values of the radial arguments, $M_0$ is an upper
bound on an eigenstate of $H$ (\ref{hamgene}). If the kinematics is non
relativistic ($T=T(\mu_k=m_k)$) and if the following conditions
$\tilde{U}_i(\nu_{i,0},r_i) \le U(r_i)$ and $\tilde{V}_{ij}(\rho_{ij,0},r_{ij})
\le V(r_{ij})$ are fulfilled $\forall i$ and $\forall j$ and for all values of
the radial argument, $M_0$ is a lower bound on an eigenstate of $H$. This
restriction about $T$ comes from the fact that the replacement of the genuine
relativistic kinetic operator by the form $T(\mu_k)$ yields upper bounds on
the eigenvalues of the genuine Hamiltonian. In the other cases, it is not
possible to obtain a relevant information about the position of $M_0$.

At this stage, approximate numerical solutions of the $N$-body problem
can be easily computed. First, the choice $P(x)=S(x)=x^2$ allows a precise
determination of the eigenvalues of Hamiltonian~(\ref{hamgene2})
(see Appendix~\ref{sec:nboh}). From now on, this choice is adopted for the
rest of this section. Second, the extremization of the eigenvalues
with respect to the real auxiliary fields is a classical numerical problem
which can be solved with a high accuracy.

\subsubsection{Identical particles}

To obtain analytical closed solutions of the eigenvalue and extremization
problems associated to Hamiltonian~(\ref{hamgene2}), it is necessary to
simplify the system. First of all, we will only consider systems with identical
particles, that is with $m_i=m$. In this case, it is reasonable to consider
identical interactions between them, namely $U_k(x)=U(x)$ and $V_{kl}(x)=V(x)$.
This means that $K_k(x)=K(x)$, $I_k(x)=I(x)$, $L_{kl}(x)=L(x)$, and $J_{kl}(x)
=J(x)$. The global variable $\bm R$ (see (\ref{eq:defjacoord})) 
is then the center of mass of the system, even
for relativistic kinematics. 

Let us denote $\hat P_{ij}$ the permutation operator exchanging
particles $i$ and $j$. We can write $\hat P_{ij}|\varphi_0\rangle =
\pm |\varphi_0\rangle$ if this state $|\varphi_0\rangle$ is completely
(anti)symmetrical. Then
\begin{equation}
P(r_{j,0}) = \langle\varphi_0|P(r_j)|\varphi_0\rangle = \langle\varphi_0|
\hat P_{ij}P(r_i)\hat P_{ij}|\varphi_0\rangle = \langle\varphi_0|P(r_i)|
\varphi_0\rangle = P(r_{i,0}).
\end{equation}
If $P(x)$ is monotonic, which is always the case in practice, then $r_{j,0}=
r_{i,0}$. Finally, if $I(x)$ is invertible, which must be the case to solve
the problem, we have $\nu_{i,0}=K(r_{i,0})=K(r_{j,0})=\nu_{j,0}$. So all
optimal values $\nu_{i,0}$ are the same. Using the same reasoning, we can
draw the same conclusion for other auxiliary fields.

Under these conditions, we can set $\mu_i=\mu$, $\nu_i=\nu$, and $\rho_{ij}=
\rho$ in the expression of the eigenenergies. We are thus led to replace the
original Hamiltonian (\ref{hamgene2}) by the following simpler Hamiltonian,
which now depends on only 3 auxiliary fields (remember that $P(x)=S(x)=x^2$)
\begin{eqnarray}\label{hamgene3}
H(\mu,\nu,\rho)&=&\frac{N}{2}\left[\mu+\frac{m^2}{\mu}\right]+N
\left[U(I(\nu))-\nu\, I(\nu)^2\right]+C_N
\left[V(J(\rho))-\rho\, J(\rho)^2\right]\nonumber\\
  &&+\sum^{N}_{i=1}\frac{\bm p^2_i}{2\mu}+\nu\sum^{N}_{i=1}r_i^2+
\rho\sum^{N}_{i<j=1} r_{ij}^2,
\end{eqnarray}
which, by virtue of~(\ref{ehonr}), has the following mass spectrum ($\bm P =
\bm 0$)
\begin{eqnarray}\label{mass}
M(\mu,\nu,\rho)&=&\frac{N}{2}\left[\mu+\frac{m^2}{\mu}\right]+N
\left[U(I(\nu))-\nu\, I(\nu)^2\right]+C_N
\left[V(J(\rho))- \rho\, J(\rho)^2\right]\nonumber\\
	&&+\sqrt{\frac{2}{\mu}(\nu+N\rho)}\ Q \\
\textrm{with} \quad Q &= &\sum^{N-1}_{i=1} (2n_i+l_i) + \frac{3}{2}(N - 1), \label{QNbody}
\end{eqnarray}
which is characterized by a high degeneracy due to the form of the principal
quantum number $Q$. 

The last step needed to get the final mass formula is to find the optimal values
$\mu_0$, $\nu_0$, and $\rho_0$ from the extremization conditions
\begin{equation}\label{minimass}
\left.\partial_\mu M(\mu,\nu,\rho)\right|_{\mu=\mu_0}=0,\ 	\left.
\partial_\nu M(\mu,\nu,\rho)\right|_{\nu=\nu_0}=0,\ 	\left.\partial_{\rho}
M(\mu,\nu,\rho)\right|_{\rho=\rho_0}=0.
\end{equation}
Writing explicitly (\ref{minimass}) and after some algebra, we arrive at
the mass formula
\begin{equation}\label{massfin1}
M(\mu_0,\nu_0,\rho_0)=N\mu_0+N\, U(I(\nu_0))+C_N V(J(\rho_0)),
\end{equation}
where the auxiliary fields are solutions of 
\begin{eqnarray}\label{mudef}
\mu_0&=&\frac{m^2}{\mu_0}+\left[\frac{2Q^2(\nu_0+N\rho_0)}{\mu_0
N^2}\right]^{1/2}, \\
\label{nubdef}
I(\nu_0)&=&\left[\frac{Q^2}{2N^2\mu_0(\nu_0+N\rho_0)}
\right]^{1/4}, \\
\label{mubdef}
J(\rho_0)&=&\left[\frac{2Q^2}{(N-1)^2\mu_0(\nu_0+N\rho_0)}
\right]^{1/4}.
\end{eqnarray}
At this stage, functions $I$ and $J$ are not known. They depend on the specific
forms of $U$ and $V$. Optimal values of $\mu_0$, $\nu_0$, and $\rho_0$ are not
known anymore. Moreover, they also depend on the state considered through the
variable $Q$.
Nevertheless, formula~(\ref{massfin1}) makes clearly appear the mean-field nature
of the AFM, because $\mu_0$ can be interpreted as the average kinetic energy
of one particle, while $U(I(\nu_0))$ and $V(J(\rho_0))$ can be respectively seen
as the average potential energy of one particle in the potential $U(r)$ and of
a pair in the potential $V(r)$. 

A general remark about the AFM can now be done: The analytical results
obtained for a given potential $U(r)$ approximated by a potential $P(r)$ can be
used as a starting point for finding analytical results for an other potential
$W(r)$, approximated this time by $U(r)$. The same procedure can be applied for
the potential $V(r)$. This considerably enlarges the
domain of applicability of this method. 

Notice that the idea of rewriting a $N$-body Hamiltonian with pairwise
interactions of the form $g(r_{ij}^2)$ as a $N$-body harmonic oscillator
has already been investigated in Ref.~\citen{hall3} within the framework of the
envelope theory. That method shares many similarities with the auxiliary
field method, as shown in Appendix~\ref{sec:et}, but the results presented
hereafter have, to our knowledge, be presented for the first time in
Ref.~\citen{silv10}. 

\subsubsection{Simplified formulations}

The four equations~(\ref{massfin1})--(\ref{mubdef}) give the AFM approximation
for the mass spectrum of a $N$-body system where the constituent
particles are identical. They can be even simplified by defining
\begin{equation}\label{defX0}
X_0=\sqrt{2\mu_0(\nu_0+N\rho_0)}.
\end{equation}
We have indeed
\begin{eqnarray}\label{massfins}
M(X_0)&=&N\sqrt{m^2+\frac{Q}{N}X_0}+N\, U\left(\sqrt{\frac{Q}{NX_0}}\right)
+ C_N V\left(\sqrt{\frac{2Q}{(N-1)X_0}}\right),
\end{eqnarray}
where the remaining auxiliary field is a solution of
\begin{equation}\label{x0def}  
X_0^2=2\sqrt{m^2+\frac{Q}{N}X_0} \left[K\left(\sqrt{\frac{Q}
{NX_0}}\right)+N\, L\left(\sqrt{\frac{2Q}{(N-1)X_0}}
\right)\right],
\end{equation}
and where $K(x)=U'(x)/(2 x)$ and $L(x)=V'(x)/(2 x)$ (see (\ref{idef})). This condition ensures
that $M(X_0)$ is extremal. In so doing, we have simplified a lot the original
formulation. The set~(\ref{mudef})-(\ref{mubdef}) is a complicated system
of non linear coupled equations, while the equivalent formulation~(\ref{x0def})
is just a transcendental equation which is \emph{a priori} easier to be solved.
As we shall see in the next sections, this task can be
achieved in several cases of interest. It is worth mentioning that the
AFM results~(\ref{massfins}) and (\ref{x0def}) are valid not only for the
ground state but also for excited states. 

For particles with null mass (ultrarelativistic limit), one obtains even
simpler formulae by simply setting $m=0$ in formulae~(\ref{massfins}) and
(\ref{x0def}).
In the nonrelativistic limit, the auxiliary field $\mu$ tends towards $m$.
In this case also the various formulae look simpler but cautions must be
taken in the limit. Explicitly, (\ref{massfins}) reduces to
\begin{eqnarray}\label{massnrfins}
M(X_0)&=&m_t+\frac{Q}{2m}X_0 
+ N\, U\left(\sqrt{\frac{Q}{NX_0}}\right)
+C_N V\left(\sqrt{\frac{2Q}{(N-1)X_0}}\right),
\end{eqnarray}
with $m_t=N m$ and where $X_0$ is the solution of
\begin{equation}\label{x0nrdef}
X_0^2=2 \, m \left[K\left(\sqrt{\frac{Q}{NX_0}}\right)+N
\, L\left(\sqrt{\frac{2Q}{(N-1)X_0}}\right)\right].
\end{equation}

Let us introduce the distance $r_0=\sqrt{N Q/X_0}$ and the momentum $p_0=Q/r_0$, and let us define $T$ by $T(\bm x)= \sqrt{\bm x^2+m^2}$ or by its nonrelativistic counterpart $m+\frac{\bm x^2}{2m}$. It is a simple algebra exercise to show that formulae~(\ref{massfins})-(\ref{x0def}) can be written as \cite{sema11b}
\begin{align}
\label{M0}
M_0&=N\, T(p_0) + N\, U \left( \frac{r_0}{N} \right) + C_N\, V \left( \frac{r_0}{\sqrt{C_N}} \right), \\
\label{pQr}
p_0&=\frac{Q}{r_0}, \\
\label{p0r0}
N\, p_0 T'(p_0) &=  N\, \frac{r_0}{N}U' \left( \frac{r_0}{N} \right) + C_N\, \frac{r_0}{\sqrt{C_N}}V' \left( \frac{r_0}{\sqrt{C_N}} \right).
\end{align}
With this formulation, an AFM eigenvalue given by (\ref{M0}) is simply the kinetic operator evaluated at the mean momentum $p_0$ plus the potential energy computed at some mean radius depending on $r_0$. This generalizes the relations found previously for two-body systems. As one could expect, the kinetic energy and the one-body potential energy are proportional to the number of particles and the two-body potential energy is proportional to the number of pairs. Formula (\ref{M0}) looks like a semiclassical approximation but this is absolutely not the case. The AFM yields an approximate $N$-body wavefunction \cite{silv10,barlnc,barlnc2}, and the relation (\ref{pQr}) between $p_0$ and $r_0$ is a full quantum link, function of the quantum numbers of the system. Lastly, the value of $r_0$ (and thus of $p_0$) is the solution of a transcendental equation (\ref{p0r0}) which is the translation into the AFM variables of the generalized virial theorem \cite{luch90} which comes from very general properties of quantum mechanics. These considerations prove that the AFM really relies on very sound physical basis. Once the system (\ref{M0})-(\ref{p0r0}) is written, it can appear finally quite natural to obtain such a result. The problem is to find a relevant link between the mean values $r_0$ and $p_0$. This is solved by the AFM. 

The eigenstates of $H(\mu_0,\nu_0,\rho_0)$ are built with
harmonic oscillator states. 
A $N$-body state with a mass $M(X_0)$ is written
\begin{equation}\label{piphi}
\psi=\prod^{N-1}_{j=1}\, \psi_{n_j l_j}(\lambda_j, \bm x_j),
\end{equation}
where $\psi_{n_j l_j}(\lambda_j, \bm x_j)$ is a three-dimensional harmonic oscillator wavefunction (see Appendix~\ref{sec:obs_OH}), depending on the Jacobi coordinate $\bm x_j$ (see Appendix~\ref{sec:Jacobcoord}) and decreasing asymptotically like ${\rm e}^{-\lambda^2_j\, \bm x_j^{\, 2}/2}$ (the magnetic quantum number is omitted). The quantum numbers are such that $\sum_{j=1}^{N-1}(2n_j+l_j)=B=Q-\frac{3}{2}(N-1)$ and the scale parameters $\lambda_j$ are given by
\begin{equation}\label{laj}
\lambda_j=\sqrt{\frac{j}{j+1}X_0}=\sqrt{\frac{j}{j+1}N Q}\frac{1}{r_0}=\sqrt{\frac{j}{j+1}\frac{N}{Q}}p_0 .
\end{equation}
The quantum number $B$ is generally called the band number.
The state~(\ref{piphi}) has neither a defined total angular momentum nor a good symmetry, but its is characterized by a parity $(-1)^B$. By combining states (\ref{piphi}) with the same value of $B$ (or $Q$), it is generally possible to build a physical state with good quantum numbers and good symmetry properties, but the task can be technically very complicated. \cite{silv85}\ Some observables can be easily computed:
\begin{align}
\label{p0}
\frac{1}{N}\left\langle \sum_{i=1}^N \bm p_i^2 \right\rangle &= p_0^2, \\
\label{r0}
N \left\langle \sum_{i=1}^N r_i^2 \right\rangle =
\left\langle \sum_{i<j=1}^N r_{ij}^2 \right\rangle &= r_0^2. 
\end{align}
Since these results only depend on the quantum numbers via $Q$, and since a physical state must be a combination of eigenstates with the same value of $Q$, formulae~(\ref{p0}) and (\ref{r0}) are also valid for a physical state.
This shows that $r_0$ can be considered as a mean radius for the system and $p_0$ as a mean momentum per particle. Indeed, (\ref{p0}) and (\ref{r0}) imply that
\begin{equation}
\label{meanvar}
\sqrt{\left\langle\bm p_i^2 \right\rangle}=p_0, \quad
\sqrt{\left\langle r_i^2 \right\rangle}=\frac{r_0}{N}, \quad
\sqrt{\left\langle r_{ij}^2 \right\rangle}=\frac{r_0}{\sqrt{C_N}},
\end{equation}
for arbitrary $i\ne j$ since the mean values are taken with completely symmetrized states. These results can also be obtained using the more general relations (\ref{muk0})-(\ref{Sr0}) relevant for $P(x)$ and $S(x)$ different from $x^2$. Identifying (\ref{massfin1}) and (\ref{M0}), it appears that
\begin{equation}
\mu_0=\sqrt{m^2+p_0^2}.
\end{equation}

In the above expression, $Q$ is the principal quantum number defined in (\ref{QNbody}).
However, it has been shown many times for the two-body problem that
a much better approximation of the exact energies can be obtained with a slight
modification of the principal quantum number. A particularly simple form which
seems to work quite well for $N$-body systems is given by
\begin{equation}
\label{eq:princnumod}
Q = \sum_{i=1}^{N-1} (\alpha\, n_i + \beta\, l_i) + \gamma(N-1).
\end{equation}
For instance, such a formula is tested numerically for baryons in 
Sect.~\ref{sec:baryoncase}; it improves substantially the results as
compared to the expression~(\ref{QNbody}).
It is worth noting that there is no systematic procedure to determine the values of
parameters $\alpha$, $\beta$ and $\gamma$, which depend on both the interaction and the
kinematics.

\subsection{Critical constants}
\label{sec:critic}    

Some interactions, as the exponential or the Yukawa potentials, admit only a finite number of bound states (see Sects.~\ref{sec:critconst} and \ref{sec:critconst2}). Let us assume that such an interaction can be written as $W(x)=-\kappa\, w(x)$, where $\kappa$ is a positive quantity which has the dimension of an energy and $w(x)$ a ``globally positive" dimensionless function such that $\lim_{x\to\infty} w(x) = 0$. The critical constant $\kappa(\{\theta\})$, where $\{\theta\}$ stands for a set of quantum numbers, is such that, if $\kappa > \kappa(\{\theta\})$, the potential admits a bound state with the quantum numbers $\{\theta\}$. The interaction energy for the state with quantum numbers $\{\theta\}$ is then just vanishing for $\kappa = \kappa(\{\theta\})$. \cite{brau1} 

Let us consider a nonrelativistic $N$-body system (no manageable calculation can be performed for a semirelativistic kinematics) with one-body potentials $U(x)=-k\, u(x)$ and two-body potentials $V(x)=-g\, v(x)$, both independent of the particle mass and both admitting only a finite number of bound states. The system (\ref{M0})-(\ref{p0r0}) for a vanishing energy gives
\begin{align}
\label{E0a}
N\, \frac{Q^2}{2\ m\, r_0^2} &= N\,k_N\, u \left( \frac{r_0}{N} \right) + C_N\,g_N\, v \left( \frac{r_0}{\sqrt{C_N}} \right), \\
\label{E0b}
N\, \frac{Q^2}{m\, r_0^2} &=-k_N\, r_0\, u' \left( \frac{r_0}{N} \right) - \sqrt{C_N}\,g_N\, r_0\, v' \left( \frac{r_0}{\sqrt{C_N}} \right),
\end{align}
where $k_N$ and $g_N$ are the critical constants for the system with $N$ particles. The elimination of the ratio $N\, Q^2/(m\, r_0^2)$ from both equations yields the equality
\begin{equation}
\label{E0ab}
2 N\,k_N\, u \left( \frac{r_0}{N} \right) + 2 C_N\,g_N\, v \left( \frac{r_0}{\sqrt{C_N}} \right)=
-k_N\, r_0\, u' \left( \frac{r_0}{N} \right) - \sqrt{C_N}\,g_N\, r_0\, v' \left( \frac{r_0}{\sqrt{C_N}} \right).
\end{equation}
When potentials $u$ and $v$ are both taken into account, nothing interesting can be said. So let us consider one type of potential at once. 

Assuming that only two-body forces are present, (\ref{E0ab}) reduces to \cite{sema11b}
\begin{equation}
\label{E0abg}
2\, \sqrt{C_N}\, v \left( \frac{r_0}{\sqrt{C_N}} \right) + r_0\, v' \left( \frac{r_0}{\sqrt{C_N}} \right) =0,
\end{equation}
where the parameter $g_N$ has disappeared. Introducing the new variable $y_0=r_0/\sqrt{C_N}$, we can rewrite 
(\ref{E0a}) and (\ref{E0b}) as
\begin{align}
\label{gQ}
&g_N  = \frac{1}{y_0^2\,v(y_0)} \frac{2}{N(N-1)^2} \frac{Q^2}{m}, \\
\label{y0gQ}
&2\, v(y_0)+y_0\, v'(y_0)=0.
\end{align}
The variable $y_0$, determined by (\ref{y0gQ}), is independent of $N$, $Q$ and $m$, and depends only on the form of the function $v(x)$. So, (\ref{gQ}) gives precise information about the dependence of the many-body critical constant $g_N$ as a function of all the characteristics of the system. With the system (\ref{gQ})-(\ref{y0gQ}), it is easy to recover previous AFM results obtained in Sects.~\ref{sec:critconst} and \ref{sec:critconst2}, and in Refs.~\citen{bsb09a,silv10}.

Within the AFM approximation, the ground state (GS) of a boson-like system is characterized by $Q=\frac{3}{2}(N-1)$. We obtain in this case the following very general relation valid, at the AFM approximation, for all pairwise potentials with a finite number of bound states
\begin{equation}
\label{gnnp1}
\frac{g_{N+1}(\textrm{GS})}{g_N(\textrm{GS})}=\frac{N}{N+1}.
\end{equation}
This ratio has previously been obtained and numerically checked for several exponential-type potentials \cite{rich94,mosz00}. Similarly, in the same general situation, 
\begin{equation}
\label{gn2}
g_N(\textrm{GS})=\frac{2}{N}g_2(\textrm{GS}),
\end{equation}
indicating that in order to bind a $N$-body system, a coupling $N/2$ times smaller than the coupling for a two-body problem is sufficient \cite{rich94,mosz00}.

Assuming that only one-body forces are present, a similar calculation gives  \cite{sema11b}
\begin{align}
\label{kQ}
& k_N = \frac{1}{y_0^2\,u(y_0)} \frac{1}{2 N^2} \frac{Q^2}{m},
\\
\label{y0kQ}
&2\, u(y_0)+y_0\, u'(y_0)=0,
\end{align}
where the change of variable $y_0=r_0/N$ has been used. Again, (\ref{kQ}) gives precise information about the dependence of the one-body critical constant $k_N$ as a function of all the characteristics of the system. 
These results are strongly different from those for pairwise forces. 

If the AFM gives upper (lower) bounds for the exact eigenvalues, the critical constants predicted by formulae above are upper (lower) bounds for the exact critical constants. In the favorable situation where the AFM gives both upper and lower bounds for the eigenvalues, it is possible to approximate from above and below the exact critical constants.

\subsection{Connection with the perturbation theory}
\label{sec:pert}

It has been shown in Ref.~\citen{bsb08b} that, for one- and two-body nonrelativistic systems, the AFM and the perturbation theory give similar results when the potential is an exactly solvable one plus a small perturbation. This result is extended here for the general Hamiltonian (\ref{hamgene}). \cite{sema11b} 

Let us first assume that each pairwise potential $V(r_{ij})$ is supplemented by a term $\epsilon \, v(r_{ij})$, with $\epsilon \ll 1$ in order that $\epsilon \, v(x) \ll V(x)$ in the physical domain of interest. In the system (\ref{M0})-(\ref{p0r0}), the potential $V(x)$ is replaced by $V(x)+\epsilon \, v(x)$. In this case, new values $r_1$  and $p_1$ for the mean radius and momentum will be the solution of the new system
\begin{align}
\label{M1}
M_1&=N\, T(p_1) + N\, U \left( \frac{r_1}{N} \right) + C_N \left[V \left( \frac{r_1}{\sqrt{C_N}} \right)+ \epsilon\, v \left( \frac{r_1}{\sqrt{C_N}} \right) \right], \\
\label{pQr1}
p_1\, r_1&=Q, \\
\label{p1r1}
N\, p_1 T'(p_1) &= r_1 U' \left( \frac{r_1}{N} \right) + \sqrt{C_N}\, r_1 \left[ V' \left( \frac{r_1}{\sqrt{C_N}} \right)+ \epsilon\,v' \left( \frac{r_1}{\sqrt{C_N}} \right) \right].
\end{align} 
Writing $r_1 = (1+\delta)r_0$, we can expect $\delta \ll 1$ since $\epsilon \ll 1$. In this case, power expansions at first order can be computed. We have $p_1 \approx (1-\delta)p_0$ from (\ref{pQr1}), and we can write $T(p_1)\approx T(p_0)-\delta\, p_0\, T'(p_0)$, $T'(p_1)\approx T'(p_0)-\delta\, p_0\, T''(p_0)$, $U(r_1/N)\approx U(r_0/N)+ \delta\, r_0\, U'(r_0/N)/N$, etc. Equation (\ref{p1r1}) reduces to an expression of the form $\delta \approx \epsilon\, h(r_0)$ where $h$ is a quite complicated function of $T'$, $U'$, $V'$ and their derivatives. It simply confirms that $\delta \sim \textrm{O} (\epsilon)$. It is then possible to perform an expansion of $M_1$ which reduces, using (\ref{M0}) and (\ref{p0r0}), simply to \cite{sema11b}
\begin{equation}
\label{M1expv}
M_1=M_0 
+ C_N\, \epsilon\, v \left( \frac{r_0}{\sqrt{C_N}} \right) + \textrm{O} (\epsilon^2).
\end{equation}
This result could seem quite obvious, but it demonstrates that the knowledge of $r_0$ is sufficient to obtain the contribution of the perturbation at the first order.

Let us now assume too that each [one-body potential $U(|\bm s_i|)$ / kinetic operator $T(|\bm p_i|)$] is supplemented by a term [$\eta \, u(|\bm s_i|)$ / $\tau \, t(|\bm p_i|)$], with [$\eta \ll 1$ / $\tau \ll 1$] in order that [$\eta \, u(x) \ll U(x)$ / $\tau \, t(x) \ll T(x)$] in the physical domain of interest. With similar calculations, we finally find
\begin{equation}
\label{M1tot}
M_1=M_0 
+ N\, \tau\, t \left( p_0 \right)
+ N\, \eta\, u \left( \frac{r_0}{N} \right) 
+ C_N\, \epsilon\, v \left( \frac{r_0}{\sqrt{C_N}} \right)
+ \textrm{O} (\tau^2,\eta^2,\epsilon^2).
\end{equation}
The parameter $\delta$ is determined at the same order by the following relation
\begin{eqnarray}
\lefteqn{N\, p_0\, \tau\, t' \left( p_0 \right)
- r_0\, \eta\, u' \left( \frac{r_0}{N} \right) 
- \sqrt{C_N}\, r_0\, \epsilon\, v' \left( \frac{r_0}{\sqrt{C_N}} \right)}  \nonumber\\
&=&\delta 
\left[ 2\, N\, p_0\, T' \left( p_0 \right) 
+ N\, p_0^2\, T'' \left( p_0 \right)
+ \frac{r_0^2}{N} U''\left( \frac{r_0}{N} \right)
+ r_0^2\, V'' \left( \frac{r_0}{\sqrt{C_N}} \right)\right].
\end{eqnarray}
Perturbed observables and wavefunctions can then be computed at first order, since $r_1 = (1+\delta)r_0$ and $p_1 = (1-\delta)p_0$ at this order.

The contribution of a perturbation at the first order can thus be very easily computed within the AFM once the unperturbed problem is solved. In order to check the quality of this approximation, let us consider a case in which the unperturbed Hamiltonian $H$ can be solved exactly by the AFM, that is $M_0$ is the exact solution. If the small perturbation potential is written $\epsilon \sum_{i<j=1}^N v(r_{ij})$, the quantum perturbation theory says that the solution $M_*$ is given by
\begin{equation}
\label{Mpert}
M_*=M_0 + C_N\, \epsilon\, \langle v(r_{ij}) \rangle + \textrm{O} (\epsilon^2),
\end{equation}
for any pair $(ij)$. The mean value is taken with a completely symmetrized eigenstate of the unperturbed Hamiltonian $H$. The comparison of (\ref{Mpert}) with (\ref{M1expv}) shows that $\langle v(r_{ij}) \rangle$ is replaced by $v \left( r_0/\sqrt{C_N} \right)$ within the AFM. This is to be compared with the exact relation $\langle S(r_{ij}) \rangle = S \left( r_0/\sqrt{C_N} \right)$ for the auxiliary potential (see~(\ref{Sr0})). So, the AFM does not give the same result as the perturbation theory. But the agreement can be very good, as shown with several examples calculated above. Similar discussions can be made for small one-body perturbation potentials or small perturbations of the kinematics. 

\subsection{Power-law potentials}
\label{sec:powlawpotNbod}

\subsubsection{General results}
\label{sec:genresplpot}

The first explicit example that will be considered below is the case of
power-law potentials, \textit{i.e.} the Hamiltonian
\begin{equation}
H_{{\rm pl}}=\sum^{N}_{i=1}\sqrt{\bm p^2_i+m^2}+a\ {\rm sgn}(\lambda)
\sum^{N}_{i=1} r^\lambda_i+b\ {\rm sgn}(\eta)\sum^{N}_{i<j=1}
r^\eta_{ij},
\end{equation}
where $\lambda$, $\eta\ge -1$ (in the nonrelativistic case, one can 
further consider $\lambda$, $\eta>-2$). When only a one-body or a two-body
interaction is present, parameters $a$ or $b$ must be positive. If both types
of potentials are present, there are less constraints on the sign provided that
a bound state can exist. Following the definitions (\ref{idef}), it is
readily computed that $K(x)=a\, |\lambda|\, x^{\lambda-2}/2$ and 
$L(x)=b\, |\eta|\, x^{\eta-2}/2$.
Then, by defining
\begin{equation}
A_\lambda=a|\lambda|\left(\frac{N}{Q}\right)^{\frac{2-\lambda}{2}},
\quad B_\eta=b|\eta|N\left(\frac{N-1}{2Q}\right)^{\frac{2-\eta}{2}},
\end{equation}
Equations~(\ref{massfins}) and (\ref{x0def}) can be recast under the form 
\begin{eqnarray}\label{masspl}
M(X_0)&=&N\sqrt{m^2+\frac{Q}{N}X_0}+Q \left(\frac{A_\lambda}
{\lambda} X_0^{-\lambda/2}+\frac{B_\eta}{\eta} X_0^{-\eta/2}\right), \\
\label{xodef}
X_0^2&=&\sqrt{m^2+\frac{Q}{N}X_0} \left[A_\lambda\ X_0^{\frac{2-\lambda}
{2}}+B_\eta\ X_0^{\frac{2-\eta}{2}}\right].
\end{eqnarray}
The sufficient condition to get a closed analytical formula for $M(X_0)$ is to
solve~(\ref{xodef}) analytically. This is possible if this last equation can be
rewritten as a polynomial equation of the fourth degree at most. If
$\lambda\neq\eta$, it can be computed that the following couples will lead to
such an analytically solvable equation
\begin{eqnarray}\label{etalam}
(\lambda,\eta)\ {\rm or}\ (\eta,\lambda)&=&(-\frac{1}{2},-1),\, (0,-1),
\, (1,-1),\, (-1,-2),\, (0,-2),\, (-\frac{1}{2},-\frac{3}{2})\, .
\end{eqnarray}
Note that the last three cases are only allowed with a nonrelativistic
kinematics. 

The problem actually becomes particularly simple when $\lambda=\eta$. In this
case, the following values lead to an analytical solution  
\begin{eqnarray}\label{lamval}
\lambda&=&-1,\, -\frac{2}{3},\, -\frac{1}{2},\, 0,\, 1,\, 2,\, -2,
\, -\frac{7}{4},\, -\frac{5}{3},\, -\frac{3}{2},\, -\frac{4}{3},\, -\frac{5}{4}.
\end{eqnarray} 
The last six values are only allowed in the case of a nonrelativistic kinematics.
The solution of~(\ref{xodef}) reads
\begin{equation}\label{x00}
X_0^{\lambda+2}=\left(A_\lambda+B_\lambda\right)^2
\left(m^2+\frac{Q}{N}X_0\right) ,
\end{equation}
and the mass formula~(\ref{masspl}) becomes
\begin{equation}\label{M00}
M(X_0)=\frac{N\lambda m^2+Q (\lambda+1)X_0}{\lambda\sqrt{m^2+\frac{Q}
{N}X_0}}.
\end{equation}
Among the values~(\ref{lamval}), three cases are of obvious physical interest:
$\lambda=-1$, 1, and 2. The case $\lambda=2$ corresponds to the harmonic
oscillator and is solved in Appendix~\ref{sec:nboh}. The Coulomb
problem, \textit{i.e.} $\lambda=-1$, will be specifically considered in
Sect.~\ref{sec:atom}, while we will discuss the case $\lambda=1$ in Sect.~\ref{sec:nbodylin}.
Notice that closed mass formulae for any $\lambda$ can be obtained in the
nonrelativistic and ultrarelativistic limits, as we also show in the following.

In the case $m=0$, (\ref{x00}) allows us to extract the $X_0$ value for
arbitrary power
\begin{equation}
\label{plpotx0}
X_0 = \left(\frac{Q}{N} (A_\lambda + B_\lambda)^2\right)^{\frac{1}
{\lambda+1}} .
\end{equation}
Inserting this value in (\ref{M00}) provides us with the expression of the
energy in the ultrarelativistic limit
\begin{equation}
\label{plpoten0}
M(X_0)=\frac{\lambda+1}{\lambda} \left[Q^{\lambda+2} N^{\lambda}
(A_\lambda + B_\lambda)^2\right]^{\frac{1}{2(\lambda+1)}}.
\end{equation}
We already know from Sect.~\ref{sec:salpcoul}, that a bound state of massless particles exists only
if $\lambda >0$.
Setting $N=2$ and $a=0$, this equation reduces to (\ref{ezerom}) with $\sigma=2$ 
(and $a$ renamed $b$). 

On the contrary, assuming a large mass, $m_t \approx M(X_0)$, it is possible
to extract $X_0$ from (\ref{x00})
\begin{equation}
\label{plpotxnr}
X_0=[m (A_\lambda + B_\lambda)]^{\frac{2}{\lambda+2}}.
\end{equation}
Except the Coulomb case ($\lambda=-1$) which will be treated subsequently, the
term $m^2$ is dominant with respect to $X_0$. Expanding the expression of the
energy (\ref{M00}) at lower order in $1/m$ leads to the nonrelativistic value
of the energy
\begin{equation}
\label{plpotennr}
M(X_0)=m_t+\frac{\lambda+2}{2 \lambda} Q \left[\frac{(A_\lambda + B_\lambda)^2}
{m^\lambda}\right]^{\frac{1}{\lambda+2}}.
\end{equation}
Setting $N=2$ and $a=0$, this equation reduces to (\ref{eq:eigenerplpot}) with the
reduced mass $m$ replaced by $m/2$ (and $a$ renamed $b$). 

\subsubsection{The linear potential}
\label{sec:nbodylin}

The case $\lambda=1$ corresponds to a linearly rising confining potential. It is
of great interest in hadronic physics since a linearly rising potential appears to
be the best way of modeling the QCD confining interaction within potential
approaches, see for example Ref.~\citen{qcd2} for more details. For the
three-body problem, the one-body potential (term in $a$) corresponds to the
so-called Y-junction, while the two-body potential (term in $b$) corresponds to
the so-called $\Delta$ approximation. For physical problems, one has the choice
of retaining one approximation or the other or both. Here for sake of generality
we present the general case including both. In the real QCD world, one
must take $N=3$. But a general $N$ value is interesting to study 
alternative approaches of QCD. \cite{barlnc,barlnc2}

For $\lambda=1$, one can write
\begin{equation}
\label{abc}
A_1 + B_1 = \sqrt{\frac{N}{Q}} c,\quad {\rm with}\quad
c=a + b \sqrt{C_N}.
\end{equation}
The solution of (\ref{x00}) when $\lambda=1$ is given by 
\begin{equation}
X_0= \frac{c}{\sqrt{3}} F_-(Y),
\end{equation}
where $F_-(Y)$ is given by (\ref{eq:rootcubeq})
and where
\begin{equation}
Y=\frac{3^{3/2}Nm^2}{2 Q c}.
\end{equation}
Introducing this value in the expression of the energy (\ref{M00}), a simple
calculation allows us to obtain the mass under the form
\begin{equation}
\label{energNlin}
M(X_0)= N m \sqrt{\frac{F_-(Y)}{2Y}}\left[F_-(Y)+\frac{3}{F_-(Y)}\right].
\end{equation}

In the ultrarelativistic limit,
where $m=0$, one obtains, from (\ref{plpoten0}), the very simple relation
\begin{equation}\label{mlin}
M(X_0)^2=4N  c\, Q.
\end{equation}
Such a linear behavior of the square mass versus the principal quantum number
is a well-known fact in hadronic physics, where light mesons and baryons are
known from experimental data to exhibit Regge trajectories.

In the nonrelativistic limit, the mass formula~(\ref{plpotennr}) can be recast
under the form
\begin{equation}
M(X_0) = m_t + \frac{3}{2} \left(\frac{N Q^2 c^2}{m}\right)^{1/3}.
\end{equation}
The second term is the ``binding energy", which is a positive quantity in case
of a positive linear potential.

In the special case of a linear potential, there is a peculiar relationship
between the mass of the two-body problem and the mass of the $N$-body
problem. More precisely, let us call $M^{(2)}(\sigma,m,b,Q_2)$ the AFM
eigenvalues of the Hamiltonian
\begin{equation}
H^{(2)} = \sigma \sqrt{\bm{p}^2+m^2}+br
\end{equation}
calculated with the natural quantum number $Q_2=2n+l+3/2$. Then, using
(\ref{mlin}) and its counterpart for $N=2$ with the value of the
quantum number $Q$ defined by (\ref{QNbody}), the mass $M^{(N)}
(N,m,a,b,Q)$ of the original Hamiltonian is given by
\begin{equation}
\label{lienlinN2}
M^{(N)}(N,m,a,b,Q) = M^{(2)}\left(N,m,a + b \sqrt{C_N},Q\right).
\end{equation}

\subsubsection{Baryonic case}
\label{sec:baryoncase}

In this section, we want to discuss a special situation which can have an
immediate application for hadronic systems, especially baryons or glueballs. \cite{barlnc,barlnc2}\
Let us consider a one-body potential with a linear shape and a two-body
potential of Coulomb type ($\lambda=1$ and $\eta=-1$). We noticed that an
analytical solution does exist, but one needs to solve a general polynomial
of degree 3 and the corresponding solution is quite involved. We prefer to
give here the solution in the ultrarelativistic limit ($m=0$). In this
special case the solution of (\ref{xodef}) is quite simple
\begin{equation}
\label{x0had}
X_0 = \frac{a}{1-b \frac{N-1}{2Q}\sqrt{C_N}},
\end{equation}
and (\ref{masspl}) can be simplified as follows
\begin{equation}
\label{m0had}
M(X_0)=2 \sqrt{a} \sqrt{Q N - b C_N^{3/2}}.
\end{equation}
Since the argument of the square root must be positive, it is readily checked
that, either there exists a maximal allowed number of particles for a fixed
value of $b$, either there exists a maximal allowed value of $b$ for a fixed
number of particles. This kind of mass formula has been 
applied to the computation of light baryon masses for various theories of QCD with 
a large number of colors. \cite{barlnc,barlnc2}

In order to test the relevance of our method, we apply it for a particular
relativistic three-body system: the light baryon composed of three massless
quarks. In the framework of constituent models, the Hamiltonian for such a
system is given by \cite{qcd1a,qcd1b}
\begin{equation}\label{hbaryon}
H^\textrm{B}=\sum^3_{i=1}\sqrt{\bm p^2_i}+\sigma_s \sum^3_{i=1} |\bm r_i-\bm R|
-\frac{2}{3}\alpha_S\sum^3_{i<j=1}\frac{1}{|\bm r_i-\bm r_j|}.
\end{equation}
The dominant interaction is a confinement by three strings with density
$\sigma_s$ meeting at the center of mass. The short-range part is given by
pairwise interactions of Coulomb nature with a strong coupling constant
$\alpha_S$. In order to allow a more realistic calculation of the correct
baryon masses, this Hamiltonian must be completed with terms that can be
computed in perturbation. \cite{qcd1a}\ Since these contributions represent no
interest for the AFM, they are not considered in this study. 

Let us denote $\tilde H^\textrm{B}$ the AFM counterpart of $H^\textrm{B}$.
Approximations of eigenmasses of the genuine Hamiltonian are given by the application
of (\ref{m0had}) with $N=3$, $a=\sigma_s$ and $b=2\alpha_S/3$, 
\begin{equation}
\label{m0bar}
M^\textrm{AFB}_0 = \sqrt{12\sigma_s \left(B+3-\frac{2 \alpha_S}{\sqrt{3}}\right)},
\end{equation}
where $B=2(n_1+n_2)+(l_1+l_2)$ is the band number (see (\ref{QNbody})). 
One can notice the strong degeneracy due to the particular
form of $B$. This formula generalizes a result obtained for a small value of
$\alpha_S$ in Ref.~\citen{qcd1a}. Following above considerations, these masses
are upper bounds on the exact masses.

In order to test the accuracy of this formula, it is necessary to compute
accurately eigenvalues of $H^\textrm{B}$. It is possible, for instance, to use
a variational method relying on the expansion of trial states with a harmonic
oscillator basis. \cite{hobasis}\ We can write
\begin{equation}
\label{hotrial}
|\psi\rangle=\sum_{B=0}^{B_\textrm{max}}\sum_{q(B)} |\phi(B,q(B))\rangle,
\end{equation}
where $B$ characterizes the band number of the basis state and where $q$
summarizes all the quantum numbers of the state (which can depend on $B$). This
procedure is specially interesting since the eigenstates of $H^\textrm{B}$ are
expanded in terms of eigenstates of $\tilde H^\textrm{B}$ (up to a length scale
factor). In practice, a relative accuracy better than $10^{-4}$ is reached with
$B_\textrm{max}=20$. Such results are denoted ``exact" in the following.

Before comparing exact masses with formula~(\ref{m0bar}), it is significant to
describe the baryon wavefunctions. Quarks are fermions of spin $1/2$ with
isospin and color degrees of freedom. The global color function is unique and
completely antisymmetrical. For total spin $S=3/2$ (isospin $T=3/2$) the spin
(isospin) wavefunction is completely symmetrical. The corresponding wave
function are of mixed symmetry for $S=1/2$ or $T=1/2$. Using degenerate
eigenstates of $\tilde H^\textrm{B}$, it is always possible to build three-quark
states completely antisymmetrical \cite{silv85,hobasis}, which are characterized
by the same mass. For instance, when $S=T=3/2$ or $S=T=1/2$, the baryon states
possess a spatial wavefunction completely symmetrical. In the real world, the
degeneracies are removed by spin- or isospin-dependent operators which can be
computed as perturbations. \cite{qcd1a}\ When an eigenstate of $H^\textrm{B}$ is
computed for given values of $S$, $T$, and the total angular momentum $L$, we
determine the band number of its dominant components. It can then be compared
with the eigenstate of $\tilde H^\textrm{B}$ with the same spin-isospin quantum
numbers and the same band number. 

In Table~\ref{tab:compbar}, exact masses are compared with the predictions
of formula~(\ref{m0bar}) for usual values of the parameters $\sigma_s$ and
$\alpha_S$. The relative error can be large but the quality of the
approximation is quite reasonable for a so simple formula. 
Due to the strong degeneracy of the harmonic oscillator,
all AFM eigenstates with the same band number have the same
energy. This approximation is better for high values of $L$. Let us note
that the relative error can be reduced by a factor around 2 when
$\alpha_S\to 0$. When $\alpha_S= 0$, the relative error becomes independent
of the energy scale factor $\sqrt{\sigma_s}$.

\begin{table}[ht]
\caption{Exact eigenmasses of the
Hamiltonian~(\ref{hbaryon}) as a function of the band number $B$ and the total
angular momentum $L$ for $\sigma_s=0.2$~GeV$^2$ and $\alpha_S=0.4$. The number
in brackets is the probability (\%) of the component with the band number $B$
in the harmonic oscillator expansion~(\ref{hotrial}). These results are
compared with the masses given by formulae~(\ref{m0bar}), (\ref{m1bar}), and
(\ref{m2bar}). The number in parenthesis is the relative error (\%) with
respect to the exact value. All masses are given in GeV.}
\label{tab:compbar} 
\begin{center}
\begin{tabular}{cccccc}
\hline\hline\noalign{\smallskip}
$B$   & $L$ & Exact & $M^\textrm{AFB}_0$~(\ref{m0bar}) &
$M^\textrm{AFB}_1$~(\ref{m1bar}) & $M^\textrm{AFB}_2$~(\ref{m2bar}) \\ [3pt]
\hline
0 & 0 & 2.128 [92.9] & 2.468 (16.0) & 2.168 (\phantom{0}1.9) & 2.168 (1.9) \\
\noalign{\smallskip}
1 & 1 & 2.606 [95.9] & 2.914 (11.8) & 2.596 (\phantom{0}0.4) & 2.596 (0.4) \\
\noalign{\smallskip}
2 & 0 & 2.739 [89.4] & 3.300 (20.5) & 2.962 (\phantom{0}8.1) & 2.811 (2.6) \\
  & 2 & 2.959 [96.0] & 3.300 (11.5) & 2.962 (\phantom{0}0.1) & 2.962 (0.1) \\
\noalign{\smallskip}
3 & 1 & 3.125 [91.7] & 3.646 (16.7) & 3.288 (\phantom{0}5.2) & 3.152 (0.9) \\
  & 3 & 3.299 [96.7] & 3.646 (10.5) & 3.288 (\phantom{0}0.3) & 3.288 (0.3) \\
\noalign{\smallskip}
4 & 0 & 3.260 [80.8] & 3.961 (21.5) & 3.585           (10.0) & 3.332 (2.2) \\
  & 2 & 3.422 [92.3] & 3.961 (15.8) & 3.585 (\phantom{0}4.7) & 3.460 (1.1) \\
  & 4 & 3.581 [96.8] & 3.961 (10.6) & 3.585 (\phantom{0}0.1) & 3.585 (0.1) \\
\noalign{\smallskip}
5 & 1 & 3.584 [86.3] & 4.253 (18.7) & 3.858 (\phantom{0}7.7) & 3.625 (1.1) \\
  & 3 & 3.716 [93.6] & 4.253 (14.5) & 3.858 (\phantom{0}3.8) & 3.743 (0.7) \\
  & 5 & 3.861 [97.0] & 4.253 (10.2) & 3.858 (\phantom{0}0.1) & 3.858 (0.1) \\
\noalign{\smallskip}
6 & 0 & 3.721 [74.4] & 4.527 (21.7) & 4.114           (10.6) & 3.782 (1.6) \\
  & 2 & 3.838 [86.4] & 4.527 (17.9) & 4.114 (\phantom{0}7.2) & 3.895 (1.5) \\
  & 4 & 3.966 [93.6] & 4.527 (14.1) & 4.114 (\phantom{0}3.7) & 4.006 (1.0) \\ 
  & 6 & 4.103 [96.9] & 4.527 (10.3) & 4.114 (\phantom{0}0.3) & 4.114 (0.3) \\ 
\noalign{\smallskip}\hline
\end{tabular}
\end{center}
\end{table}

We have shown many times that it is possible to improve two-body mass formulae
by changing the structure of the principal quantum numbers $Q$. By fitting another
form on exact eigenvalues, a very high accuracy can sometimes be reached. Here,
we will proceed differently and will try to use analytical results to find
the best shape for the mass formula. Using a good upper bound on the ground state given
by the trial state of (\ref{hotrial}) at $B_\textrm{max}=0$, 
a modified AFM formula was proposed
\begin{equation}
\label{m1bar}
M^\textrm{AFB}_1 = \sqrt{\frac{32}{\pi}\sigma_s\left( B+3-\sqrt{3}
\alpha_S\right)}.
\end{equation}
One can see in Table~\ref{tab:compbar} that the masses are greatly improved, but
the variational character of the formula~(\ref{m1bar}) cannot longer be
guaranteed: some masses are now below the exact ones. At last, using 
information coming from the WKB method \cite{brau00}, the problem of
the strong degeneracy has been improved by a new formula
\begin{equation}
\label{m2bar}
M^\textrm{AFB}_2 = \sqrt{\frac{32}{\pi}\sigma_s\left( B'+3-\sqrt{3}
\alpha_S\right)}
\quad \textrm{with} \quad B'=\frac{\pi}{2}(n_1+n_2)+(l_1+l_2).
\end{equation}
Details about these procedures are given in Ref.~\citen{silv10}. One
can see in Table~\ref{tab:compbar}, that the relative error is now
around 1\% and generally below. Despite its simplicity and its nonvariational
character, (\ref{m2bar}) is then a very good mass formula for the
eigenstates of the Hamiltonian~(\ref{hbaryon}). It is not sure that the
procedure used here to improve the mass formula for baryons could work so
well for other Hamiltonians. But, this shows that an improvement is possible,
at least in some particular cases.

\subsubsection{Atomic systems}
\label{sec:atom}

We turn now our attention to atom-like systems, \textit{i.e.} systems
described by the following Hamiltonian 
\begin{equation}\label{hat}
H_{{\rm at}}=\sum^{N}_{i=1}\sqrt{\bm p^2_i+m^2}-\alpha\sum^{N}_{i=1}
\frac{1}{r_i}+\bar\alpha \sum^{N}_{i<j=1}\frac{1}{r_{ij}}.
\end{equation}
For this section, we remain completely general, the only hypothesis being
that the parameters $\alpha$ and $\bar \alpha$ should be real positive numbers.
The obvious application could correspond in a first approximation to $N$
identical electrons, with charge $e$, feeling the attraction of a central
static source (the nucleus) and their own repulsion. In this particular situation,
one must make the identification $\alpha = N e^2$ and $\bar \alpha = e^2$.
At this stage, we point out that our formalism is spin-independent and thus that
this model cannot be directly applied to a real physical atom. Moreover, one
should be very careful in applying the formulae already given. Two cautions are
in order:
\begin{itemize}
\item In (\ref{hat}), $r_i$ must be the distance between the particle $i$
and the center of mass of the system of $N$ electrons. In the atomic
interpretation, this center of mass must coincide with the nucleus. This is not
always the case but this prescription should be valid in case of a spherical atom.
\item Electrons are fermions and the total wavefunction (space and spin) must
be completely antisymmetrical. Since the spatial wavefunction cannot be
completely symmetrical, all possible values are not allowed for $Q$ given
by (\ref{honriden}). For closed shell atoms, $Q_{\textrm{AGS}}$
(\ref{honrfond2}) should be used to estimate the ground-state energy only.
\end{itemize}

Nevertheless, solving Hamiltonian~(\ref{hat}) has an intrinsic interest since,
to our knowledge, no corresponding analytical mass formula is known so far.
A general solution can moreover be found with the relativistic kinematics,
starting from~(\ref{x00}) and (\ref{M00}). Applying them to the case
$\lambda = -1$, $a = \alpha$, $b = - \bar \alpha$ and defining
\begin{equation}
\label{defDat}
D=\frac{1}{Q} \left[\alpha N - \frac{\bar \alpha}{N}
C_N^{3/2}\right],
\end{equation}
the solution of equation~(\ref{x00}) is given by
\begin{equation}
X_0=\frac{N m^2 D^2}{Q(1-D^2)}.
\end{equation}
The mass formula~(\ref{M00}) then reads
\begin{equation}\label{matsr}
M(X_0) = m_t \sqrt{1 - D^2}.
\end{equation}

In the ultrarelativistic limit, this expression gives the value $M = 0$.
This property is well known in the two-particle system and prolongates to the
$N$-body problem. It is quite obvious that a system of massless particles cannot be characterized
by a (negative) binding energy since the resulting mass would be negative.
Moreover, with a Coulomb-type potential, the only energy scale of the problem
is the particle mass. So, if this mass vanishes, no bound state can exist. In
the nonrelativistic regime, one must assume that the $D$ quantity is small in
(\ref{matsr}). In this case
\begin{equation}
M^{{\rm nr}}= m_t - \frac{1}{2} m_t D^2.
\end{equation}

It is clear from formula~(\ref{matsr}) that the value of the parameters cannot
be arbitrary. One must satisfy the condition $D < 1$. Explicitly this means
\begin{equation}
\label{condD}
\alpha N^2 - \bar \alpha C_N^{3/2}
< Q N.
\end{equation}
Let us emphasize that this condition is independent of the mass of the
particle. \cite{lieb88}\ Let us notice that, for $\alpha = N e^2$ and
$\bar \alpha = e^2$, a maximal value of $N$ exists if $Q$ increases
less rapidly that $N^2$.

\subsection{Duality relations}
\label{sec:dualrel}

\subsubsection{Generalities}
\label{sec:gendual}

In this section, we will show that AFM is a very powerful method to give
relationships between the energies of states (excited or ground states) for 
two different systems. We call the relations between the energies of both
systems ``duality relations''. The case for which one of the two systems
contains 2 particles is specially interesting
since the corresponding eigenenergies are rather easy to obtain. These relations
are \textbf{exact} if we consider the energies obtained in the AFM; they are no
longer exact for the true eigenstates, but if AFM gives a good precision on the
true states, we hope that the duality relations are good approximations of the
physical situations and gives, at least, general trends for the exact
levels. A rather complete treatment of duality relations can be found in
Ref.~\citen{bsb11}. We present here a simplified version where we restrict our
discussion to systems of $N$ identical particles interacting only through
a two-body potential $V(r)$. Readers interested in more specific details may
have a glance to Ref.~\citen{bsb11} for further informations.

Our original Hamiltonian is thus
\begin{equation}
\label{eq:orhamil}
H = \sum_{i=1}^N \sqrt{\bm p_i^2 + m^2} +  \sum_{i<j = 1}^N V(|\bm r_i - \bm r_j |).
\end{equation}
In this expression $\bm r_i$ is the position operator for particle $i$, $\bm p_i$ its
conjugate momentum and $m$ the common mass of all particles.
The AFM mass of the system is given by (\ref{massfins}) which is written now as
\begin{eqnarray}\label{massfins2}
M(X_0)&=&N\sqrt{m^2+\frac{Q}{N}X_0}+ C_N V\left(\sqrt{\frac{2Q}{(N-1)X_0}}\right),
\end{eqnarray}
while the $X_0$ quantity is solution of the transcendental equation (\ref{x0def})
which turns out to be now
\begin{equation}\label{x0def2}  
X_0^2=2N\,\sqrt{m^2+\frac{Q}{N}X_0}\, L\left(\sqrt{\frac{2Q}{(N-1)X_0}}\right),
\end{equation}
and where $L(x)=V'(x)/(2 x)$ as usual (see (\ref{idef})).

In general, the potential $V(r)$ depends on physical parameters $\{ \tau_1,\tau_2,
\ldots, \tau_p \} \equiv \{ \tau \}$ and should be noted more precisely
$V(\{ \tau \};r)$. The other parameters of the problem are the number of
particles $N$ and the mass $m$ of the particles.  Lastly, we want to have a
description of the whole spectrum, so that the principal quantum number $Q$
also enters the game. In consequence, a complete notation for the eigenmasses
would be $M^{(N)}(\{ \tau \};m,Q)$.

Let us stress now a very important point: for realistic potentials, the
parameters $\{ \tau \}$ could depend on $N$ or/and $m$. We do not consider such
a particular behavior here. Thus we assume that \textbf{the parameters of the
potentials are independent of $N$ and $m$}. This means that the $m$ dependence
of $H$ is only through the kinetic energy and its $N$ dependence through the
numbers of terms in the summation. We suppose that the potentials are given
once for all and consider that their form and parameters do not vary for
all studied systems. Therefore we are finally interested only by the $N$, $m$,
$Q$ dependence of the eigenmasses and use the simplified notation $M^{(N)}(m,Q)$.
What we call a ``duality relation'' is just a relation between
$M^{(N)}(m,Q)$ and $M^{(p)}(m',Q')$.

\subsubsection{General case}
\label{sec:gencasedual}
Instead of the $X_0$ quantity, let us define the new variable
\begin{equation}
\label{eq:defr02}
s_0 = \sqrt{\frac{2 Q}{(N-1)X_0}}.
\end{equation}
Then, the transcendental equation (\ref{x0def2}) and the mass
expression (\ref{massfins2}) are replaced by the following ones
\begin{eqnarray}
\label{eq:trans4N}
\frac{2Q}{(N-1)\sqrt{C_N}} & = & s_0^2 \sqrt{1+C_N (m\, s_0/Q)^2}\;V'(s_0), \\
\label{eq:mass4N}
M^{(N)}(m,Q) & = & C_N \left[ \frac{2Q}{(N-1)\sqrt{C_N}} \frac{1}{s_0}\sqrt{1+C_N (m\,s_0/Q)^2}
                  + V(s_0) \right].
\end{eqnarray}

A little algebra on those equations allows to arrive at the following duality
relation between the $N$-body and the $p$-body systems with particles interacting
via the \textbf{same two-body interaction}
\begin{equation}
\label{eq:dualpN4}
M^{(N)}(m,Q) = \frac{C_N}{C_p} M^{(p)}\left( \frac{p-1}{N-1}m,
                \frac{p-1}{N-1}\sqrt{\frac{C_p}{C_N}}\;Q \right).
\end{equation}
In this case, the spectrum of the $N$-body system is the same as the $p$-body
system (with the same two-body potential) provided we consider different particle
masses in both situations and different excitation states.

\subsubsection{Ultrarelativistic limit}
\label{sec:ultralimdual}

The case of ultrarelativistic systems, characterized by a vanishing mass $m=0$,
presents some very specific and interesting features. The case of
systems composed of gluons or/and light quarks can be well represented in this
scheme. The Hamiltonian of the system is then
\begin{equation}
\label{eq:urhamil}
H = \sum_{i=1}^N \sqrt{\bm p_i^2} + \sum_{i<j = 1}^N V(|\bm r_i - \bm r_j |).
\end{equation}
Indeed, the formulation is simpler for this particular
situation. Putting the value $m=0$ in (\ref{massfins2}) and (\ref{x0def2}), one
gets a new set of equations.
The transcendental equation looks like
\begin{equation}
\label{eq:transur}
\sqrt{\frac{N}{Q}} X_0^{3/2} = 2 N L \left( \sqrt{\frac{2Q}{(N-1)X_0}} \right),
\end{equation}
while the corresponding AFM eigenmass is given by
\begin{equation}
\label{eq:urmass}
M_u(X_0)=\sqrt{N Q X_0} + C_N  V \left( \sqrt{\frac{2Q}{(N-1)X_0}} \right).
\end{equation}
The eigenmass, depending now only on $N$ and $Q$, will be noted
$M_u^{(N)}(Q)$ (the index $u$ stands for ``ultrarelativistic''). 

With the same definition (\ref{eq:defr02}) for $s_0$,
the transcendental equation (\ref{eq:transur}) reduces to
\begin{equation}
\label{eq:transredu2}
\frac{2Q}{(N-1) \sqrt{C_N}} = s_0^2 V'(s_0).
\end{equation}
Defining the $D(x)$ and $F(x)$ function as in (\ref{eq:deffuncD}) and
(\ref{eq:deffuncF}) calculated with the potential $V(x)$, one has
\begin{eqnarray}
\label{eq:rofC2}
s_0 & = & D \left( \frac{2Q}{(N-1) \sqrt{C_N}} \right), \\
\label{eq:massF2}
M_u^{(N)}(Q) & = & C_N F \left( \frac{2Q}{(N-1) \sqrt{C_N}} \right).
\end{eqnarray}
As we already mentioned, the $D$ and $F$ functions are \textbf{universal}.
From expression (\ref{eq:massF2}), one deduces immediately the duality relation
\begin{equation}
\label{eq:dualuN2}
M_u^{(N)}(Q) = \frac{C_N}{C_p} M_u^{(p)} \left( \frac{p-1}{N-1}
\sqrt{\frac{C_p}{C_N}}Q\right),
\end{equation}
which is the special case of (\ref{eq:dualpN4}) with $m=0$.

\subsubsection{Nonrelativistic limit}
\label{sec:nrlimitdual}

Another interesting limit of the theory is the nonrelativistic one, valid when
the mass of the particles is large compared to the mean potential. In this case,
the considered Hamiltonian is simply
\begin{equation}
\label{eq:orhamilnr}
H = \sum_{i=1}^N \frac{\bm p_i^2}{2m} + \sum_{i<j = 1}^N V(|\bm r_i - \bm r_j |).
\end{equation}
Instead of dealing with the total mass $M^{(N)}$, it is better to consider the
binding energy obtained by removing the total rest mass: $E^{(N)} = M^{(N)} - Nm$.
The AFM approximation $E^{(N)}(m,Q)$ of the binding energy is given by the
following equations,
\begin{equation}
\label{eq:transnr}
X_0^2 = 2 m  N L \left( \sqrt{\frac{2Q}{(N-1)X_0}} \right)
\end{equation}
and
\begin{equation}
\label{eq:nrmass}
E^{(N)}(X_0)= \frac{QX_0}{2m}+ C_N V \left( \sqrt{\frac{2Q}{(N-1)X_0}} \right).
\end{equation}
We will see that, in this nonrelativistic limit, one obtains additional
interesting properties. 

Introducing $s_0$ by (\ref{eq:defr02}) as above, the transcendental equation
(\ref{eq:transnr}) reduces to
\begin{equation}
\label{eq:transredn2}
\frac{N Q^2}{m C_N^2} = s_0^3 V'(s_0).
\end{equation}
Defining the $\mathcal{D}(x)$ and $\mathcal{F}(x)$ function as in
(\ref{eq:defcalD}) and (\ref{eq:defcalF}) calculated with the potential $V(x)$,
one has
\begin{eqnarray}
\label{eq:rofD2}
s_0 & = & \mathcal{D}\left( \frac{N Q^2}{m C_N^2} \right), \\
\label{eq:EF2}
E^{(N)}(m,Q) & = & C_N \mathcal{F} \left( \frac{N Q^2}{m C_N^2} \right).
\end{eqnarray}
As the case studied in the previous section, the $\mathcal{D}$ and
$\mathcal{F}$ functions are \textbf{universal}. 
For a nonrelativistic system, an additional
property appears. Remarking that $E^{(N)}$ depends only on the ratio
$Q^2/m$, one has the general duality relation
\begin{equation}
\label{eq:dualgennr}
E^{(N)}(m,Q) = E^{(N)}(\beta^2 m, \beta Q),
\end{equation}
valid for any value of the real parameter $\beta$.

From expression (\ref{eq:EF2}), one deduces immediately the duality relation
\begin{equation}
\label{eq:dualnN3b}
E^{(N)}(m,Q) = \frac{C_N}{C_p} E^{(p)} \left( \frac{p-1}{N-1}m, \frac{p-1}{N-1}
\sqrt{\frac{C_p}{C_N}}Q\right),
\end{equation}
which is identical to the general case (\ref{eq:dualpN4}). One can
use the property (\ref{eq:dualgennr}) to obtain many other possibilities. For
example, choosing the value $\beta = \sqrt{p(N-1)/(N(p-1))}$, one has the
alternative simpler duality relation
\begin{equation}
\label{eq:dualnN4}
E^{(N)}(m,Q) = \frac{C_N}{C_p} E^{(p)} \left(\frac{p}{N} m, \frac{C_p}{C_N} Q \right).
\end{equation}
In this last relation, let us choose $\beta=\sqrt{N/p}$, one arrives at the relation
\begin{equation}
\label{eq:dualnN5}
E^{(N)}(m,Q) = \frac{C_N}{C_p} E^{(p)} \left( m, \frac{p-1}{N-1}\sqrt{\frac{p}{N}} Q \right).
\end{equation}
In this expression, we decide to keep the same mass for systems with $N$ and $p$ particles.
The duality relation leads to a link between different excited states of both systems.
Choosing the value $\beta=C_N/C_p$ in equation~(\ref{eq:dualnN4}), one obtains
the alternative expression
\begin{equation}
\label{eq:dualnN6}
E^{(N)}(m,Q) = \frac{C_N}{C_p} E^{(p)} \left( \frac{(N-1)C_N}{(p-1)C_p} m, Q \right).
\end{equation}
In this expression we decide to maintain a one to one correspondence in the spectrum
but for systems with different particle masses.

\subsubsection{Passing from nonrelativistic to ultrarelativistic limits}
\label{sec:passnrur}

We showed that both the ultrarelativistic limit and the nonrelativistic limit
for the eigenmasses share the property of being expressed in terms of universal
functions. The $F$ function for the ultrarelativistic case and the $\mathcal{F}$
function for the nonrelativistic case are independent of the system and of the
excitation quantum numbers, but depends only on the form of the potential under
consideration.

If the same form of potential is used in both situations, the expressions of the
$F$ function and of the $\mathcal{F}$ function are not identical so that the
corresponding spectra are quite different. However, we showed in Ref.~\citen{bsb11} that
if the potentials are different but linked by a certain relationship, one can
arrive at very interesting conclusions.

We report here the main conclusions, skipping all the rigorous proofs that can be
found in the previous reference.
Let us consider a system described by a nonrelativistic treatment based on a
two-body potential $V(r)$ whose binding energy is $E^{(N)}(m,Q)$. One knows that
\begin{equation}
\label{eq:EF5}
E^{(N)}(m,Q) = C_N \mathcal{F} \left( R \right)  \quad \textrm{with}
\quad R = \frac{N Q^2}{m C_N^2}.
\end{equation}
If one uses rather the potential $W(r) = V(\alpha \sqrt{r})$ , one has
\begin{equation}
\label{eq:EF6}
E^{(N)}(m,Q) = C_N F (S)  \quad \textrm{with} \quad S = \frac{R}{2 \alpha^2}.
\end{equation}

If we take the special value $S = 2Q/((N-1)\sqrt{C_N})$, the value $C_N F(S)$
represents the ultrarelativistic mass $M_u^{(N)}(Q)$ of the same system but
obtained with the potential $W(r)$. The condition on $S$ and the link between $S$
and $R$ leads to the condition defining the value of $\alpha$, namely
\begin{equation}
\label{eq:defalpha1b}
Q = 2m \sqrt{C_N} \alpha^2.
\end{equation}
In consequence, one can state the following theorem:
\begin{quote}
If $E^{(N)}(m,Q)$ is the binding energy of a nonrelativistic system governed
by the two-body potential $V(r)$ and if $M_u^{(N)}(Q)$ is the mass of the related
ultrarelativistic system governed by the two-body potential $W(r)$ defined by
$W(r) = V \left( \sqrt{Qr/(2m \sqrt{C_N})} \right)$, then one has the general
property $M_u^{(N)}(Q) = E^{(N)}(m,Q)$.
\end{quote}
From its definition, the potential $W(r)$ depends on $m$ so that the notation
$M_u^{(N)}(Q)$ which we have used up to now, depends indirectly on $m$, as imposed
by the theorem.

\subsubsection{Applying duality relations to exact levels}
\label{sec:appdualrel}

All the duality relations presented above are exact for the AFM solutions of
quantum systems.  One can wonder to what extent these constraints are
satisfied for the corresponding  exact solutions. We discuss briefly this point
and, again, refer the reader to Ref.~\citen{bsb11} for additional information.
We are interested essentially to the case of nonrelativistic systems.

Let us call $\epsilon^{(N)}(m;\{n_i,l_i\})$ the \textbf{exact eigenenergy} for 
state of the Hamiltonian (\ref{eq:orhamil}) with
$\{n_i,l_i\} = \{n_1,l_1,n_2,l_2,\ldots,n_{N-1},l_{N-1}\}$, while $\epsilon(m;n,l) = \epsilon^{(2)}(m;\{n,l\})$
gives the exact spectrum of the corresponding two-body problem.
The first thing to do is to connect the exact level $\epsilon^{(N)}(m;\{n_i,l_i\})$
to the corresponding AFM approximation $E^{(N)}(m,Q)$. Of course, one can always
impose an equality between both
\begin{equation}
\label{eq:eqexafm}
\epsilon^{(N)}(m;\{n_i,l_i\}) = E^{(N)}(m,Q(m;\{n_i,l_i\})).
\end{equation}
This equality defines in practice the value of the principal quantum number $Q$
that must be used in the AFM solution. The problem is that the value obtained
in that way does depend explicitly on the mass $m$, in addition to the basic
quantum numbers $\{n_i,l_i\}$. This is contrary to the philosophy of the AFM
approach for which $Q$ depends on $\{n_i,l_i\}$ only. To apply AFM formulas,
one must give up the dependence on $m$ for $Q$ and choose a dependence on quantum
numbers only, $Q(\{n_i,l_i\})$. With such a constraint, the equality
(\ref{eq:eqexafm}) does not hold anymore and is replaced by an approximate
value
\begin{equation}
\label{eq:aproxexafm}
\epsilon^{(N)}(m;\{n_i,l_i\}) \approx E^{(N)}(m,Q^{(N)}(\{n_i,l_i\})).
\end{equation}
The cleverness of the physicist is to guess the form for $Q^{(N)}(\{n_i,l_i\})$
which makes the previous approximation as precise as possible. The choice
(\ref{QNbody}) is the most natural one since it follows directly from the
construction of the AFM levels. However, the corresponding approximation
is not always good and, moreover, it is plagued by degeneracies which are
absent in the real physical spectrum. A choice like the one proposed above
(\ref{eq:princnumod}) could be much better in this respect, but very often
the determination of the parameters entering it needs a preliminary study
of the problem. Let us suppose that we have the technical ability to choose
a practical and correct form for the principal quantum number.

The idea of the method is to approximate the exact values by the AFM ones
and to take the exact duality relations on the AFM results to extend them
to the true states. This can be done in two steps
\begin{enumerate}
\item the link between the excited state of the $N$-body problem to
the ground state of another $N$-body problem;
\item the link between the ground state of the $N$-body problem to
the ground state of a two-body problem which can be solved easily.
\end{enumerate}
These two steps are the consequence of the duality relations given above
and their proofs can be found in Ref.~\citen{bsb11}. Here we focus essentially
on the practical result.

\begin{figure}[htb]
\centerline{
\includegraphics*[height=6cm]{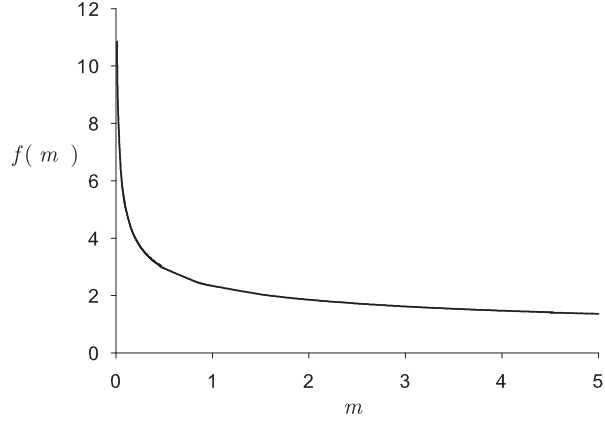}
}
\caption{Universal function $f(m)$ as a function of the mass $m$ for
the ground state of the two-body problem with $V(r)=r$.}
\label{fig:fm}
\end{figure}

The two-body problem can be solved exactly (at least numerically) for
two particles of mass $m$ interacting through the potential $V(r)$. Thus,
given the potential $V$, the ground sate energy is function of the mass
only. Let us define the $f$ function by
\begin{equation}
\label{eq:deffuncf}
f(m) = \epsilon(m;0,0).
\end{equation}
This function is \textbf{universal} in the sense that it depends only on the
form of the potential $V(r)$ and it can be computed once for all. An example
of the function $f(m)$ for the Hamiltonian $H=\bm p^2/m+r$ is given in
Fig.~\ref{fig:fm}.

The conclusion of the two steps indicated previously can be summarized
by the following approximation
\begin{equation}
\label{eq:exprexfM}
\epsilon^{(N)}(m;\{n_i,l_i\}) \approx C_N f(M(m;\{n_i,l_i\})
\end{equation}
with the following expression for the mass $M$ entering as the argument in
the $f$ function:
\begin{equation}
\label{eq:defMassnl}  
M(m;\{n_i,l_i\}) = \frac{2m}{N} \left( \frac{C_N Q^{(2)}(0,0)}
{Q^{(N)}(\{n_i,l_i\})} \right)^2.
\end{equation}
The conclusion of the relation (\ref{eq:exprexfM}) is very strong. It proves that
\textbf{the whole spectrum of all systems} can be obtained \textbf{approximatively}
by the calculation of a universal function, $f(m)$, corresponding to the ground
state of the 2-body system with the same potential, and for arguments
$M(m;\{n_i,l_i\})$ given by (\ref{eq:defMassnl}). Getting $f(m)$ is a very easy task. 
Solving the 2-body system can be performed with a great accuracy for any potential
(let us recall that this potential must not depend on $m$ and $N$); moreover, obtaining
the ground state energy is free from possible numerical complications arising
for excited states. Let us note also that the duality relation (\ref{eq:exprexfM}),
which is obviously an approximation, concerns the exact eigenvalues only; the AFM
values, which were very convenient intermediate quantities in our demonstration, have
completely disappeared.
The accuracy of this duality relation for exact states has been tested with success
in Ref.~\citen{bsb11} for some systems.

\section{Conclusions}
\label{secp:conclus}

In this paper, we presented a detailed review of the auxiliary field method (AFM)
to solve eigenequations 
in the context of quantum mechanics. We showed that this method is very powerful
to obtain closed analytical approximate expressions for the properties of many
physically interesting systems.

The idea of the method is to replace a Hamiltonian $H$ for which analytical solutions are
not known by another one $\tilde H$ for which they are known. For instance, a potential $V(r)$ 
not solvable is replaced by another one $P(r)$ more familiar, or a semirelativistic
kinetic part is replaced by an equivalent nonrelativistic one.
The bridge between both Hamiltonians is ensured by a special function 
including one or more auxiliary fields. These fields can be determined
by an extremization procedure and Hamiltonian $\tilde H$ reduces to Hamiltonian $H$. 
In so doing the original problem is
completely equivalent to the new one. The approximation comes from the replacement
of the auxiliary fields by pure real constants. The approximant solutions for $H$ are then
obtained by the solutions of $\tilde H$ in which the auxiliary parameters are eliminated
by an extremization procedure for the eigenenergies.

The great advantage of this method
is that it is able to predict the behavior of the observables in terms of the
various parameters entering into the problem and also in terms of 
the quantum numbers. We presented the principles of
the method, its characteristics, its major properties and we discussed the
quality of the results for many relevant situations appearing in various domains
of physics. The only restriction is the fact that, for relativistic systems or
many-body problems, the computation is manageable only for
systems composed of identical particles.

Although we focused essentially on the search for closed formulae of the eigenenergies, 
we showed that the method is also able to provide good results
for the eigenstates and for different types of observables. In this
context the harmonic oscillator wavefunctions and the hydrogen-like 
wavefunctions are particularly important. As an example, we proved that the AFM is very 
simple to implement for the Schr\"{o}dinger equation with a linear potential and gives
very satisfactory results.

The main part of the paper deals with the approximate values coming from the AFM
concerning the binding energies of the Schr\"{o}dinger equation (nonrelativistic)
and the eigenmasses of the spinless Salpeter equation (semirelativistic). We
proved a number of very important and pleasant general features of the method:
\begin{itemize}
  \item The AFM is essentially a kind of mean field approximation. In particular,
using a power-law form for $P(r)$,
one can define a ``mean radius" $r_0$ and a ``mean momentum" $p_0$ which
allows to express the final AFM eigenvalue in the very simple form $T(p_0)+V(r_0)$, where 
$T(|\bm p|)$ and $V(|\bm r|)$ are respectively the kinetic and potential part of $H$. 
A very simple link exists between $p_0$ and $r_0$. This last parameter is the 
solution of a transcendental equation which looks like a semiclassical version 
of a generalized virial theorem.
  \item Even in complicated situations (for example $N$-body systems with relativistic
kinematics, and presence of both one and two-body interactions), the AFM can always be
brought to the resolution of a transcendental equation, a procedure quite easy from
the numerical point of view. In many cases, this transcendental equation can be
solved analytically and the corresponding results have been presented in this paper.
  \item We discovered, and demonstrated, an amazing property of the AFM solutions that we
called ``universality of the form". This means that, provided a power-law potential 
is chosen for the function $P(r)$, the AFM gives an universal expression of the 
resulting energies for any potential $V(r)$ under consideration, whatever
the exponent chosen for power-law potential. The only remnant of this exponent
is entirely contained in the form of a principal quantum number $Q(n,l)$. This
property of universality is valid not only for the eigenvalues but also for 
the parameter $r_0$ and is
maintained passing from a nonrelativistic treatment to a semirelativistic one.
  \item The AFM is fully equivalent to the envelope theory (ET) and, as such, it can
benefit of all the properties already demonstrated in the framework of this method
which was introduced a long time ago. In particular, in many occasions, one can
deduce that the AFM energies are upper or lower bounds on the exact energy depending
on the convexity of a certain function linked to the potential. As in the ET, the
energy function (potential and/or kinetics) appearing in $\tilde H$ is, in a sense, 
the best possible energy function tangent to the exact one. But, in the AFM, 
we also pushed the theory in tracks that have not been explored by ET.
  \item If we consider only the AFM approximation for the energy, we were able to
prove several exact duality relations between different types of problems.
For instance, a nonrelativistic problem with a potential $V(r)$ can be reduced to the
ultrarelativistic ($m$ = 0) problem with a related potential $W(r) = V(\alpha
\sqrt{r})$, and the mass of a $N$-body system with identical particles is related to the mass
of a two-body system with a simple renormalization of the parameters. Moreover, these 
duality relations are shown approximately verified for exact solutions.
\end{itemize}

The 2-body Schr\"{o}dinger equation with a power-law potential was the prototype 
for the starting potential $P(r)$. The resulting AFM energies are particularly simple. Using the
``universality of the form" we proposed expressions for the principal quantum numbers
$Q(n,l)$ which allow to improve drastically the quality of the results. 
In some cases, a sophisticated expression for $Q$ gives a relative 
accuracy of the order of $10^{-4}$ over tens of the lowest sates of the spectrum, while even a very simple
prescription is already able to give an accuracy of the order of $10^{-2}$. Since the AFM
can be used recursively, the power-law potential is very often chosen as the basic
potential $P(r)$.

In the framework of a nonrelativistic approach, we applied the AFM to a great number
of 2-body problems with potentials of various forms: sum of powers, square root,
exponential potentials. In each case the AFM reproduces in a very simple way the basic properties
of the solutions, even when the starting interaction $P(r)$ is very far from the
genuine potential $V(r)$. We were able to give a closed form
for the energies of two particles interacting via a funnel potential. This situation is quite
realistic in hadronic systems and, to our knowledge, it is the first time that an
analytical expression for the corresponding eigenenergies is given.

The AFM gives a very general formula for the critical coupling constants of nonrelativistic Hamiltonians with a finite number of bound states. The dependence on the quantum numbers, the mass $m$ of the particles, the number $N$ of particles, and the structure of the potential are predicted. Different $N$ behaviours are obtained depending on the one-body or pairwise character of the interaction. If the AFM gives upper (lower) bounds for the exact eigenvalues, the critical coupling constants predicted are upper (lower) bounds for the exact critical coupling constants. 

The 2-body Salpeter equation is also tested for a variety of potentials: power laws,
square root, funnel. We showed that, at the limit of a large mass, the
nonrelativistic expression is recovered in each case. The ultrarelativistic limit
leads to particularly simple expressions. The expression for the funnel potential is
really astonishing of simplicity in the case of massless particles.

Once a problem is solved within the AFM (quantities $p_0$ and $r_0$ found), it is very easy to compute the contribution of a small perturbation at the first order. It is given by the perturbation Hamiltonian evaluated at the mean momentum $p_0$ for a kinetic energy or at a function of the mean radius $r_0$ for a potential. The result does not coincide with the one obtained by the quantum perturbation theory, but the agreement can be very good. 

Lastly the AFM is able to provide analytical results for the $N$-body problems even for quite
sophisticated types of potential: Coulomb, linear, funnel interactions are presented. Although the accuracy
would probably become worse and worse with increasing values of $N$, we showed that the spectrum of
a realistic 3-body system can be reproduced with a $1\%$ accuracy if a good definition
of the principal quantum number $Q$ is adopted. Let us mention also that the AFM $N$-body results 
were recently applied to the computation of light baryon masses for various theories of QCD with 
a large number of colors. \cite{barlnc,barlnc2}

In summary, the auxiliary field method is easy to use, can be applied to a great varieties
of problems in quantum mechanics regardless the number of particles, 
exhibits remarkable properties and is able to provide closed
analytical expressions of sometimes very impressive accuracy.

\appendix

\section{The envelope theory}
\label{sec:et}

\subsection{Presentation of the envelope theory}
\label{sec:presET}

As the AFM, the envelope theory (ET) \cite{env0} is a method aiming to get approximate 
analytical energy formulae from an arbitrary Hamiltonian. Initially, it has been
introduced to obtain bounds on the eigenenergies of the $N$-body problem (see
Ref.~\citen{hall1980}). Later, it has been explored and refined to simpler systems and to
other purposes.
We only present here its key features and refer the reader to 
Refs.~\citen{env0,env4,env,env2} for a detailed discussion about the 
basis and applications of ET.

Let us set $V(r)=v\, f(r)$ in Hamiltonian~(\ref{eq:formpartH}). Then the energy 
spectrum of this Hamiltonian is formally given by $E=F(v)$, where the 
dependence on the usual quantum numbers $n$ and $l$ will be dropped for 
simplicity. The function $F(v)$ is concave but not necessarily monotonic. 
This allows to define a so-called kinetic potential $k(S)$ by using the 
Legendre transformation (here 
the prime denotes the derivative with respect to $v$)
\begin{equation}\label{legendre}
    k(S)=F'(v)\quad {\rm and}\quad S=F(v)-v\, F'(v).
\end{equation}
This transformation can be understood as follows. 
$\left|\Psi\right\rangle$ being the eigenstates of 
Hamiltonian~(\ref{eq:formpartH}), one can define $S=\left\langle 
\Psi\right|T(\bm p^2)\left|\Psi\right\rangle$ and rewrite formally 
the energy spectrum as $F(v)=S+v\, \left\langle 
\Psi\right|f(r)\left|\Psi\right\rangle\equiv S+v\, k(S)$. The 
transformation~(\ref{legendre}) follows from these relations. One is 
consequently led to the exact formula
\begin{equation}
    E = F(v)=\min_{S>0}\left[S+v\, k(S)\right].
\end{equation}

What can now be done to go a step further in ET is to assume that 
$V(r)=g(P(r))$, where $P(r)$ is a potential for which the solution of 
the eigenequation
\begin{equation}\label{appro}
    \left[T(\bm p^2)+v\, P(r)\right] 
\left|\Psi_A\right\rangle=\epsilon_A(v)\, \left|\Psi_A\right\rangle
\end{equation}
is analytically known. Then,  
\begin{equation}
    s=\left\langle\Psi_A\right| T(\bm p^2)\left|\Psi_A\right\rangle
\end{equation}
can be analytically computed. It can moreover be shown that the kinetic 
potential corresponding to $V(r)$, namely $K(s)$, is given approximately by
\begin{equation}
    K(s)\approx g(k_A(s)),
\end{equation}
where $k_A(s)$ is the kinetic potential associated to $P(r)$. One then 
obtains an approximate form for the eigenenergies that reads \cite{env0}
\begin{equation}\label{ET1}
E\approx{\cal E} = \min_{s>0}\left[s+g(k_A(s))\right].    
\end{equation}
The variable $s$ actually plays the role of a variational parameter. 
But, thanks to (\ref{appro}), the following equalities hold 
\begin{equation}
\epsilon_A(v)=s+v\, k_A(s),\quad \epsilon_A'(v)=k_A(s),
\end{equation}
and another approximate energy formula coming from the rewriting of 
(\ref{ET1}) is
\begin{equation}\label{ET2}
    {\cal E}=\min_v 
\left[\epsilon_A(v)-v\,\epsilon_A'(v)+g(\epsilon_A'(v))\right].
\end{equation}
This last formula is called the principal envelope formula in 
Refs.~\citen{env,env2}.

It is possible to understand (\ref{ET2}) as follows. If $V(r)=g(P(r))$, with
$g(x)$ a smooth function of $x$, then we can define
the ``tangential potential" $V^t(r)$ at the point $r=t$ as 
\begin{eqnarray}\label{ET4}
	V^t(r)&=&a(t)\, P(r)+g(P(t))-a(t)\, P(t) \nonumber\\
	&& \textrm{with}\quad a(t)=\frac{V'(t)}{P'(t)}=g'(P(t)).
\end{eqnarray}
Such a particular form is obtained by demanding that $V^t(r)$ and its derivative
agree with $V(r)$ and $V'(r)$ at the point of contact $r=t$. If $\varepsilon\ll 1$,
one has indeed
\begin{equation}\label{vapp}
	V(t+\varepsilon)-V^t(t+\varepsilon)=\frac{ \varepsilon^2}{2}P'(t)^2\, g''(P(t))+O(\varepsilon^3).
\end{equation}
The eigenenergies of Hamiltonian $H^t=T(\bm p^2)+V^t(r)$, denoted by ${\cal E}(t)$, are given by
\begin{equation}\label{ET3}
	{\cal E}(t)=\epsilon_A(a(t))+g(P(t))-a(t)\, P(t).
\end{equation}
Let us now set 
\begin{equation}
	t=a^{-1}(v).
\end{equation}
It can be computed from (\ref{ET4}) that $a^{-1}(v)=P^{-1}(A(v))$ with $A(v)=g'^{-1}(v)$,
and (\ref{ET3}) becomes
\begin{equation}\label{enenv2}
	{\cal E}(v)=\epsilon_A(v)+g(A(v))-v\, A(v).
\end{equation}
The final energy spectrum has to be extremized with respect to $v$, so we have
\begin{equation}
	\left.\partial_v {\cal E}(v)\right|_{v=v_0}=0\Rightarrow A(v_0)=\epsilon'_A(v_0)
\end{equation}
and the physical energy reads 
\begin{equation}
	{\cal E}(v_0)=\epsilon_A(v_0)+g(\epsilon_A'(v_0))-v_0\, \epsilon_A'(v_0),
\end{equation}
that is nothing else than the principal envelope formula~(\ref{ET2}). 

We have just shown that ET can lead to analytical approximate energy formulae, namely
(\ref{ET1}) and (\ref{ET2}), which are both equivalent. Moreover, it has been shown in
Ref.~\citen{env0}, as suggested by (\ref{vapp}), that ${\cal E}$ is a lower (upper) bound on
the exact energy if the function $g$ is convex (concave), that is if $g''>0$ $(g''<0)$.
Let us note that $H$ and $H^t$ have the same kinetic part. The tangential potential
indeed always underestimates (overestimates) the exact potential in this case. A clear
interest of ET is thus that it allows to know the variational or antivariational nature
of the approximation that is performed. In practice, this information can be obtained
only for a nonrelativistic kinematics, since it is necessary to know exactly
$\epsilon_A(v)$ in (\ref{appro}). 

\subsection{Equivalence between AFM and ET}
\label{sec:equivAFMET}

The similarity of the starting points of ET and the AFM is obvious: In both cases, a
potential for which no analytical solution is known is ``approximated" by an other potential
for which analytical solutions exist. It suggests that a connection between both approaches
should exist; and it will indeed be established in this section. Let us apply the AFM as
described above with $V(r)=g(P(r))$. We find the following expression for the
energy~(\ref{eq:enerprop}) 
\begin{equation}\label{enenv}
	E(\nu)=\epsilon_A(\nu)+g\left(P(J(\nu))\right)-\nu\, P(J(\nu)),
\end{equation}
the function $J(x)=K^{-1}(x)$ being computed from the relation~(\ref{eq:funcK}). Remarkably,
this AFM formula is equal to the ET one (\ref{enenv2}) since $J(x)=P^{-1}(g'^{-1}(x))$.
Consequently, the AFM and the ET lead to the same final energy formula (\ref{ET1}). The link
between both approaches is given by 
\begin{equation}
	\nu=a(t).
\end{equation}
Moreover, with the point $r_0$ defined by the relation $r_0=J(\nu_0)$, the potential
$\tilde V(r)$ takes the form
\begin{equation}
	\tilde V(r)=K(r_0)\, \left( P(r)-P(r_0)\right) + V(r_0). 
\end{equation}
It is then easy to see that $\tilde V(r_0)=V(r_0)$ and that $\tilde V'(r_0)=V'(r_0)$. So,
the potential $\tilde V(r)$ is tangent to the potential $V(r)$. An explicit example is
presented in Ref.~\citen{afmenv}.

The function $J(x)$ can be defined if the function $K(x)$ can be inverted. In order to
fulfill this condition, it is sufficient that $K(x)$ is monotonic,
that is to say that $K'(x)$ has a constant sign. But, from the definitions above, we have
$K(x)=g'(P(x))$, which implies that $K'(x)=g''(P(x))\, P'(x)$. Since $K(x)$ must be monotonic,
the convexity of the function $g$ is well defined if $P(x)$ is also monotonic. This is the
case if $P(x)$ is a power-law potential, for instance. In these conditions, the convexity
of the function $g$ can also be used to determine the variational character of the AFM. 

Let us summarize our results. The auxiliary field $\nu$ can be introduced as an operator
in the Hamiltonian~(\ref{eq:formpartH}), and leads to an equivalent formulation of this
Hamiltonian. If one considers it as a variational parameter rather than an operator, as in
the AFM, the results are approximate but can be analytical. We have shown in this section
that the auxiliary field, when seen as a variational parameter, is nothing else than the
function $a(t)$ generating the tangential potential in ET. This shows that, although
obtained in different ways, the AFM and the ET lead to the same results. In this way,
some formulae about the power-law potentials obtained in Ref.~\citen{env3} by the ET were
rediscovered with the AFM in Ref.~\citen{bsb08a}, but supplementary results are given in this
last reference. 

Taking this equivalence into account, we can now better understand the meaning of the
variational parameter $v$ in the ET: Its optimal value can be seen as given by a mean field
approximation since $Z(v_0)=\left\langle Z(g'(P(r_0)))\right\rangle$ (see (\ref{Zrho})).
Moreover, the properties of the AFM that have been proven in Refs.~\citen{bsb08a,bsb08b} also hold
for ET. Finally, we can now have an \emph{a priori} knowledge of the (anti)variational nature
of the AFM energy formulae provided that we express $V(r)$ as $g(P(r))$ and compute whether
$g$ is convex or concave. Equivalently, since the potential $\tilde V(r,\nu_0)$ is tangent to
the potential $V(r)$ at $r=r_0$, the approximation $E(\nu_0)$ is an upper (lower) bound on
the exact energy if $\tilde V(r,\nu_0) \ge V(r)$ ($\tilde V(r,\nu_0) \le V(r)$) for all
values of $r$. \cite{afmenv}\ All these results holds provided that $H$ and $\tilde H(\nu_0)$
have the same kinetic part. Several examples are presented above. 

\section{Reduced equations}
\label{sec:Redeq}

Finding analytical energy formulae for the potentials that we study in this
work requires an analytical knowledge of the roots of particular cubic and
quartic polynomial equations. In each case of interest, algebraic manipulations
allow to transform the original equations to one of the following reduced
equations. We sum up these equations in this appendix and put their roots
in a form that is as convenient as possible to deal with.

\subsection{Third order equation}
\label{sec:Thirdeq}
We begin by the cubic equation ($Y \ge 0$)
\begin{equation}
\label{eq:redcubeq1}
x^3 \pm 3x - 2Y =0,
\end{equation}
for which there exists only one positive root given by
\begin{equation}
\label{eq:rootcubeq}
F_\pm(Y) = \left(Y + \sqrt{Y^2\pm 1} \right)^{1/3} \mp \left(Y + \sqrt{ Y^2\pm1}
\right)^{-1/3}.
\end{equation}
Written in the above form, it seems that $F_{-}(Y)$ is not properly defined
for $Y < 1$. But, for this range of $Y$ values, one can show that
\begin{equation}
F_{-}(Y) = 2 \cos\left( \frac{1}{3} \arccos Y \right).
\end{equation}
So $F_{-}(Y)$ is well defined for all positive values of its argument. 
It can be checked that the following approximate forms hold
\begin{eqnarray}
\label{eq:behYcsmall1}
F_+(Y) \approx \frac{2Y}{3}, \quad F_-(Y) \approx \sqrt 3+\frac{Y}{3} \quad 
&\textrm{if}& \quad Y \ll 1, \\
\label{eq:behYcsmall3}
F_\pm(Y) \approx (2Y)^{1/3} \quad &\textrm{if}& \quad Y \gg 1.
\end{eqnarray}

\subsection{Fourth order equation}
\label{sec:Fourtheq}

The quartic equation which gives the most pleasant form for the roots is
($Y \ge 0$)
\begin{equation}
\label{eq:redcubeq2}
4 x^4 \pm 8x - 3Y =0.
\end{equation}
There exists only one positive root given by
\begin{equation}
\label{eq:rootquarteq}
G_{\pm}(Y) = \mp \frac{1}{2} \sqrt{V(Y)} + \frac{1}{2} \sqrt{ 4 (V(Y))^{-1/2}
- V(Y)},
\end{equation}
with
\begin{equation}
\label{eq:defVY}
V(Y)=\left(2 + \sqrt{4 + Y^3} \right)^{1/3} -  Y\left(2 + \sqrt{4 + Y^3}
\right)^{-1/3}.
\end{equation}
The following approximate expressions can also be useful 
\begin{eqnarray}
\label{eq:behYqsmall1}
G_{+}(Y) \approx \frac{3Y}{8}, \quad G_{-}(Y) \approx 2^{1/3} + \frac{Y}{8} 
\quad &\textrm{if}& \quad Y \ll 1, \\
\label{eq:behYqsmall3}
G_{\pm}(Y) \approx \left( \frac{3Y}{4}\right)^{1/4} \quad &\textrm{if}& \quad Y \gg 1.
\end{eqnarray} 

\section{Lambert function}
\label{sec:lambert}

\begin{figure}
\centerline{
\includegraphics*[width=8cm]{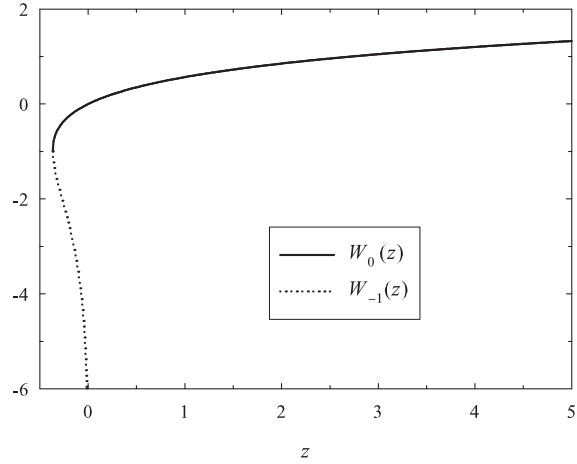}
}
\caption{Plot of the two branches of the Lambert function,
namely $W_0(z)$ (solid line) and $W_{-1}(z)$ (dotted line). }
\label{fig1} 
\end{figure} 

Let us briefly recall some points concerning the Lambert function
(also called Omega function or product-log), that we will denote $W(z)$.
Complements of information can be found in Ref.~\citen{corl96}. First of all, $W(z)$
is defined as the inverse function of $z\, \textrm{e}^z$. Consequently,
it has the following properties:
\begin{equation}
\label{defLamb}
	W(z)\, \textrm{e}^{W(z)}=W\left(z\, \textrm{e}^z\right)=z,
\end{equation}
\begin{equation}
\label{derivLamb}
	\partial_z\, W(z)=\frac{W(z)}{z\left[1+W(z)\right]}.
\end{equation}
But, it is readily observed that the inverse function of $z\, \textrm{e}^z$ is
multivalued. Two branches of the Lambert function thus exist \footnote{Notice
that the two branches $W_0(x)$ and $W_{-1}(x)$ of the Lambert function $W(x)$
are known by the software \emph{Mathematica} package as {\ttfamily
ProductLog[0,x]} and {\ttfamily ProductLog[-1,x]} respectively.}, respectively
denoted as $W_0(z)$, defined for $z\geq-1/\textrm{e}$, and $W_{-1}(z)$, defined
for $-1/\textrm{e}\leq z\leq0$ They are plotted in figure~\ref{fig1}.
Obviously, both branches of the Lambert function share the same properties
(\ref{defLamb}) and (\ref{derivLamb}) as $W(z)$; they meet in
$W_{-1}(-1/\textrm{e})=W_0(-1/\textrm{e})=-1$. It can moreover be checked that
$W_0(|x|\ll 1)\approx x$, $\lim_{x\to 0}W_{-1}(x)=-\infty$, and
$\lim_{x\to \infty}W_{0}(x)=\infty$.

For our purpose, it is worth mentioning that the following equation in which
$a$, $b$, $\rho$ are real numbers such that $a\, z+b$ is positive
\begin{equation}\label{resol}
	(a\, z+b)^\rho\, \textrm{e}^{-z}=\theta
\end{equation}
is analytically solvable with the Lambert function. One has 
\begin{eqnarray}\label{solulamb}
	z&=&-\frac{b}{a}-\rho\, W\left[-\frac{1}{a\, \rho}
(\textrm{e}^{-b/a}\theta)^{1/\rho}\right],
\end{eqnarray}
where either $W_0$ or $W_{-1}$ has to be chosen following the range
of $\theta$ and $z$ that the function $z(\theta)$ has to cover.

\section{Extended virial theorem}
\label{sec:genvirialth}

In this section and in the following ones, we will denote $\langle O \rangle =
\langle n,l| O |n,l\rangle$ and $\langle n| O |n \rangle = \langle n,0| O |n,0\rangle$,
$|n,l\rangle$ being an eigenstate of $H$ with a radial quantum number $n$ and an
orbital quantum number $l$. If $A$ is an arbitrary operator, it follows that
\begin{equation}\label{meanHA}
\langle [H,A] \rangle=0,
\end{equation}
due to the hermiticity of $H$. Note that the mean value is calculated on the
\emph{exact} eigenstate of $H$. If $p_r$ is the radial momentum (with $[r,p_r]=i$)
and $f(r)$ an arbitrary function depending on $r = |\bm r|$ only, the computation
of (\ref{meanHA}) with $A=p_r f(r)$ yields to the following relation, called the
extended virial theorem
\begin{equation}\label{vir1}
\left\langle 2 E f'(r)-V'(r) f(r) - 2 V(r) f'(r) 
+ \frac{f'''(r)}{4 m} - \frac{l(l+1)}{m r} \left( \frac{f(r)}{r} \right)'
\right\rangle =0,
\end{equation}
One recovers the usual virial theorem (see Ref.~\citen{lucha}) for the
special choice $f(r)=r$. But new interesting relations can be obtained for other
choices. In particular, if $f(r)=r^{s+1}$ the previous equation becomes
\begin{equation}\label{vir3}
2 (s+1) E \langle r^s \rangle - 2 (s+1) \langle r^s V(r) \rangle - \langle r^{s+1}
V'(r) \rangle 
+ \frac{s}{4 m} \left( s^2-1-4 l (l+1) \right) \langle r^{s-2} \rangle = 0.
\end{equation}
Finally, if, in addition, $V(r)=\textrm{sgn}(\lambda) a \,r^\lambda$ for
$\lambda \ne 0$, (\ref{vir3}) reduces to
\begin{equation}\label{vir5}
2 (s+1) E \langle r^s \rangle - a\, \textrm{sgn}(\lambda)(2 s+\lambda+2)
\langle r^{\lambda+s} \rangle  
+ \frac{s}{4 m} \left( s^2-1-4 l (l+1) \right) \langle r^{s-2} \rangle = 0.
\end{equation}
This recurrence relation is particularly useful to compute $\langle r^k
\rangle$ mean values for potentials with integer power, choosing $s$ as an integer.

\section{Observables with harmonic oscillator functions}
\label{sec:obs_OH}
The eigenenergies of the harmonic oscillator Hamiltonian \cite{flug}
\begin{equation}\label{HOH}
H = \frac{\bm p^2}{2 m}+  \nu \, r^2
\end{equation}
are given by
\begin{equation}\label{EOH}
E = \sqrt{\frac{2 \nu}{m}} Q_{HO} \quad \textrm{with} \quad Q_{HO}=2 n +l+\frac{3}{2}
\end{equation}
and the normalized eigenvectors by
\begin{equation}\label{psiOH}
\psi_{nl\mu}(\bm r)=\lambda^{3/2}\sqrt{\frac{2 \; n!}{\Gamma(n+l+3/2)}}\left( 
\lambda r \right)^l e^{-\lambda^2 r^2/2}L_{n}^{l+1/2}(\lambda^2 r^2) Y^l_\mu(\hat{\bm{r}}),
\end{equation}
with $\lambda=\left( 2 m \nu \right)^{1/4}$. $L_{\alpha}^{\beta}$
is a Laguerre polynomial and $Y^l_\mu$ a spherical harmonic. 
At the origin, the S-states are such that 
\begin{equation}\label{psi0OH}
|\psi_{n,0}(0)|^2=\lambda^3 \frac{2 \Gamma(n+3/2)}{\pi^2 n!}.
\end{equation}
Mean values $\langle r^k \rangle$ can be computed by performing directly the integrals
or by using (\ref{vir5}). One obtains
\begin{equation}\label{rkHOgen}
\langle r^k \rangle = \frac{1}{\lambda^k}\frac{\Gamma(n+l+3/2)}{n!}\sum_{p,q=0}^n
(-1)^{p+q}C_n^p C_n^q \frac{\Gamma(l+p+q+(k+3)/2)}{\Gamma(p+l+3/2) \Gamma(q+l+3/2)},
\end{equation}
where $C_\alpha^\beta$ is the usual binomial coefficient.
One can also write, with $L = l(l+1)$,
\begin{alignat}{3}\label{rkOH}
&\langle n|r |n\rangle = \frac{1}{\lambda} \frac{4 \Gamma(n+3/2)}{\pi n!}, \quad
&&\langle n,l|r^2 |n,l\rangle = \frac{Q_{HO}}{\lambda^2},  \nonumber \\
&\langle n|r^3 |n\rangle = \frac{1}{\lambda^3} \frac{8 (4 n+3) \Gamma(n+3/2)}
{3 \pi n!}, \quad
&&\langle n,l|r^4 |n,l\rangle = \frac{1}{4 \lambda^4} \left( 6 Q_{HO}^2 - 2 L
+ \frac{3}{2} \right).
\end{alignat}
Using the Fourier transform of a harmonic oscillator, it is easy to show that 
\begin{equation}\label{pkOH}
\langle p^k \rangle = \lambda^{2 k} \langle r^k \rangle .
\end{equation}

\section{Observables with hydrogen-like functions}
\label{sec:obs_Hy}
The eigenenergies of a hydrogen-like Hamiltonian \cite{flug}
\begin{equation}\label{HHy}
H = \frac{\bm p^2}{2 m} -  \frac{\nu}{r},
\end{equation}
are given by
\begin{equation}\label{EHy}
E = -\frac{m \nu^2}{2 Q_{C}^2}  \quad \textrm{with} \quad Q_{C}= n+l+1
\end{equation}
and the normalized eigenvectors by
\begin{equation}\label{psiHy}
\psi_{nl\mu}(\bm r)=\left(2 \gamma_{nl}\right)^{3/2}\sqrt{\frac{n!}
{2(n+l+1)(n+2 l+1)!}}\left( 2 \gamma_{nl} r \right)^l e^{-\gamma_{nl} r}
L_{n}^{2 l+1}(2 \gamma_{nl} r) Y^l_\mu(\hat{\bm{r}}),
\end{equation}
with $\gamma_{nl}= \eta/(n+l+1)$ and $\eta= m \nu$. At the origin,
the S-states are such that 
\begin{equation}\label{psi0Hy}
|\psi_{n,0}(0)|^2=\frac{\eta^3}{\pi (n+1)^3}.
\end{equation}
Mean values $\langle r^k \rangle$ can be computed by performing directly the
integrals or by using (\ref{vir5}). One obtains
\begin{equation}\label{rkHygen}
\langle r^k \rangle = \frac{(n+l+1)^{k-1}}{2(2\eta)^k}\frac{(n+2 l+1)!}{n!}
\sum_{p,q=0}^n (-1)^{p+q}C_n^p C_n^q \frac{(p+q+k+2 l+2)!}{(p+2 l+1)!(q+2 l+1)!}.
\end{equation}
One can write also, with again $L = l(l+1)$, 
\begin{alignat}{3}\label{rkHy}
&\left\langle \frac{1}{r} \right\rangle = \frac{\eta}{Q_{C}^2}, \quad
&&\left\langle \frac{1}{r^2} \right\rangle = \frac{2 \eta^2}{(2 l+1)Q_{C}^3},
\nonumber \\
&\langle r \rangle   = \frac{1}{2 \eta} \left( 3 Q_{C}^2 - L \right), \quad
&&\langle r^2 \rangle = \frac{Q_{C}^2}{2 \eta^2} \left( 5 Q_{C}^2 - 3 L + 1
\right), \nonumber \\
&\langle r^3 \rangle = \frac{Q_{C}^2}{8 \eta^3} \left( 35 Q_{C}^4 + 5 Q_{C}^2
(5 -6 L) + 3 L (L-2) \right), &&\nonumber \\
&\langle r^4 \rangle = \frac{Q_{C}^4}{8 \eta^4} \left( 63 Q_{C}^4 + 35 Q_{C}^2
(3 -2 L) + 5 L (3 L-10) +12 \right).
\end{alignat}
Using the virial theorem and the square of the Hamiltonian, it is easy to show that 
\begin{equation}\label{pkHy}
\langle p^2 \rangle = \frac{\eta^2}{Q_{C}^2}, \quad
\langle p^4 \rangle = \eta^4 \frac{8 n + 2 l+5}{(2 l+1)Q_{C}^4}.
\end{equation}

\section{Observables with Airy functions}
\label{sec:obs_Ai}

The eigensolutions with $l=0$ of the Hamiltonian
\begin{equation}\label{HAi}
H = \frac{\bm p^2}{2 m}+  a \, r
\end{equation}
are analytically known in terms of the Airy function $\textrm{Ai}$. \cite{abra}\ 
The eigenenergies can be written in terms of the (negative) zeros $\alpha_n$ of
this function, namely 
\begin{equation}\label{EAi}
E = -\left( \frac{a^2}{2 m} \right)^{1/3} \alpha_n,
\end{equation}
and the normalized eigenvectors $\psi_{n0}(\bm r)=\langle \bm r |n\rangle$ are given by
\begin{equation}\label{psiAi}
\psi_{n0}(\bm r)=\frac{\sqrt{\kappa}}{\sqrt{4 \pi} \left|
\textrm{Ai}'(\alpha_n) \right| r}
\textrm{Ai} \left( \kappa r + \alpha_n \right),
\end{equation}
with $\kappa=( 2 m a )^{1/3}$.
An approximate form for $\alpha_n$ is given by \cite{abra} 
\begin{equation}\label{betan}
\alpha_n = -\beta_n \left( 1+ \frac{5}{48} \beta_n^{-3} - \frac{5}{36} \beta_n^{-6} +
O(\beta_n^{-9}) \right) \quad \textrm{with} \quad \beta_n =\left[ \frac{3 \pi}{2}
\left( n + \frac{3}{4}\right) \right]^{2/3},
\end{equation}
the series converging very rapidly with $n$. At the origin, the square of the
wavefunction reduces to
\begin{equation}\label{psi0Ai}
|\psi_{n,0}(0)|^2=\frac{m a}{2 \pi}.
\end{equation}
The remarkable fact is that it does not depend on the radial quantum number. This
property is specific to the linear potential.

Mean values $\langle r^k \rangle$ can be computed by performing directly the integrals
or by using (\ref{vir5}). One obtains
\begin{alignat}{3}\label{rkAi}
&\langle n|r |n\rangle = \frac{2 |\alpha_n|}{3\kappa}, \quad
&&\langle n|r^2 |n\rangle = \frac{8 |\alpha_n|^2}{15\kappa^2}, \nonumber \\
&\langle n|r^3 |n\rangle = \frac{16 |\alpha_n|^3+15}{35\kappa^3}, \quad
&&\langle n|r^4 |n\rangle = 16 \frac{8 |\alpha_n|^4+25 |\alpha_n|}{315\kappa^4}.
\end{alignat}
Using the virial theorem and the square of the Hamiltonian, it is easy to show that 
\begin{equation}\label{pkAi}
\langle n|p^2 |n\rangle = \kappa^2\frac{|\alpha_n|}{3}, \quad
\langle n|p^4 |n\rangle = \kappa^4\frac{|\alpha_n|^2}{5}.
\end{equation}

\section{Overlap with dilated functions}
\label{sec:overdil}

The scalar product of two radial functions $R_{n,l}(r)$ and $R_{n',l}(r)$
of a set of orthonormal states is simply given by $\delta_{nn'}$. When one
of these functions is scaled by a positive factor $a$, the overlap 
\begin{equation}\label{ovgen}
F_{n,n',l}(a)=a^{3/2}\int_0^\infty R_{n,l}(x)R_{n',l}(a x)x^2 dx
\end{equation}
satisfies the following properties \cite{sema95}:
\begin{eqnarray}\label{ovprop}
&&\lim_{a\to 1} F_{n,n',l}(a)=\delta_{nn'}, \nonumber \\
&&|F_{n,n',l}(a)| \le 1, \nonumber \\ 
&&F_{n,n',l}(1/a)=F_{n',n,l}(a), \nonumber \\
&&\lim_{a\to 0} F_{n,n',l}(a) = \lim_{a\to \infty} F_{n,n',l}(a)=0.
\end{eqnarray}
The first relation stems from the definition~(\ref{ovgen}), the second one from
the Schwarz inequality, and the others are due to scaling properties. 

Using the dilation properties of the Laguerre polynomials and the various
existing recurrence relations \cite{grad80}, it is possible to compute
analytically the formula $F_{n,n',l}(a)$ for hydrogen-like systems and
harmonic oscillators.

In the first case (hydrogen-like systems), one obtains
\begin{eqnarray}\label{FHy}
F^{Hy}_{n,n',l}(a)&=&(-1)^{n+n'}\sqrt{a\, n!\, (N+l)!\, n'!\, (N'+l)!}
\left( 4 a N N' \right)^N \frac{Q(a)^{n'-n}}{S(a)^{N'+N+1}} \nonumber \\
&&\times\sum_{k=0}^n (-1)^k \left(\frac{Q(a)^2}{4 a N N'}\right)^k
\frac{1}{k!\, (n-k)!\, (N-k+l)!\, (n'-n+k+1)!} \nonumber \\
&&\times \left( 2(N-k)(n'-n+k+1) + (n-k)(N-k+l)\frac{Q(a)}{2 a N}\right.  \nonumber \\
&&+\left.(n'-n+k)(n'-n+k+1)\frac{2 a N}{Q(a)} \right)
\end{eqnarray}
with $N=n+l+1$, $N'=n'+l+1$, $Q(a)=a N-N'$, and $S(a)=a N+N'$.

In the last case (harmonic oscillators), the formula is given by
\begin{eqnarray}\label{FOH}
F^{HO}_{n,n',l}(a)&=&\sqrt{n!\, n'!\, \Gamma(n+l+3/2)\, \Gamma(n'+l+3/2)}
(2 a)^{2 n+l+3/2}
\frac{\left( 1-a^2 \right)^{n'-n}}{\left( 1+a^2 \right)^{n+n'+l+3/2}}
\nonumber \\
&&\times\sum_{k=0}^n (-1)^k \frac{\left( 1-a^2 \right)^{2 k}}{(2 a)^{2 k}
k!\, (n-k)!\,
(n'-n+k)!\, \Gamma(n-k+l+3/2)}.
\end{eqnarray}

\section{Jacobi coordinates}
\label{sec:Jacobcoord}
Usually the many-body problem in the nonrelativistic framework is treated
starting with the shell-model or variants. In this method the degrees of
freedom are simply the positions $\bm r_i$ of the various particles
($i = 1,\ldots,N$). But, in this case, the motion of the center of mass
is not separated correctly and this leads to spurious components which spoil
the results. On the contrary, in the few-body problem, one introduces new
degrees of freedom which allow to solve exactly this important drawback. 
Among the various possible new degrees of freedom, the Jacobi coordinates are
of common use. First, we choose a reference mass $m$ (it can be that of a
particle or the total mass of the system for example) and define the
dimensionless quantities $\alpha_i=m_i/m$, $\alpha_{12\dots i}=\sum^i_{k=1}
\alpha_k$ and $\alpha = \alpha_{12\ldots N}$. Then, the standard Jacobi
coordinates, $\bm x_i$, can be expressed as 
\begin{equation}
\label{eq:defjacoord}
  \bm x_i=\frac{\sum^i_{k=1}\alpha_k\, \bm r_k}{\alpha_{12\dots i}}-\bm r_{i+1},
  \quad i=1,\ldots,N-1, \quad {\rm and}\quad \bm x_{N}=\bm R =
  \frac{1}{\alpha} \sum^i_{k=N}\alpha_k\, \bm r_k.
\end{equation}
For convenience, the center of mass (for a nonrelativistic system) 
coordinate $\bm R$ is relegated as the last
Jacobi coordinate $\bm x_{N}$ whereas $\bm x_i$ represents the vector
joining the center of mass (for a nonrelativistic system)
of the first $i$ particles to the particle $i+1$.
Using the matrix notation $\bm x_i=\sum^{N}_{j=1}U_{ij}\bm r_j$, one is
led to the following definition of the $U$-matrix
\begin{eqnarray}
U_{ij}&=&\frac{\alpha_j}{\alpha_{12\ldots i}} \quad {\rm if}\quad  j \leq i 
\nonumber \\
U_{ii+1}&=&-1 \\ 
U_{ij}&=&0 \quad {\rm if}\quad  j > i+1. \nonumber
\end{eqnarray}
It is not difficult to calculate the inverse matrix $B=U^{-1}$ whose matrix
elements are given by
\begin{eqnarray}
\label{defB}
B_{kl}&=&\frac{\alpha_{l+1}}{\alpha_{12\ldots l+1}} \quad {\rm if}
\quad  k \leq l <  N  \nonumber \\
B_{l+1 l}&=&- \frac{\alpha_{12\ldots l}}{\alpha_{12\dots l+1}}
\quad {\rm if} \quad  l <  N \\
B_{kl}&=&0 \quad {\rm if}\quad  k > l+1  \nonumber \\
B_{k N}&=&1 \quad \forall k. \nonumber
\end{eqnarray}
Denoting $\bm p_i$ and $\bm \pi_i$ the conjugate momenta associated
respectively to $\bm r_i$ and $\bm x_i$, it is easy to prove that
\begin{equation}
\bm \pi_i=\sum^{N}_{j=1}B_{ji}\bm p_j \quad {\rm and}
\quad \bm p_i=\sum^{N}_{j=1}U_{ji}\bm \pi_j. 
\end{equation}
$\bm \pi_N$ = $\bm P$ = $\bm p_1+\bm p_2+\ldots+\bm p_N$ is the
total momentum of the system.

\section{The $N$-body harmonic oscillator}
\label{sec:nboh}
The case of a quadratic potential (the $N$-body nonrelativistic
harmonic oscillator) is practically the only one for which an exact solution
is reachable in three dimensions, at least formally. This section is devoted to this problem, since it is
the basic ingredient of the $N$-body AFM.

\subsection{Nonrelativistic general case}
Let us start with the most general harmonic-oscillator-like Hamiltonian,
corresponding to $N$ particles of arbitrary masses, a one-body and a
two-body quadratic potentials with arbitrary spring constants, so that the
Hamiltonian looks like
\begin{equation}\label{honr}
 H_{{\rm ho}}=\sum^{N}_{i=1}\frac{\bm p^2_i}{2m_i}+\sum^{N}_{i=1}
a_i (\bm r_i-\bm R)^2+\sum^{N}_{i<j=1} b_{ij} (\bm r_i-\bm r_j)^2.
\end{equation}

The kinetic energy operator, expressed in terms of Jacobi variables, allows
the correct separation of the center of mass motion and appears decoupled in
the various variables (this is in fact the justification of the form of Jacobi
coordinates)
\begin{equation}
\label{Tcinoh}
T=\sum^{N}_{i=1}\frac{\bm p^2_i}{2m_i} = \frac{\bm P^2}{2 m_t}+
\sum^{N-1}_{i=1} \frac{\lambda_i^2}{2m} \bm \pi_i^2
\end{equation}
where $m_t = \alpha m = m_1+m_2+\cdots m_N$ is the total mass of the system.
The kinematical quantities $\lambda_i$ are calculated as
 
\begin{equation}
\label{deflambi}
\lambda_i = \left(\frac{\alpha_{12\ldots i+1}}{\alpha_{i+1}
\alpha_{12\ldots i}}\right)^{1/2}.
\end{equation}
For further convenience, it is judicious to switch from the standard Jacobi
variables to renormalized conjugate Jacobi coordinates defined as
\begin{equation}
\bm y_i = \frac{\bm x_i}{\lambda_i}; \quad \bm \rho_i = \lambda_i \bm \pi_i.
\end{equation}
Working in the center of mass frame ($\bm P= \bm 0$) and using these new
variables, the kinetic energy operator has a very simple form
\begin{equation}
T = \frac{1}{2m} \sum^{N-1}_{i=1} \bm \rho_i^2.
\end{equation}

With these new variables, the one-body operator is written
\begin{equation} 
V_1 = \sum^{N}_{i=1} a_i (\bm r_i-\bm R)^2 =
\sum^{N-1}_{l,m=1} F_{lm}\, \bm y_l \cdot \bm y_m,
\end{equation}
where the symmetric definite positive matrix $F$ is defined by
\begin{equation}
\label{defF} 
F_{lm} = \lambda_l \lambda_m \sum^{N}_{i=1} a_i B_{il} B_{im},
\end{equation}
the matrix $B$ being given by (\ref{defB}). In the very same way, the
two-body operator is written
\begin{equation} 
V_2 = \sum^{N}_{i<j=1} b_{ij} (\bm r_i-\bm r_j)^2 =
\sum^{N-1}_{l,m=1} G_{lm}\, \bm y_l \cdot \bm y_m
\end{equation}
where the symmetric definite positive matrix $G$ is defined by
\begin{equation}
\label{defG} 
G_{lm} = \lambda_l \lambda_m \sum^{N}_{i<j=1} b_{ij} (B_{il} - B_{jl})
(B_{im} - B_{jm}).
\end{equation}
Introducing the matrix $J=F+G$, the total potential $V=V_1+V_2$ is expressed as
\begin{equation} 
V = \sum^{N-1}_{l,m=1} J_{lm}\, \bm y_l \cdot \bm y_m.
\end{equation}
For arbitrary masses or/and spring constants there is no reason why the matrix
$J$ should be diagonal. However, this matrix being a symmetric definite
positive matrix, it can be diagonalized with help of a unitary matrix (in fact
an orthogonal one since all the quantities are real). Thus
\begin{equation} 
J = O^{-1}DO,\quad {\rm with} \quad O^{-1}=\tilde{O}.
\end{equation}
The elements of the diagonal matrix $D$ are all positive (due to definiteness)
and are chosen under the form $d_i = m \omega_i^2/2$.

The last step is an ultimate change of conjugate variables
\begin{equation}
\bm z_l = \sum^{N-1}_{j=1} O_{lj}\, \bm y_j ; \quad \bm \sigma_l =
\sum^{N-1}_{j=1} O_{lj}\, \bm \rho_j.
\end{equation}
Expressed with these new variables, the original Hamiltonian (\ref{honr})
appears to be the sum of $N-1$ decoupled harmonic oscillators
\begin{equation}
H_{{\rm ho}}=\sum^{N-1}_{i=1} \left[ \frac{\bm \sigma_i^2}{2m} +
\frac{1}{2} m \omega_i^2 \bm z_i^2 \right],
\end{equation}
and consequently the energy of the system is given by
\begin{equation}
\label{Eohgen}
E_{{\rm ho}}=\sum^{N-1}_{i=1}  \omega_i (2n_i+l_i+3/2),
\end{equation}
where $n_i$ and $l_i$ are respectively the radial and orbital quantum numbers
associated to the coordinate $\bm z_i$. The problem is now completely solved.
Moreover, this result is valid for any excited state. Even if the expression
of $\omega_i$ is in general not analytical, the result (\ref{Eohgen}) is exact
and could be calculated with a high accuracy. Notice that
formula~(\ref{Eohgen}) extends a previous result \cite{schw79,ma}, where an equivalent
mass formula is obtained in the case $a_i=0$. An explicit result 
for the case $N=3$ is given in Ref.~\citen{silv10}.

\subsection{Case of identical particles}\label{sec:caseofidpart}

It is of interest to rewrite the solution in the case where all the particles
are identical, implying that they have the same mass $m_i=m$ and the same
spring constants $a_i=a$, $b_{ij}= b$. Hamiltonian~(\ref{honr}) indeed reads
in this case 
\begin{equation}\label{honr3}
H_{{\rm ho}}=\sum^{N}_{i=1}\frac{\bm p^2_i}{2m}+a\sum^{N}_{i=1}
(\bm r_i-\bm R)^2+b\sum^{N}_{i<j=1} (\bm r_i-\bm r_j)^2,
\end{equation}
and it is easy to see that the $J$ matrix is already of diagonal from the very
beginning, so that its eigenvalues are analytically
known (they all read $a+Nb$). Consequently, the eigenenergies of
the system are also analytical. Explicitly they are given by
\begin{equation}\label{ehonr}
E_{{\rm ho}}=\sqrt{\frac{2}{m}(a+Nb)}\ Q	,
\end{equation}
where $Q$ is the total principal number
\begin{equation}\label{honriden}
Q=\sum^{N-1}_{i=1} (2n_i+l_i) + \frac{3}{2}(N - 1).
\end{equation}

Wavefunction with good global quantum numbers and various symmetries can be
built by appropriate linear combinations of states characterized by the same
value of the principal quantum number $Q$. \cite{silv85}\ All symmetrized states
have then the same energy than the nonsymmetrized states. For spatial wave
functions completely symmetrical, the ground state is obtained for the values
$n_i=l_i=0$ $\forall i$ so that the principal quantum number is simply
\begin{equation}\label{honrfond}
Q_{\textrm{SGS}}=\frac{3}{2}(N - 1).
\end{equation}
For mixed symmetry or completely antisymmetrical spatial wavefunctions, the
situation is much more involved. An estimation of the ground-state energy for
the completely antisymmetrical case can be computed by choosing different values
for the quantum numbers and piling the states ($d$ identical values of the same
$n_j$ and $l_j$ per state in order to take care of a possible 
degeneracy due to internal degrees of freedom) 
up to the Fermi level. By considering only particle number insuring
a saturated Fermi level (closed shell), one obtains 
\begin{equation}\label{honrfond2}
Q_{\textrm{AGS}}=\frac{3}{4}(N - 1)(B_f+2),
\end{equation}
where $B_f$, the band number of the Fermi level, is the real positive solution
of
\begin{equation}\label{Bfermi}
N - 1=\frac{d}{6}(B_f+1)(B_f+2)(B_f+3).
\end{equation}
Asymptotically, we have
\begin{equation}\label{honrfond3}
\lim_{N \gg 1} Q_{\textrm{AGS}}= \left(\frac{81}{32}\right)^{1/3}
\frac{N^{4/3}}{d^{1/3}}.
\end{equation}

\subsection{Relativistic case}

Let us now treat the relativistic generalization of~(\ref{honr3}), namely
\begin{equation}\label{hosr}
H_{{\rm ho}}=\sum^{N}_{i=1}\sqrt{\bm p^2_i+m^2}+a\sum^{N}_{i=1}
(\bm r_i-\bm R)^2+b\sum^{N}_{i<j=1} (\bm r_i-\bm r_j)^2,
\end{equation}
by the AFM. 
Extremization formula~(\ref{x00}) can be written $4X^4-8X-3Y=0$ with
\begin{equation}\label{ho1}
Y=\frac{4m^2}{3}
\left(\frac{2N^2}{(a+Nb) Q^2}\right)^{2/3}.
\end{equation}
The only positive root of this last equation is given by $X=G_{-}(Y)$
(see Sect.~\ref{sec:Fourtheq}). Finally, mass formula~(\ref{M00}) becomes
\begin{eqnarray}\label{mhosr1}
M_{{\rm ho}}(\mu_0)&=&\sqrt{\frac{3}{Y}}\frac{m_t}{2 G_-(Y)^2}(4G_-(Y)+Y)\nonumber\\
&=& \frac{2m_t}{\sqrt{3Y}}\left[\frac{1}{G_-(Y)}+G_-(Y)^2\right].
\end{eqnarray}
Notice the simple
ultrarelativistic limit
\begin{equation}\label{mhosr2}
\lim_{m\to 0}M_{{\rm ho}}(\mu_0)=\frac{3}{2}\left[2N
(a+Nb)Q^2\right]^{1/3}.
\end{equation} 
The duality relations presented in Sect.~\ref{sec:dualrel} can be easily tested with 
relations ~(\ref{ehonr}), (\ref{mhosr1}) and (\ref{mhosr2}).

\end{document}